Untersuchungen zur Modellierung und Schaltungsrealisierung von synaptischer Plastizitaet

Christian Georg Mayr

Habilitation Thesis

# Abstract

This manuscript deals with the analysis and VLSI implementation of adaptive information processing derived from biological measurements. There is currently significant attention centered on this research area by national and international funding agencies, e.g. in the framework of the DARPA-financed SYNAPSE project, in the research framework program of the EC, in the German Bernstein network, etc. The main focus of this manuscript is on the information processing functions carried out at the coupling points (synapses) between cortical neurons. Synaptical processing functions on various timescales are investigated singly and with respect to their interaction, starting from dynamic transfer functions (short term plasticity) through learning (long term plasticity) up to modulation of learning on long timescales (metaplasticity). An efficient, biology-centric overall circuit implementation of these processing functions in CMOS is presented.

The customary mathematical model describing short-term dynamic filtering of pulse signals at the input of a synapse is analytically transformed into an expression which is significantly simpler than current literature. The novel, simplified filter expression enables a very parsimonious circuit implementation of this dynamic transfer process, which still exhibits a high degree of accuracy with respect to the biological archetype. This is especially important when trying to replicate dynamical processing functions of groups of neurons (e.g. attractor networks). In addition, the simplified filter expression is shown to provide a biologically realistic short term plasticity component for models of long term plasticity.

Learning at a synapse describes the deterministic changes in synaptic transfer behaviour occuring on a timescale of hours to weeks. These changes can effect the connection structure between neurons, which is for example employed in the visual cortex for pattern learning. With respect to synaptic learning, a novel model is introduced in the manuscript which derives its learning as a function of the local state variables at the synapse, in contrast to conventional learning rules, which are formulated in a more abstract, descriptive way. The novel learning rule is thus situated very close to biological mechanisms, while still being as economical with its parameters and defining equations as conventional abstract models. The high relevancy of the novel learning rule with respect to various current experimentally shown biological learning paradigms is proven theoretically and via simulations. In addition, the novel learning rule offers access points (=model parameters) for modulation of its learning (i.e. metaplasticity) which are in close agreement with biological evidence.

The last chapter of the work is composed of a documentation of a test circuit implementing the above plasticity models. Since the plasticity at the synapse is only dependent on local state variables such as the transmission profile of the incoming pulse and the pulse arriving at the receiving neuron, the complexity of the novel learning rule is significantly reduced and transferred away from the synapse. Thus, a very efficient and space-saving circuit realization of the synaptic functionality can be achieved. The successful transfer of the neurobiological functionality of the models as described above to CMOS circuits is shown. In addition, due to the detailed configuration options of the test circuit in combination with the metaplastically adjustable parameters offered by the novel learning rule, this will be the first implementation of plasticity in a neuromorphic circuit which allows a detailed test of metaplasticity.

# References: Cumulative

# References: Own

# Technische Universität Dresden

## Untersuchungen zur Modellierung und Schaltungsrealisierung von synaptischer Plastizität

**Dr.-Ing. Christian Georg Mayr**

von der Fakultät Elektrotechnik und Informationstechnik

der Technischen Universität Dresden

zur Erlangung des akademischen Grades

**Doktoringenieur habilitatus**

(Dr.-Ing. habil.)

genehmigte Habilitation

Vorsitzender: Prof. Dr. phil. nat. habil. R. Tetzlaff

Gutachter: Prof. Dr.-Ing. habil. R. Schüffny     Vortrag mit Kolloquium:     04.04.2012

Prof. Dr.-Ing. U. Rückert

Prof. Dr. A. Braun

# Kurzfassung

Diese Arbeit befasst sich mit der Analyse und mikroelektronischen Abbildung von aus biologischen Messungen abgeleiteter adaptiver Informationsverarbeitung. Dieses Forschungsgebiet liegt aktuell stark im Fokus der nationalen und internationalen Förderagenturen, z.B. im Rahmen des von der DARPA geförderten SYNAPSE-Projektes, in den Forschungsrahmenprogrammen der EU, im deutschlandweiten Bernstein-Netzwerk, etc. Das Hauptaugenmerk der vorliegenden Arbeit liegt dabei speziell auf den Kopplungsstellen (Synapsen) zwischen kortikalen Neuronen. An den Synapsen werden Vorgänge auf allen Zeitskalen von kurzzeitigen dynamischen Übertragungsvorgängen (Kurzzeitplastizität) über Lernen (Langzeitplastizität) bis hin zu einer Modulation des Lernens über lange Zeiträume (Metaplastizität) einzeln und in ihren Wechselwirkungen untersucht und in eine effiziente, biologienahe Gesamtimplementierung in CMOS überführt.

Das kurzzeitige dynamische Filterverhalten für Pulssignale am Eingang einer Synapse wird analytisch in eine gegenüber bisheriger Literatur stark vereinfachte Beschreibungsform gebracht. Die vereinfachte Beschreibungsform ermöglicht eine mit wenig Schaltungsaufwand durchführbare biologiegetreue Implementierung des dynamischen Übertragungsverhaltens, wie es insbesondere für die Modellierung dynamischer Verarbeitungsfunktionen von Gruppen von Neuronen (z.B. in Attraktornetzwerken) wichtig ist. Darüberhinaus wird gezeigt, dass die vereinfachte Beschreibungsform als biologisch realistische Kurzzeitplastizitätskomponente in Modellen für Langzeitplastizität verwendet werden kann.

Lernvorgänge an einer Synapse bezeichnen Änderungen des synaptischen Übertragungsverhaltens im Stunden- bis Wochenbereich, bei denen sich deterministisch Verschaltungen zwischen Neuronen ändern, womit beispielsweise optische Muster im visuellen Kortex gelernt werden. Hier wurde eine neuartige Lernregel erstellt, welche die synaptischen Lernvorgänge in Abhängigkeit von lokalen Zustandsvariablen an der Synapse darstellt, im deutlichen Kontrast zu bisherigen, eher deskriptiv bzw. phänomenologisch formulierten Modellen. Die neuartige Lernregel ist damit sehr nahe an biologischen Mechanismen angesiedelt, jedoch ähnlich sparsam in Bezug auf Parameter und definierende Gleichungen wie bisherige abstrakte Modelle. Es wird theoretisch und simulativ die sehr hohe Relevanz der neuartigen Lernregel für verschiedenste aktuell in Experimenten nachgewiesene biologische Lernparadigmen gezeigt. Die neuartige Lernregel bietet darüberhinaus Eingriffspunkte (=Modellparameter) für eine Modulation des Lernverhaltens (d.h. Metaplastizität) in sehr guter Übereinstimmung mit biologischen Erkenntnissen.

Den Abschluss der Arbeit bildet eine Dokumentation eines Testschaltkreises zu den oben geschilderten Modellen. Durch die Abhängigkeit des Lernverhaltens von lokalen Zustandsvariablen, insbesondere vom Übertragungsprofil des eingehenden Pulses sowie des weitergeleiteten Pulses auf dem empfangenden Neuron, wird die Komplexität der neuartigen Lernregel deutlich reduziert und von der Synapse wegverlagert. Dadurch ergibt sich eine sehr effiziente, platzsparende Schaltungsimplementierung der synaptischen Funktionalität. Der erfolgreiche Transfer der geschilderten neurobiologischen Funktionalität der Modelle in CMOS Schaltungstechnik wird gezeigt. Darüberhinaus wird durch die weitreichenden Konfigurationsmöglichkeiten des Testschaltkreises in Verbindung mit den metaplastisch anpassbaren Parametern der neuartigen Lernregel erstmals ein detaillierter Test von Metaplastizität auf einem neuromorphen VLSI Schaltkreis möglich sein.

# Vorwort

> *"I made this letter longer than usual because I lack the time to make it short."*
>
> *Blaise Pascal*

# Inhaltsverzeichnis

# Verwendete Abkürzungen

| | |
|---|---|
| ABS | "Artola-Bröcher-Singer", nach den Autoren von (Artola et al, 1990) benannte spannungsbasierte Lernregel |
| ADC | "Analog-Digital-Converter", Analog-Digital Wandler |
| ADPLL | "All-digital Phase Locked Loop", Frequenzsyntheseschaltkreis bestehend aus Digitalgattern |
| AER | "Address Event Representation", Wiedergabe eines APs als Adresse (und Zeitpunkt) des zugehörigen Neurons |
| AFE | "Analog Frontend", Kombination aus Signalvorverstärker und ADC |
| AHP | "Afterhyperpolarization", verlängerte Refraktärszeit in bestimmten Neuronentypen bzw. Arbeitsbereichen des Neurons |
| AP | "Action Potential", Aktionspotential/Ausgangspuls eines Neurons |
| bAP | "Backpropagating Action Potential", Entlang des Dendriten zurücklaufendes Aktionspotential eines Neurons |
| BCM | "Bienenstock Cooper Munroe", Modell für pulsratenbasiertes Lernen, benannt nach (Bienenstock et al, 1982) |
| BrainScaleS | "Brain-inspired multiscale computation in neuromorphic hybrid systems", EU-Projekt, Nachfolge von FACETS |
| CMOS | "Complementary Metal Oxide Semiconductor", Halbleiterfertigungsprozess mit n- und p-Kanal Feldeffekttransistoren |
| CORONET | "Coupling Activity Dynamics across biomimetic Brain Interfaces with neuromorphic VLSI",EU-Projekt zur Schaffung eines bidirektionalen, adaptiven Schnittstelle VLSI $\longleftrightarrow$ Nervengewebe |
| DAC | "Digital-Analog Converter", Digital-Analog Wandler |
| DC | "Direct Current", zeitinvariantes Strom- oder Spannungssignal |
| DNC | "Digital Network Chip", Chip für digitale Pulskommunikation und Konfiguration des neuromorphen Wafers in FACETS |
| DSM | "Delta Sigma Modulator", überabtastender ADC |
| FACETS | "Fast Analog Computing with Emergent Transient States", EU-Projekt zur Erforschung dynamischer neuronaler Verarbeitung |
| FACETS-ITN | "FACETS Initial Training Network", EU-finanzierte Marie Curie Graduiertenschule zur Unterstützung von FACETS/BrainScaleS |
| FLANN | "Final Learning Attractor Neural Network", Exemplar einer neuromorphen Chipfamilie (Camilleri et al, 2010) |
| FSM | "Finite State Machine", Zustandsautomat der Synapsenmatrix im MAPLE |
| GND | Masse, negative Versorgungsspannung, in den Varianten GNDA und GNDD (analoge und digitale Masse) |
| HICANN | "High Input Count Analog Neural Network", Einzelschaltkreis mit Neuronen und Synapsen auf dem neuromorphen Wafer im FACETS Waferscale System |
| JTAG | "Joint Test Action Group", Low-Level Konfigurationsschnittstelle für Digitalschaltkreise |
| LCP | "Local Correlation Plasticity", neuartige synaptische Lernregel basierend auf der Korrelation lokaler synaptischer Zustandsvariablen |
| LCP mit SRM | LCP-Regel mit SRM-Rekonstruktion des postsynaptischen Neurons |
| LCP mit LIAF | LCP-Regel mit LIAF-Rekonstruktion des postsyn. Neurons |
| LIAF | "Leaky Integrate-and-Fire Neuron", Vereinfachtes Neuronenmodell mit Leckterm, Integratorverhalten und fester Feuerschwelle |
| LTP | "Long Term Potentation", über größeren Zeitraum wirkende Verstärkung des synaptischen Gewichtes |
| LTD | "Long Term Depression", über größeren Zeitraum wirkende Abschwächung des synaptischen Gewichtes |
| MAPLE | "Multiscale Plasticity Experiment", Chipimplementierung von neuronalem Lernen auf multiplen Zeitskalen |
| NMDA | "N-Methyl-D-Aspartat", Ionenkanal in der postsynaptischen Zellmembran, vermuteter Sitz des STDP-Koinzidenzdetektors |
| OSR | "Oversampling Ratio", Quotient aus Arbeitstakt eines DSM durch die Nyquistfrequenz (d.h. $2*$ die maximale Signalfrequenz) |
| OTA | "Operational Transconductance Amplifier", Transkonduktanzverstärker |
| PCNN | "Pulse Coupled Neural Network", technische Realisierung eines durch Pulse miteinander kommunizierenden neuronalen Netzes |
| PLL | "Phase Locked Loop", Schaltkreis zur Taktsynthese |
| PPD | "Paired Pulse Depression", Verminderung des zweiten von zwei kurz aufeinander folgenden präsynaptischen APs an einer Synapse |
| PSC | "Postsynaptic Current", durch ein eingehendes AP an einer Synapse hervorgerufener postsynaptischer Strom |
| PSP | "Postsynaptic Potential", durch einen PSC auf der postsynaptischen Membran- bzw. Dendritenkapazität hervorgerufenes charakteristisches Spannungsprofil |
| SF | Skalierungsfaktor |
| SRM | "Spike Response Model", idealisierte Gleichungen für die Membranspannung eines Neurons um den Pulszeitpunkt |
| STDP | "Spike Timing Dependent Plasticity", Lernen bedingt durch Pulszeitpunkte |
| TDC | "Time to Digital Converter", Umsetzung eines Ereigniszeitpunktes (z.B. AP) in eine digitale Zeitmarke |
| VDD | Positive Versorgungsspannung, in den Varianten VDDA: analoge Versorgungsspannung, in UMC 180 nm gleich 3.3 V, VDDD, in UMC 180 nm gleich 1.8 V |
| VLSI | "Very Large Scale Integration", Höchstintegration in einem Halbleiterprozess |
| ZK | Zeitkonstante |

# Verwendete Formelzeichen und Signalnamen

| | |
|---|---|
| $A$ | Absoluter synaptischer Wirkungsgrad im Quantalmodell |
| $A_+$ | Skalierungsfaktor für das LTP-Fenster bei STDP |
| $A_-$ | Skalierungsfaktor für das LTD-Fenster bei STDP |
| $\alpha_{att}$ | Verringerung der Fläche des APs in Abhängigkeit der postsynaptischen Rate |
| $\alpha_{att,pre}$ | Verringerung der Höhe des PSCs in Abhängigkeit der präsynaptischen Rate |
| $b$ | Tastverhältnis |
| $B$ | Proportionalitätskonstante im LCP-Modell |
| $c(t)$ | postsynaptische Aktivitätsvariable bei der BCM-Regel |
| $C_{att}$ | zentrale Kapazität in der Schaltungsumsetzung der postsynaptischen AP-Adaption |
| $C_{mem}$ | Membrankapazität des LIAF-Neurons |
| $C_{weight}$ | Kapazität, auf der die Gewichtsspannung $U_{weight}$ in der Schaltungsrealisierung der LCP-Synapse gespeichert wird |
| $C_{\tau_{R,\lambda}}$ | $\tau_{R,\lambda}$-basiertes Diskriminanzkriterium im nicht-iterativen Quantalmodell |
| $C_{\tau_{u,\lambda}}$ | $\tau_{u,\lambda}$-basiertes Diskriminanzkriterium im nicht-iterativen Quantalmodell |
| $d(t)$ | präsynaptische Aktivitätsvariable bei der BCM-Regel |
| $\Delta t$ | Zeitabstand zwischen prä- und postsynaptischem Puls (z.B. im STDP-Kontext) |
| $\Delta R_n$ | Änderung von $R_n$ für jeden weiteren präsynaptischen Puls |
| $\Delta u_n$ | Änderung von $u_n$ für jeden weiteren präsynaptischen Puls |
| $f_m$ | Modulationsfrequenz für die frequenzmodulierte Pulsfolge in der nicht-iterativen Version des Quantalmodells |
| $f_{m,opt}$ | optimale Modulationsfrequenz zwischen $\lambda_1$ und $\lambda_2$ für höchstes $\overline{uR}$ |
| $g(t)$ | Konduktanz der präsynaptisch aktivierten Ionenkanäle im LCP-Modell |
| $\hat{G}$ | maximale Konduktanz der präsynaptischen Ionenkanäle |
| $I_{psc}$ | Auf die verteilten Stromspiegel skalierte Version von $I_{syn}$ in der Schaltungsrealisierung der LCP-Lernregel, in drei verschieden skalierten Varianten von der PSC-Schaltung geliefert: $I_{psc}$ für die Gewichtsberechnung zwischen APs, $I_{psc\_spk}$ während APs und $I_{psc\_weight}$ für die Gewichtswirkung Richtung Neuron, in der Synapse jeweils noch in einer Version von den unterhalb der Matrix liegenden PSC-Schaltungen ($I_{psc\_down}$, $I_{psc\_spk\_down}$, $I_{psc\_weight\_down}$) und in einer von den oberhalb liegenden PSC-Schaltungen ($I_{psc\_up}$, $I_{psc\_spk\_up}$, $I_{psc\_weight\_up}$), gruppenweise selektierbar über das SOURCE_SEL Bit der Synapse |
| $I_{spktime}$ | Die maximale AP-Dauer bestimmender Strom in der CMOS-Realisierung der postsynaptischen Adaption der LCP-Regel |
| $I_{syn}$ | Durch $g(t)$ hervorgerufener postsynaptischer Strom |
| $I_{refr}$ | Strom, mit dem im Leckstrom-OTA des LIAF die Zeitkonstante $\tau_{refr}$ eingestellt werden kann |
| $I_{refr'}$ | Strom zur Einstellung der Zeitkonstanten der Schaltungsumsetzung der postsynaptischen Adaption, separat von $I_{refr}$ einstellbar |
| $\lambda$ | Ausgangspulsrate eines Neurons, spezifisch: $\lambda_1,\lambda_2$ hohe bzw. niedrige Pulsrate in der nicht-iterativen Version des Quantalmodells |
| $n$ | Synapsenanzahl zwischen zwei Neuronen |
| $N$ | Pulsanzahl im Burst |
| $p$ | Wahrscheinlichkeit einer Neurotransmitterausschüttung für ein eingehendes AP |
| $\overline{PSC}_{xy}$ | mittlerer PSC zwischen Punkt x und y im nicht-iterativen Quantalmodell |
| $q$ | Neurotransmitterausschüttungsmenge |
| $R_c$ | Konvergenzwert des Erholungsparameters $R_n$ für eine feste präsynaptische Rate |
| $R_n$ | Erholung des Reservoirs an Neurotransmittern zum Zeitschritt $n$ |
| $R(t)$ | Nicht-iterative Version von $R_n$ |
| $SPK$ | Digitalsignal, high während eines APs in der Schaltungsrealisierung der LCP-Regel, auch als negiertes Signal $S\bar{P}K$ oder $SPK_{delayed}$, d.h. eine um einige Nanosekunden verzögerte Version von $SPK$ |
| $\tau_+$ | Zeitkonstante für das LTP-Fenster bei STDP |
| $\tau_-$ | Zeitkonstante für das LTD-Fenster bei STDP |
| $\tau_{all}$ | Inverse Zusammenfassung von $\tau_g$ und $\tau_{refr}$ in der LCP-Analyse |
| $\tau_{facil}$ | Iterative Zeitkonstante von $u_n$ im Quantalmodell |
| $\tau_g$ | Zeitkonstante von $g(t)$ in der LCP-Regel |
| $\tau_{rec}$ | Iterative Zeitkonstante von $R_n$ im Quantalmodell |
| $\tau_{refr}$ | Zeitkonstante von $u(t)$ (d.h. Membranzeitkonstante) in der LCP-Regel |
| $\tau_{refr'}$ | Über $I_{refr'}$ einstellbare Zeitkonstante der postsynaptischen Adaption in der Schaltungsrealisierung der LCP-Regel, separat von $\tau_{refr}$, üblicherweise jedoch identisch gewählt, da die postsyn. Adaption im idealen LCP-Modell ebenfalls mit $\tau_{refr}$ operiert. |
| $\tau_{R,\lambda}$ | Absolute Zeitkonstante von $R(t)$ im Quantalmodell |
| $\tau_{u,\lambda}$ | Absolute Zeitkonstante von $u(t)$ im Quantalmodell |
| $t_{spk,n}$ | Pulsdauer des n-ten Pulses |
| $t_n^{post}$ | Zeitpunkt des n-ten postsynaptischen Pulses, auch als $t^{post}$ für einen einzelnen Puls |
| $t_n^{pre}$ | Zeitpunkt des n-ten präsynaptischen Pulses, auch als $t^{pre}$ für einen einzelnen Puls |
| $T_{high}$ | Dauer des Intervalls mit hoher Rate im Quantalmodell |
| $\Theta_M$ | Ratenschwellwert für das Lernen der BCM-Regel |
| $\Theta_u$ | Spannungsschwellwert für das Lernen beim neuartigen LCP-Modell |
| $u(t)$ | Momentanwert des Membranpotentials, im Quantalmodell auch die nicht-iterative Version von $u_n$ |
| $u_c$ | Konvergenzwert des Verwendungsparameters $u_n$ für eine feste präsynaptische Rate |
| $u_n$ | Verwendung des Reservoirs an Neurotransmittern zum Zeitschritt $n$ |
| $U$ | Verwendung des Reservoirs an Neurotransmittern an einer in Ruhe befindlichen Synapse |
| $U_p$ | Fläche des APs des Neurons im LCP-Modell |
| $U_{PSP}$ | Amplitude eines durch einen einzelnen präsynaptischen Puls hervorgerufenen postsynaptischen Potentials |
| $\overline{uR}_{xy}$ | mittlere $u \cdot R$-Antwort zwischen Punkt x und y. Durch Wichtung mit Rate und Tastverhältnis entsteht $\overline{PSC}_{xy}$ |
| $U_{refr}$ | Amplitude des Zurücksetzens (Refraktärsperiode) des Neurons im LCP-Modell |
| $U_{low}$ | Untere Spannung in der CMOS-Realisierung der postsynaptischen Adaption der LCP-Regel |
| $U_{C_{att}}$ | Als Spannung kodierte zentrale Zustandsvariable (Zeitfensterrealisierung) in der CMOS-Realisierung der postsynaptischen Adaption der LCP-Regel |
| $U_{high}$ | Obere Spannung in der CMOS-Realisierung der postsynaptischen Adaption der LCP-Regel |
| $U_{\alpha att}$ | Spannung zur Bestimmung der Adaptionsstärke in der CMOS-Realisierung der postsynaptischen Adaption der LCP-Regel |
| $U_{mem}$ | Spannung auf der Membrankapazität des LIAF-Neurons |
| $U_{thr}$ | Feuerschwellwert des LIAF-Neurons |
| $U_{rest}$ | Ruhespannung (Bezugswert des passiven Leitwertes/Leckstroms im LIAF-Neuron) |
| $U_{reset}$ | Rücksetzspannung des LIAF-Neurons (üblicherweise unter der Ruhespannung, d.h. der Abstand $U_{reset}$ bis $U_{rest}$ ist gleich $U_{refr}$) |
| $U_{weight}$ | Spannung proportional zum Gewicht der LCP-Synapse, gespeichert auf der Gewichtskapazität $C_{weight}$ |
| $U_{weight\_out}$ | Version von $U_{weight}$ nach dem Sourcefolger der Synapse (d.h. um eine Schwellwertspannung angehoben) |
| $U_{weight\_reset}$ | Spannung, auf die $U_{weight}$ nach einer Veränderung des LSBs des digitalen Gewichtsspeichers zurückgesetzt wird |
| $U_{weight\_high}$ | Spannung, mit der $U_{weight\_out}$ verglichen wird, um eine Entscheidung über ein Increment des digitalen Gewichtsspeichers zu treffen |
| $U_{weight\_low}$ | Spannung, mit der $U_{weight\_out}$ verglichen wird, um eine Entscheidung über ein Decrement des digitalen Gewichtsspeichers zu treffen |
| $w$ | Synaptisches Gewicht |

# 1 Einleitung

Mittels der Nachbildung von Lern- und Signalübertragungsvorgängen im kortikalen Nervengewebe sollen biologische Verarbeitungsvorgänge für simulative und mathematische Analysen zugänglich gemacht werden (Gerstner and Kistler, 2002; Kandel, 1995; Koch, 1999). Ein Verständnis dieser Vorgänge kann zum einen dazu verwendet werden, in Nervengewebe eingebundene Prothesen wie z.B. Cochleaimplantate weiterzuentwickeln (Dahmen and King, 2007). Zum anderen sollen durch die Modelle und schaltungstechnischen Abbildungen der kortikalen Vorgänge neue, biologieorientierte Verarbeitungsparadigmen gefunden werden (D'Souza et al, 2010; Nessler et al, 2009).

Information liegt im kortikalen Nervengewebe in der Regel in Form von sogenannten Aktionspotentialen (AP) vor (siehe Abbildung 1.1A(oben)), d.h. stereotypen Spannungspulsen, deren Zeitpunkt und Ort als informationstragendes Medium postuliert wird (Koch, 1999)[1]. Diese Spannungspulse werden von den Einzelbausteinen des Nervengewebes, den Neuronen, verarbeitet und weitergeleitet. Neuronen bestehen aus einem weitverzweigten Netzwerk an Zellausläufern zur Sammlung eingehender Pulse (den sogenannten Dendriten), dem eigentlichen Zellkörper und einem weiteren Ausläufer, welcher die APs an nachfolgende Neuronen weitergibt, dem sogenannten Axon (siehe Abbildung 1.1B).

Die Kopplungsstelle zwischen Axon und Dendriten, an dem ein von einem Neuron ausgesandtes AP weitergegeben wird, ist die sogenannte Synapse. In biologischer Terminologie wird das nach der Synapse liegende Neuron als postsynaptisches Neuron bezeichnet, das vor der Synapse liegende als präsynaptisches. Ein entlang des Axons ankommendes (d.h. präsynaptisches) AP führt an der Synapse zu einer Ausschüttung von Neurotransmittern aus ihren Speicherblasen, den Vesikeln, siehe Abbildung 1.1B. Die Neurotransmitter docken an Rezeptoren an der Aussenhaut des Dendriten an und verursachen einen postsynaptischen Strom (postsynaptic current, PSC) im empfangenden Neuron (Koch, 1999). Die Integration dieses Stromes auf der intrinsischen (Membran-)Kapazität des postsynaptischen Neurons (siehe Abbildung 1.1B) wiederum führt zu charakteristischen Spannungssignalen, den postsynaptischen Potentialen (PSP, Abbildung 1.1A zeigt den typischen Verlauf, siehe auch Abbildung 5.14B für eine Sequenz von PSPs). Wenn auf der Membrankapazität eine gewisse Anzahl an PSPs integriert wurden, d.h. ein Spannungsschwellwert erreicht wird (siehe Abbildung 1.1A(oben)), wird im Zellkörper wiederum ein AP generiert und durch

---

[1] Puls und Aktionspotential werden im weiteren Verlauf der Arbeit als Synonyme verwendet

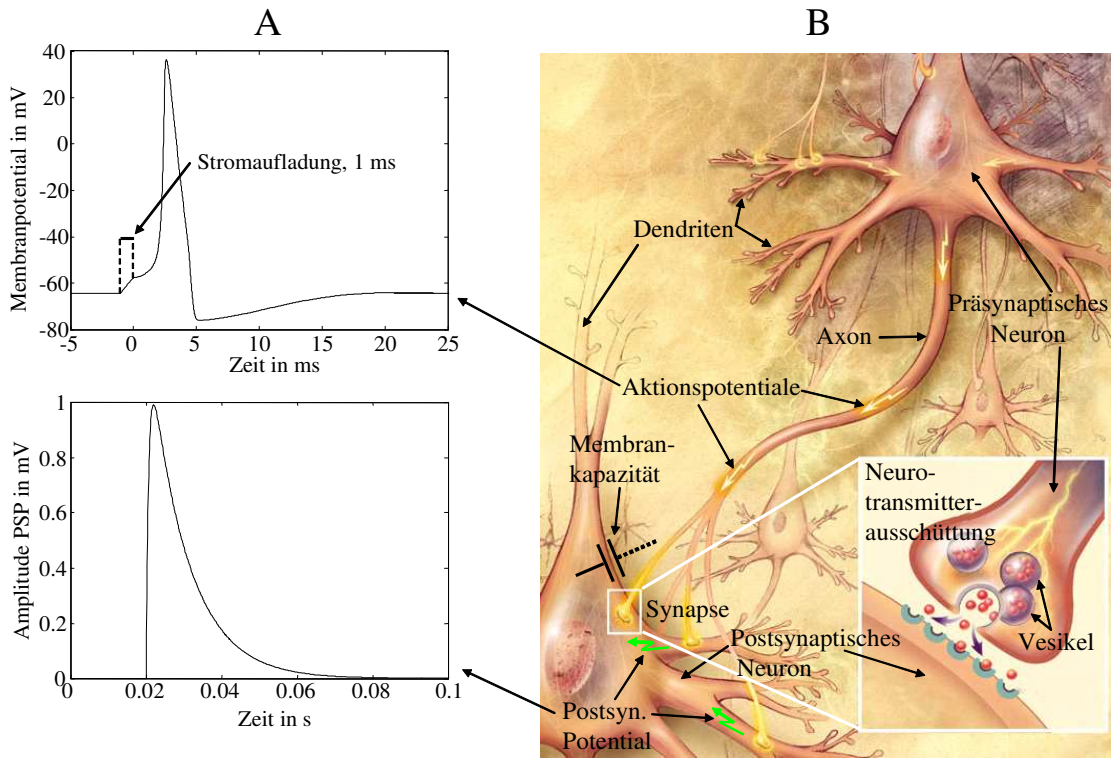

Abbildung 1.1: (A) Oben: Typisches Aktionspotential eines Neurons nach Aufladung der Membrankapazität durch einen Konstantstrom (Simulation eines Hodgkin-Huxlex-Neuronenmodells (Koch, 1999), Parameter und Ladestrom gemäß (Mayr, 2008)); nach dem AP durchläuft die Spannungskurve die sogenannte Hyperpolarisation, d.h. ein Potential unterhalb der Ruhespannung, das mit der Membranzeitkonstante wieder auf die Ruhespannung abklingt. Das Neuron zeigt in dieser Phase Refraktärsverhalten, d.h. eine verminderte Anregbarkeit (Koch, 1999). Unten: Durch ein präsynaptisches Aktionspotential hervorgerufenes postsynaptisches Potential (Rekonstruktion als 'Double Exponential' mit Parametern gemäß (Pfister et al, 2006)); (B) Informationsverarbeitung im kortikalen Nervengewebe, Bild adaptiert aus http://en.wikipedia.org/wiki/Neuron.

das Axon weitergeleitet[2].

Die Amplitude der AP-Übertragung an der Synapse (in Form des PSCs bzw. PSPs) wird gemeinhin als ein dimensionsloser Skalar, das sogenannte synaptische Gewicht $w$, model-

---

[2]Biologische Neuronen zeigen neben dieser Aufladecharakteristik auch deutlich komplexere Pulsfolgen als Antwort auf einen anliegenden Stimulus, siehe (Izhikevich, 2004). In der vorliegenden Arbeit sind jedoch nur das angeführte einfache Integratorverhalten und ein Rücksetzen auf ein Potential unterhalb der Ruhespannung nach einem AP relevant (das Refraktärsverhalten, siehe (Koch, 1999) bzw. die Membranspannungskurve in Abbildung 1.1A). Aus dem Repertoire komplexer Neuronenantworten wird zusätzlich nur das Burstverhalten herausgegriffen, d.h. es wird angenommen, dass bestimmte Neuronentypen auf anliegende Stimuli mit einer Gruppe hochfrequenter Ausgangspulse antworten, gefolgt von einer Pause (Izhikevich, 2004). Mithin kann Informationsübertragung in dieser vereinfachten Sichtweise als einzelner Puls oder als Burst stattfinden.

liert. Das Gewicht lässt sich als eine Funktion von drei verschiedenen Variablen darstellen (Koch, 1999):

$$w = f(n, p, q) \tag{1.1}$$

Der Faktor $n$ bezeichnet die Anzahl an Neurotransmitterausschüttungsstellen an einer einzigen Synapse bzw. die Gesamtanzahl der Synapsen zwischen zwei Neuronen (Koch, 1999). Mechanismen, die $n$ verändern, werden durch sehr langsam wirkende Wachstumsprozesse beeinflusst und sind nicht Bestandteil dieser Arbeit. Mit dem Proportionalitätsfaktor $q$ wird der Bezug zwischen der Neurotransmittermenge und dem maximalen daraus resultierenden PSP hergestellt (vorstellbar als Effizienz der postsynaptischen Rezeptoren bzw. als Maß für die Gesamtneurotransmittermenge). Für jeden der Neurotransmitter enthaltenden Vesikel in Abbildung 1.1 existiert eine bestimmte Wahrscheinlichkeit $p$, mit der für ein einkommendes AP eine Neurotransmitterausschüttung ausgelöst wird. Für einen hohen Anteil an Synapsen im Kortex gilt, dass entweder die Anzahl an Neurotransmitterausschüttungsstellen (Koch, 1999) oder die Gesamtanzahl der Synapsen zwischen zwei Neuronen relativ groß ist (Hellwig, 2000), so dass gemäß Gleichung 1.1 jedes präsynaptische AP eine näherungsweise dem Erwartungswert $n \cdot p$ entsprechende Menge an Neuotransmittern aktiviert[3]. Für die Gesamtanzahl der durch ein AP ausschüttbaren Neurotransmitter werden kurzzeitige Erschöpfungs- und Regenerierungsvorgänge (Zeitskala 100-1000ms) angenommen, welche die präsynaptische Pulsfolge einer komplexen Filterung, der sogenannten Kurzzeitplastizität, unterziehen (Abbott et al, 2004; Markram et al, 1998), (Mayr et al, 2009a).

Neben der Kurzzeitplastizität werden $p$ (und vermutlich $q$) auch in Zeiträumen von Stunden und Tagen moduliert (Senn, 2002; Sjöström et al, 2008), wodurch sich deterministisch die Kopplungsstärke zwischen Neuronen ändert. Dieser Vorgang wird Langzeitplastizität genannt, manifestiert als Verstärkung (Long Term Potentation, LTP) und Abschwächung (Long Term Depression, LTD) der mittleren Ausschüttungsmenge bzw. des mittleren PSP (Bi and Poo, 1998; Froemke et al, 2006; Sjöström et al, 2008). Spezifisch werden bei LTP und LTD Korrelationen zwischen prä- und postsynaptischen Pulsmustern extrahiert, womit beispielsweise optische Muster im visuellen Kortex gelernt werden (Senn, 2002). Im Kontrast dazu sind die o.a. kurzzeitigen Filtervorgänge in der Regel entweder nur präsynaptisch oder nur postsynaptisch wirksam, d.h. die Erschöpfungs- und Regenerierungsvorgänge 'arbeiten' jeweils nur auf Grundlage der ihnen gerade vorliegenden Pulsfolge, während LTP und LTD an der Synapse Zugriff auf beide (prä- und postsynaptische) Pulsfolgen haben (Bienenstock et al, 1982; Koch, 1999; Markram et al, 1997).

In Zeiträumen von Wochen bis Monaten findet darüberhinaus wiederum eine Modulation des Lernens (d.h. von LTP und LTD) statt, die sogenannte Metaplastizität. Diese kann Synapsen je nach Kontext entweder eine Tendenz zu einer homeostatischen Bewahrung der aktuellen Gewichte (Bienenstock et al, 1982; Lebel et al, 2001) vorgeben oder Synapsen für verstärktes Lernen 'vorspannen' (Disterhoft and Oh, 2006).

Die Gesamtverarbeitungsfunktion einer Neuronenpopulation wird maßgeblich durch die hochselektive Art dieser Pulsweitergabe und durch das Wechselspiel dieser Selektion auf verschiedenen Zeitskalen beeinflußt (D'Souza et al, 2010; Lazar et al, 2007; Mejias and Tor-res, 2009; Senn, 2002). Das aktuelle Forschungsrahmenprogramm der EU (FP7) hat diesen Erkenntnissen Rechnung getragen, indem die Analyse und Abbildung neuronaler Verarbeitung über mehrere Zeitskalen ('multiscale') hinweg explizit in die Zielsetzung des Rahmenprogramms aufgenommen wurde. Das in dieser Habilitation vorgestellte Lernmodell basierend auf lokaler synaptischer Korrelation (Local Correlation Plasticity, LCP) und die darauf beruhende Chipimplementierung Multiscale Plasticity Experiment (MAPLE) sind sehr gut für eine derartige zeitskalenübergreifende Emulation geeignet. Von LCP werden in guter Übereinstimmung mit biologischen Erkenntnissen Lernmechanismen auf verschiedenen Zeitskalen bereit gestellt, angefangen bei kurzzeitigen Filtervorgängen über LTP/LTD bis hin zu Eingriffspunkten (=anpassbare, Biologie-analoge Modellparameter) für verschiedene Formen von Metaplastizität (Mayr and Partzsch, 2010).

---

[3]Es gibt jedoch auch Synapsen, bei denen sehr wenige Vesikel vorhanden sind, so dass die Übertragung eines präsynaptischen AP zum postsynaptischen Neuron insgesamt stochastisch wird (Koch, 1999)

Die vorliegende kumulative Habilitation, welche die Resultate dieser Plastizitätsforschung und deren Implementierung dokumentiert, beruht im Speziellen auf folgenden Publikationen:

- Die in (Mayr et al, 2009b) sowie in (Mayr et al, 2009a) zur Modellierung bzw. Analyse von Kurzzeitplastizität vorgestellten Erkenntnisse.
- Das in (Mayr et al, 2008b) eingeführte Modell zur Langzeitplastizität sowie seine Erweiterung und Diskussion in (Mayr and Partzsch, 2010).
- Die in (Mayr et al, 2010c) diskutierten Mängel bisheriger Modelle bei der Erklärung biologischer Messdaten.
- Die in (Mayr et al, 2010b; Noack et al, 2010) eingeführte Systemarchitektur und schaltungstechnische Umsetzung der Kurzzeitplastizität auf dem MAPLE sowie die in (Mayr et al, 2010a) vorgestellte Schaltung zur Langzeitplastizität.
- Die in (Mayr et al, 2008a) geschilderten Konzepte zur Ansteuerung und Versorgung von in einer Matrix angeordneten Pixelzellen, welche sich direkt auf die hier entwickelten Synapsenzellen anwenden lassen.
- Der in (Ellguth et al, 2009) entworfene Vorverstärker für Test und Überwachung zukünftiger Versionen des MAPLE.
- Der Taktsyntheseschaltkreis aus (Eisenreich et al, 2009) zur Taktversorgung neuromorpher Schaltkreise und zur selektiven Verzögerungsgenerierung.
- Der vorverstärkende Delta-Sigma Modulator aus (Mayr et al, 2010d) zur Beobachtung von Populationsdynamiken eines neuronalen Netzes.

Auf Grundlage dieser Publikationen wird in den Abschnitten 2.1 und 2.2 die Modellierung bzw. Erklärungskraft eines neuen Modellansatzes zur synaptischen Kurzzeitplastizität thematisiert. In Abschnitt 2.3 wird die Kompatibilität dieses Modells mit der für Langzeitplastizität relevanten Paired-Puls-Depression (PPD, (Froemke et al, 2006; Koch, 1999)) nachgewiesen. Darauf aufbauend wird in Abschnitt 2.4 eine Umformulierung dieses Kurzzeitplastizitäts-Modells zur verbesserten Integration in Modelle für Langzeitplastizität vorgenommen, was insbesondere relevant ist für die schaltungstechnisch implementierte Variante von Kurzzeitplastizität in Abschnitt 5.3. Im Kapitel 3 wird die neuartige LCP-Lernregel vorgestellt, wobei in Abschnitt 3.1 zunächst ein umfangreicher Vergleich mit aktuellen verwandten Arbeiten aus der Literatur durchgeführt wird. In Abschnitt 3.2 wird gezeigt, dass das in Abhängigkeit von lokalen Zustandsvariablen an der Synapse formulierte

Lernen in der LCP-Regel sehr nahe an biologischen Mechanismen angesiedelt ist, jedoch ähnlich sparsam in Bezug auf Parameter und definierende Gleichungen ist wie bisherige abstrakte Modelle. Es wird theoretisch und simulativ die sehr hohe Relevanz der neuartigen Lernregel für verschiedenste aktuell in Experimenten nachgewiesene biologische Lernparadigmen gezeigt (Abschnitt 3.3 und 3.4). Modulation des Lernverhaltens (d.h. Metaplastizität) wurde in bisherigen Lernregeln nicht thematisiert oder auf eine extrem vereinfachte Form reduziert (Benuskova and Abraham, 2007; Brader et al, 2007). Demgegenüber bietet LCP als eine der ersten Lernregeln differenzierte, biologisch realistische Eingriffspunkte für mehrere parallel arbeitende Formen von Metaplastizität (Kapitel 4).

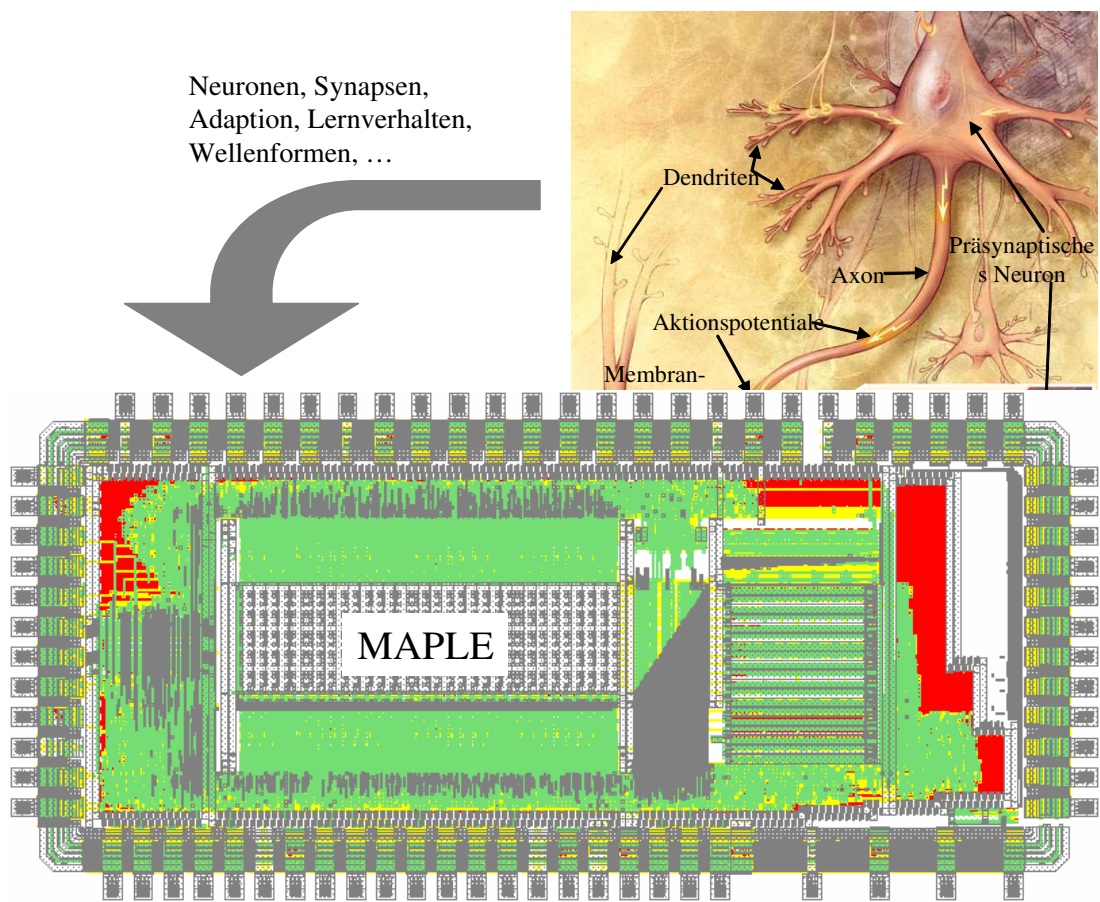

Abbildung 1.2: Darstellung des MAPLE (16 Neuronen mit postsynaptischer Kurzzeitplastizität, 512 Synapsen mit Langzeit- und Metaplastizität, 64 Schaltungen zur Emulierung präsynaptischer Kurzzeitplastizität) in einer UMC 180 nm CMOS Technologie. Für eine detaillierte Beschreibung der Baugruppen sei auf Abbildung 5.1 verwiesen, ein Layout mit markierten Baugruppen ist in Abbildung 5.12 enthalten.

Den Abschluss der Arbeit bildet eine Beschreibung des Testschaltkreises MAPLE, der die oben geschilderten Modellen umsetzt (Kapitel 5, siehe auch Abbildung 1.2). Topologische Maßnahmen zur verbesserten Abbildung von Netzwerkmodellen bei gleichzeitiger Optimierung des notwendigen Schaltungsaufwandes werden in Abschnitt 5.1 vorgestellt. Durch die Abhängigkeit des Lernverhaltens von lokalen Zustandsvariablen, insbesondere vom Übertragungsprofil des eingehenden Pulses (Abschnitt 5.3) sowie des weitergeleiteten Pulses auf dem empfangenden Neuron (5.2), wird die Komplexität der neuartigen Lernregel deutlich reduziert und von der Synapse wegverlagert. Dadurch ergibt sich eine sehr effiziente, platzsparende Schaltungsimplementierung der synaptischen Funktionalität (Abschnitt 5.4). Der erfolgreiche Transfer der geschilderten neurobiologischen Funktionalität der Modelle in CMOS Schaltungstechnik wird gezeigt (Abschnitt 5.6). Darüberhinaus wird durch die weitreichenden Konfigurationsmöglichkeiten des Testschaltkreises in Verbindung mit den Metaplastizitätseingriffspunkten der neuartigen Lernregel erstmals der Test von Metaplastizität auf einem neuromorphen VLSI Schaltkreis möglich sein (Abschnitt 5.5). In den Abschnitten 5.7 und 5.8 wird ein Ausblick auf die erweiterte Funktionalität zukünftiger Iterationen des MAPLE gegeben. Kapitel 6 gibt eine Zusammenfassung der erreichten Ergebnisse und eine Einbettung in den erweiterten wissenschaftlichen Kontext.

# 2 Kurzzeitplastizität

## 2.1 Das nicht-iterative Quantalmodell

In Markram et al (1998) wurde ein auf biologischen Messungen und Mechanismen beruhendes deskriptives Modell für die Plastizität der Neurotransmitterausschüttungsmenge eingeführt (im Folgenden als *Quantalmodell* bezeichnet). Das Quantalmodell wurde seit seiner Einführung umfangreich untersucht, beispielsweise in Bezug auf Informationsübertragung (Abbott et al, 2004; Fuhrmann et al, 2002; Matveev and Wang, 2000; Natschläger and Maass, 2001) oder mit Hinblick auf mögliche Wechselwirkungen mit Langzeitlernen ((Abbott et al, 2004; Farajidavar et al, 2008)). Diese Untersuchungen wurden größtenteils simulativ durchgeführt, da die iterative, puls-basierte Formulierung des Modells eine geschlossene analytische Untersuchung nahezu verhindert. Insbesondere ein direkter Bezug zwischen dem Verhalten des Modells und seinen Parametern, sichtbar etwa in Abbildung 2.1, kann mit Simulationen wie den in (Natschläger and Maass, 2001) durchgeführten nur indirekt hergestellt werden. Dieser Bezug wäre allerdings wichtig zum einen, um die Verarbeitungseigenschaften der Synapse zu klären und zum anderen, um beispielsweise für eine Kopplung mit Modellen für Langzeitplastizität für die Synapse eine Übertragungsfunktion auf kurzen Zeitskalen abzuleiten. Diese kann dann mit einer aus der Langzeitplastizität abgeleiteten zweiten Übertragungsfunktion zu einem Gesamtübertragungsverhalten überlagert werden. Im Folgenden wird deshalb ein Versuch unternommen, eine derartige Übertragungsfunktion herzuleiten. Der Fokus liegt dabei auf den zwei Hauptarbeitsgebieten der Synapse, welche in Abbildung 2.1 bzw. in (Natschläger and Maass, 2001) unterschieden werden können. Im ersten Arbeitsgebiet werden die Pulse in kurze Zeiträume gruppiert um einen maximalen mittleren PSC hervorzurufen (Natschläger and Maass, 2001), während für Synapsen mit anderen Eigenschaften (d.h. anderen Parametern des Quantalmodells) eine gleichmäßige Verteilung der Pulse über die Zeit einen höheren mittleren PSC bewirkt.

Ein erster Ansatz basierend auf diesen quasi-statischen Betriebsmodi (eine konstante Rate oder eine konstante Frequenzmodulation[1]) wurde bereits in (Mayr, 2008) vorgestellt. In (Mayr et al, 2009b) wurde dieser Ansatz weiterentwickelt und darauf aufbauend synaptische Wirkmechanismen diskutiert. Mayr et al (2009a) erforscht die Aussagekraft der im analytischen Ausdruck gewonnenen absoluten Zeitkonstanten zur Charakterisierung des

---

[1] Rate bezeichnet in dieser Arbeit äquidistant über ein bestimmtes Intervall verteilte Pulse, d.h. eine Rate von $\lambda$=20 Hz, die für ein Intervall von einer Sekunde z.B. am Quantalmodell anliegt, wären 20 Pulse im Abstand von jeweils 50 ms. Desgleichen bedeutet Frequenzmodulation im Rahmen dieser Arbeit eine Pulsfolge, die mit der Modulationsfrequenz $f_m$ und dem zugehörigen Tastverhältnis $b$ zwischen einer Rate $\lambda_1$ und $\lambda_2$ hin- und herschaltet, d.h. eine Frequenzmodulation mit einem Rechtecksignal. In Zahlen wäre dies bei $f_m$=2 Hz, $b$=0.4, $\lambda_1$=100 Hz und $\lambda_2$=10 Hz ein Intervall $1/f_m \cdot b$=200 ms, in dem $\lambda_1$ anliegt, d.h. 20 Pulse im Abstand von jeweils $1/\lambda_1$=10 ms, danach ein Intervall $1/f_m \cdot (1-b)$=300 ms, in dem $\lambda_2$ anliegt, d.h. 3 Pulse im Abstand 100 ms, in dieser Abfolge weiter fortgesetzt.

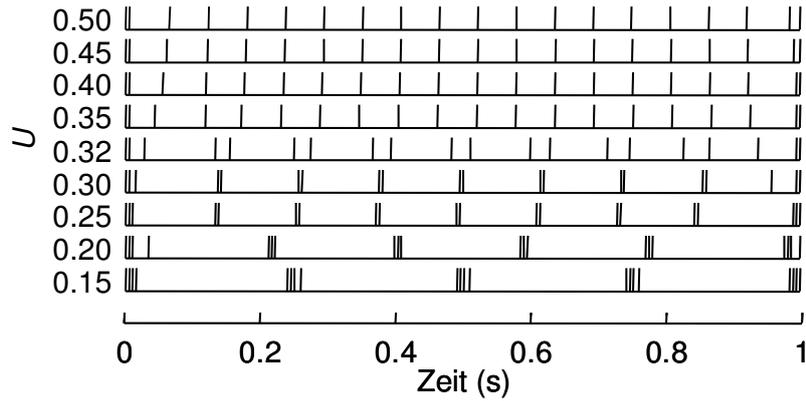

Abbildung 2.1: Optimales Pulsmuster (d.h. maximaler durch die Pulse im Quantalmodell im Mittel hervorgerufener PSC) in Abhängigkeit von $U$, adaptiert aus Abbildung 5A von (Natschläger and Maass, 2001), restliche Parameter $\tau_{\text{facil}} = 62\text{ms}$, $\tau_{\text{rec}} = 144\text{ms}$.

Synapsenverhaltens.

Das Quantalmodell (Markram et al, 1998) basiert auf einer math. Formulierung eines Erschöpfungseffektes der Neurotransmitterausschüttung an einer Synapse. Die mögliche Ausschüttungsmenge $R_n$ (Recovery, normiert auf die maximal mögliche Menge, d.h. $R_n$ liegt einheitenlos im Bereich 0..1) für Puls $n$ ist dabei abhängig von der beim vorhergehenden Puls ausgeschütteten Menge $R_{n-1}$ sowie über die Zeitkonstante $\tau_{\text{rec}}$ von der zurückliegenden Zeit. Gleichzeitig existiert eine multiplikative Skalierung von $R_n$ mittels eines zweiten Prozesses, der in einer verstärkenden Abhängigkeit von einer zweiten Zeitkonstanten $\tau_{\text{facil}}$ die für einen präsynaptischen Puls verwendete Menge $u_n$ (Utilization, ebenfalls einheitenlos im Bereich 0..1) an Neurotransmittern modelliert. Die iterativen Gleichungen, welche die zeitliche Änderung von $u_n$ und $R_n$ beschreiben, sind wie folgt (Markram et al, 1998)[2]:

$$u_{n+1} = u_n e^{-\frac{t^{pre}_{n+1}-t^{pre}_n}{\tau_{\text{facil}}}} + U \cdot \left(1 - u_n e^{-\frac{t^{pre}_{n+1}-t^{pre}_n}{\tau_{\text{facil}}}}\right) \qquad (2.1)$$

$$R_{n+1} = R_n(1-u_n)e^{-\frac{t^{pre}_{n+1}-t^{pre}_n}{\tau_{\text{rec}}}} + 1 - e^{-\frac{t^{pre}_{n+1}-t^{pre}_n}{\tau_{\text{rec}}}}, \qquad (2.2)$$

wobei $t^{pre}_{n+1} - t^{pre}_n$ die zwischen Puls $(n)$ und $(n+1)$ vergangene Zeit beschreibt. Die Anfangswerte für Gleichung 2.1 und 2.2 werden aus der maximal verwendbaren Neurotransmittermenge/Utilization $U$ einer Synapse in Ruhe zu $u_1 = U$ und $R_1 = 1$ berechnet (Markram et al, 1998). Die durch einen präsynaptischen Puls hervorgerufene PSC-Amplitude wird somit wie oben bereits erwähnt durch ein Produkt aus $u_n$ und $R_n$ gebildet, multipliziert mit einer Skalierungskonstanten $A$, welche das Verhältnis zwischen Neurotransmitterausschüttung und resultierender PSC-Amplitude beschreibt ($A$ hat damit die Einheit eines Stromes, üblicherweise im Bereich 0.5-5 pA (Markram et al, 1998)):

$$\text{PSC}_n = A \cdot R_n \cdot u_n . \qquad (2.3)$$

---

[2]Während die Herleitung in (Mayr, 2008) auf der ursprünglichen Formel für $R_n$ aus (Markram et al, 1998) beruht, wird hier wie auch bereits in (Mayr et al, 2009a,b) die Indexkorrektur aus (Natschläger and Maass, 2001) verwendet, wodurch sich die Formel für $\tau_{R,\lambda}$ signifikant ändert.

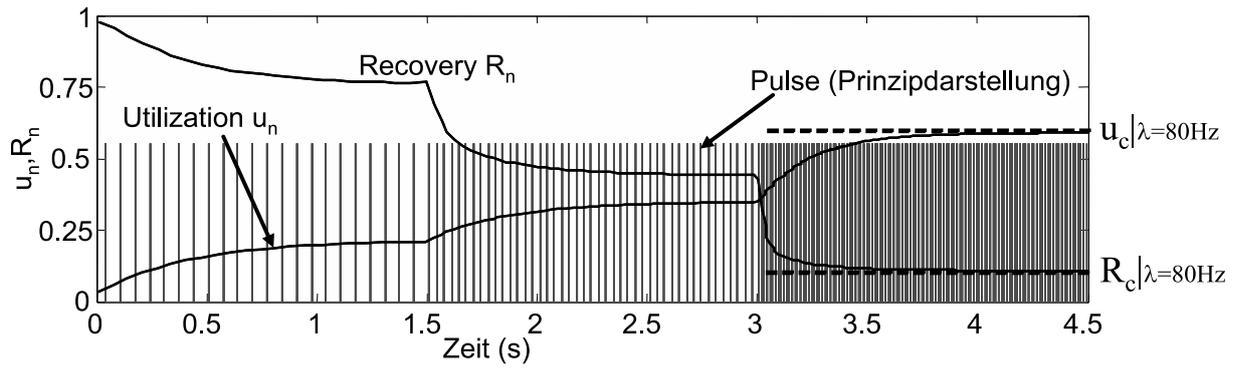

Abbildung 2.2: Verhalten des Quantalmodells, Protokoll ähnlich wie in (Markram et al, 1998, Abbildung 4), aber Wechsel zwischen abschnittsweise festen Raten statt Poissonverteilungen, Frequenzsprung nach 1.5s und 3s, Pulsraten $15\text{s}^{-1} \to 30\text{s}^{-1} \to 80\text{s}^{-1}$, synaptische Eingangspulse, resultierende Utilization $u_n$ sowie Recovery $R_n$, Parameter wie in (Markram et al, 1998, Abbildung 4), d.h. $\tau_{\text{facil}} = 530\text{ms}$, $\tau_{\text{rec}} = 130\text{ms}$, $A = 1540\text{pA}$, $U = 0.03$. Gestrichelt eingezeichnet sind die Konvergenzwerte $u_c$ bzw. $R_c$ von $u_n$ bzw. $R_n$ für $\lambda$=80 Hz.

Die Auswirkung der o.a. Adaption im Quantalmodell kann am ehesten als eine Übertragung von Transienten beschrieben werden, d.h. Änderungen in der präsynaptischen Pulsrate werden mit der vollen Dynamik zum postsynaptischen Neuron übertragen, während für gleichförmig anliegende Stimuli die Antwort auf einen kleineren Dynamikbereich eingeschränkt wird (siehe Abbildung 2.2, sowie die PSC-Antwort zum gleichen Stimulus in (Mayr et al, 2009b, Abbildung 2.2)). Dass neuartige Stimuli gegenüber statischen höhere Beachtung finden, scheint ein allgemeines Merkmal der Informationsverarbeitung in biologischen neuronalen Netzen zu sein (Abbott et al, 2004; Koch, 1999). In Erweiterung dieser Überlegung wird im Folgenden, wie oben bereits erwähnt, das Übertragungsverhalten des Quantalmodells für vollständig transiente Stimuli (d.h. modulierte Pulsfolgen) untersucht, siehe Abbildung 2.3.

Für derartige abschnittsweise konstante Pulsraten kann $u_n$ aus Gleichung 2.1 des ursprünglichen, iterativen Quantalmodells in einer geschlossenen Herleitung in einen zeitkontinuierlichen exponentiellen Abfall bzw. Anstieg $u(t)$ (mit $u_n$ gleich $u(t_n)$) umgeschrieben werden (Mayr et al, 2009a):

$$u(t) = (u_0 - u_c)\mathrm{e}^{-\frac{t}{\tau_{u,\lambda}}} + u_c \ . \tag{2.4}$$

Der Konvergenzwert $u_c$ für eine gegebene feste Pulsrate $\lambda$ ergibt sich aus der absolutzeitbezogenen Umschreibung in (Mayr et al, 2009a) zu[3]

$$u_c(\lambda) = \frac{U}{1 - (1-U) \cdot \mathrm{e}^{-\frac{1}{\lambda \cdot \tau_{\text{facil}}}}} \ . \tag{2.5}$$

Abbildung 2.3 zeigt den Zeitverlauf von $u(t)$ qualitativ für eine modulierte Pulsfolge. In

---

[3] Er kann auch durch Gleichsetzen von $u_n$ and $u_{n+1}$ in (2.1) berechnet werden (Markram et al, 1998).

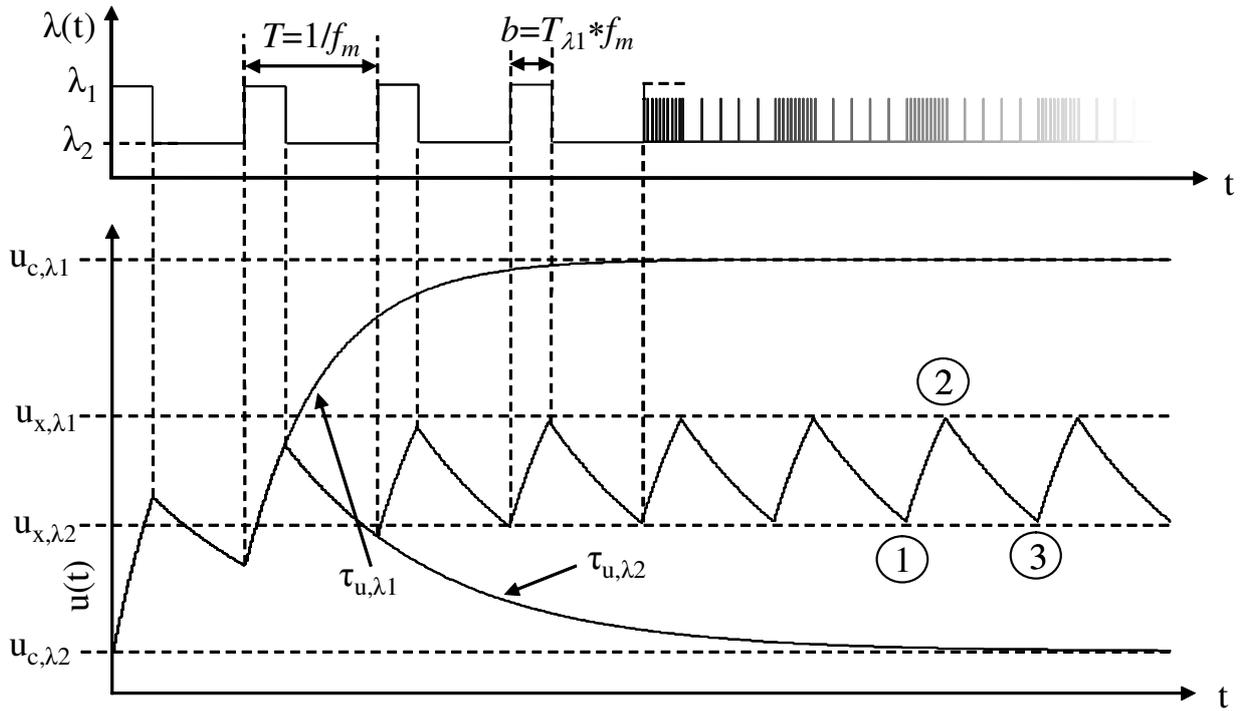

Abbildung 2.3: Zeitverlauf von $u(t)$ und die entsprechenden Abhängigkeiten von der Modulationsfrequenz $f_m$ sowie den Konvergenzwerten $u_c$ für hohe bzw. niedrige Pulsraten $\lambda_1$ and $\lambda_2$. Das Tastverhältnis der Modulation wurde in diesem Beispiel auf $b = 0.3$ gesetzt. Die Darstellung $\lambda(t)$ zeigt zuerst den Zeitverlauf der Modulationswellenform, danach symbolisch einen Übergang zu der entstehenden Pulsfolge

Abhängigkeit vom Vorzeichen des Terms $(u_0 - u_c)$ beschreibt Gleichung 2.4 einen abklingenden bzw. ansteigenden Teil des Zeitverlaufs. Der Wert von $u(t)$ oszilliert im eingeschwungenen Zustand in einem festen Intervall $[u_{x,\lambda_2}, u_{x,\lambda_1}]$, welches von der Modulationsfrequenz $f_m$, dem Tastverhältnis $b$, den Konvergenzwerten $u_c$ für niedrige und hohe Pulsraten, sowie den Zeitkonstanten $\tau_{u,\lambda_1}$ und $\tau_{u,\lambda_2}$ der nicht-iterativen Umschreibung in (Mayr et al, 2009a) abhängt:

$$\tau_{u,\lambda} = \frac{1}{\lambda \cdot \ln\left(\frac{1}{1-U}\right) + \frac{1}{\tau_{\text{facil}}}} \ . \tag{2.6}$$

Diese absolutzeitbezogene Zeitkonstante $\tau_{u,\lambda}$ beinhaltet im signifikanten Unterschied zu der ursprünglichen iterativen Zeitkonstante $\tau_{\text{facil}}$ auch $\lambda$ und $U$. Dies berücksichtigt die Abhängigkeit des absoluten Zeitverlaufs von $u_n$ bzw. von $u(t)$ von allen Parametern des Modells, nicht nur den ursprünglichen Zeitkonstanten.

Die Herleitung von Gleichungen 2.4, 2.5 und 2.6 für das korrespondierende $R(t)$ erfolgt analog (Mayr et al, 2009a). Nach Berechnung der Intervallgrenzen $[u_{x,\lambda_2}, u_{x,\lambda_1}]$ kann dann die mittlere $u \cdot R$-Antwort der Synapse für den eingeschwungen-oszillierenden Zustand von Punkt 1 zu Punkt 2 ($\overline{uR}_{12}$) und weiter zu Punkt 3 ($\overline{uR}_{23}$) aufintegriert werden (siehe Abbildung 2.3). Eine Multiplikation mit der Zeitdauer der einzelnen Pulse $T_{\text{pulse}}$ sowie mit den Raten $\lambda_1$ bzw. $\lambda_2$ sorgt für die Gewichtung der $\overline{uR}$-Antwort mit der Zeitdauer, in der durch präsynaptische Pulse wirklich eine Neurotransmitterausschüttung stattfindet. ($T_{\text{pulse}} \cdot \lambda_1 \cdot \overline{uR}_{12}$, siehe auch Gleichung 14 in (Mayr et al, 2009a)). Für den PSC muss die $\overline{uR}$-Antwort noch wie in Gleichung 2.3 mit der Skalierungskonstanten $A$ multipliziert werden. Um daraus den mittleren PSC über die gesamte Strecke von Punkt 1 bis Punkt 3 zu berechnen, muss außerdem das Tastverhältnis $b$ berücksichtigt werden, d.h. der Anteil, für den der jeweilige PSC gilt:

$$\overline{PSC} = b \cdot \overline{PSC}_{12} + (1-b) \cdot \overline{PSC}_{23} = A \cdot T_{\text{pulse}} \cdot \left( b\lambda_1 \overline{uR}_{12} + (1-b)\lambda_2 \overline{uR}_{23} \right) . \quad (2.7)$$

## 2.2 Quantalmodell: Analyse und Ergebnisse

Im Folgenden wird die Umschreibung des Quantalmodells aus (Mayr et al, 2009a,b) dazu verwendet, die Analyse zum bevorzugten Pulsmuster einer Synapse aus (Natschläger and Maass, 2001) zu erweitern und insbesondere die Vorhersagekraft der absoluten Zeitkonstanten zu testen. Eine der möglichen Charakteristiken, an denen sich der Übergang zwischen den beiden Hauptarbeitsgebieten aus Abbildung 2.1 festmachen lässt, ist die Modulationsfrequenz $f_m$. Mit dem expliziten Ausdruck für den mittleren PSC (Gleichung 2.7), kann mit $f_m$ ein alternativer Ansatz zu dem Kriterium aus (Natschläger and Maass, 2001) verfolgt werden, welches nur die optimale Verteilung einer bestimmten Anzahl Pulse in einem Zeitraum beinhaltet. Zur besseren Vergleichbarkeit mit (Natschläger and Maass, 2001) wird das Tastverhältnis $b(\lambda_1, \lambda_2)$ so eingestellt, dass sich über den gesamten $f_m$-Bereich eine mittlere Pulsrate von $\overline{f} = 20$Hz einstellt, was mit der optimalen Verteilung von 20 Pulsen in einer Sekunde in (Natschläger and Maass, 2001) korrespondiert. Die Maximierung des mittleren PSC über $f_m$ entspricht in der hier durchgeführten Analyse der Suche nach dem optimalen Pulsregime einer Synapse. In diesem Sinne kann die optimale Modulationsfrequenz $f_{m,\text{opt}}$ aus folgender notwendigen Bedingung abgeleitet werden:

$$0 = \frac{\partial \overline{PSC}}{\partial f_m} = AT_{\text{pulse}} \cdot \left( b\lambda_1 \frac{\partial \overline{uR}_{12}}{\partial f_m} + (1-b)\lambda_2 \frac{\partial \overline{uR}_{23}}{\partial f_m} \right) . \quad (2.8)$$

Im Allgemeinen kann dieser Ausdruck nicht in eine explizite Form gebracht werden. Näherungsweise explizite Formeln können angegeben werden, wenn beispielsweise $f_m \tau_{u,\lambda} \gg 1$ angenommen wird, allerdings sind diese Näherungen nicht über den gesamten Parameterraum und den Optimierungsbereich gültig. Deshalb wird Gleichung 2.8 im Folgenden numerisch gelöst. Modulierte (d.h. pulsgruppierte) Pulsfolgen werden trivialerweise nur für ein $f_m$ im Bereich $(0, \frac{1}{2}\lambda_1)$ generiert; andere Werte ergeben eine gleichmässig verteilte Pulsfolge. Wenn also die partielle Ableitung $\frac{\partial \overline{PSC}}{\partial f_m}$ in diesem Intervall nicht das Vorzeichen ändert, d.h. es existiert dort kein lokales Maximum, dann wird die Synapse für diesen Parametersatz für alle Randbedingungen nur für eine reguläre Pulsfolge einen maximalen PSC erzeugen (siehe 2.4A).

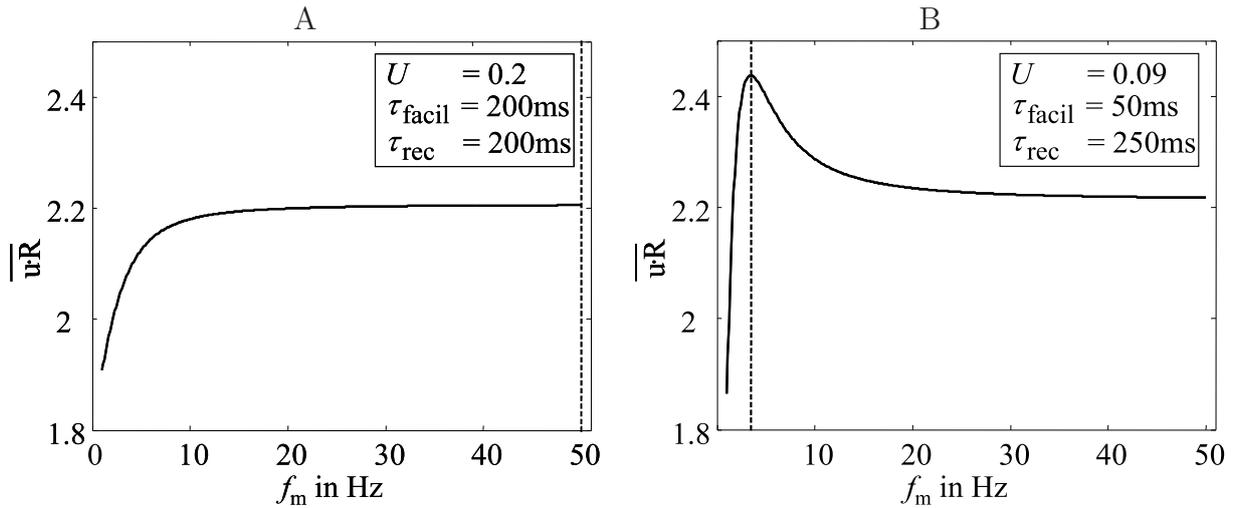

Abbildung 2.4: Verlauf von Gleichung 2.8 über der Modulationsfrequenz $f_m$ für einen Parametersatz ($\tau_{\text{facil}} = 200$ ms, $\tau_{\text{rec}} = 200$ ms, $U = 0.2$), bei dem eine reguläre Pulsfolge bevorzugt wird (A) bzw. für einen Parametersatz ($\tau_{\text{facil}} = 50$ ms, $\tau_{\text{rec}} = 250$ ms, $U = 0.09$), für den eine modulierte Pulsfolge einen höheren PSC erzeugt (B). (A) enthält im Bereich $(0, \frac{1}{2}\lambda_1)$ kein Maximum, sondern steigt stetig an, während in (B) deutlich das Maximum bei der optimalen Modulationsfrequenz zu sehen ist. (Restliche Parameter: $\lambda_1$=100 Hz, $\lambda_2$=5 Hz, $b$ wird für jedes $f_m$ zur besseren Vergleichbarkeit der Ergebnisse untereinander und mit (Natschläger and Maass, 2001) so nachgeführt, dass die mittlere Rate über das gesamte Intervall bei 20 Hz liegt)

Wenn andererseits wie in Abbildung 2.4B ein definiertes Maximum[4] für Gleichung 2.7 existiert, wird von der Synapse generell ein gruppiertes Pulsmuster bevorzugt, was eine wesentlich allgemeinere Aussage darstellt als die nur punktuell gültige Analyse aus (Natschläger and Maass, 2001).

Die in (Natschläger and Maass, 2001) dokumentierten Ergebnisse zeigen, dass ein modulierter Arbeitsbereich der Synapse für niedrige Werte von $U$ und $\tau_{\text{facil}}$ optimal ist, wohingegen für höhere Werte ein reguläres, gleich verteiltes Pulsmuster bevorzugt wird. Dies wird in einem Parametersweep über die Grundparameter des Quantalmodells $U$, $\tau_{\text{rec}}$ und $\tau_{\text{facil}}$ in (Mayr et al, 2009a, Abbildung 8) ebenfalls gezeigt. Dank des verminderten Aufwandes der analytischen Lösung ist dieser deutlich ausführlicher als in (Natschläger and Maass, 2001), was eine verbesserte Verallgemeinerung der Ergebnisse zulässt.

Da die optimale Modulationsfrequenz $f_{m,\text{opt}}$ ausser von den Grundparametern auch von den Charakteristiken ($b$, $\lambda_1$, $\lambda_2$) der modulierten Pulsfolge abhängt, wird in einer zweiten Analyse, ebenfalls ausgehend von Gleichung 2.8, diese optimale Frequenz für einen einzelnen synaptischen Parametersatz ($U$, $\tau_{\text{rec}}$, $\tau_{\text{facil}}$) über ($b$, $\lambda_1$, $\lambda_2$) dargestellt (Abbildung 2.5). Es kann eine gute Übereinstimmung zwischen der Simulation des ursprünglichen ite-rativen Quantalmodells und dem mittleren $\overline{uR}$ aus der Analyse des vorherigen Abschnitts beobachtet werden.

---

[4] Es handelt sich bei diesem Extremwert immer um ein Maximum, d.h. entweder ist der PSC für eine bestimmte Modulationsfrequenz maximal und fällt danach wieder ab oder es existiert der in Abbildung 2.4A gezeigte Kurvenverlauf, bei dem aufgrund des mit der Zunahme von $f_m$ abnehmenden Anteils an längeren Intervallen niedriger Pulsrate der PSC stetig zunimmt

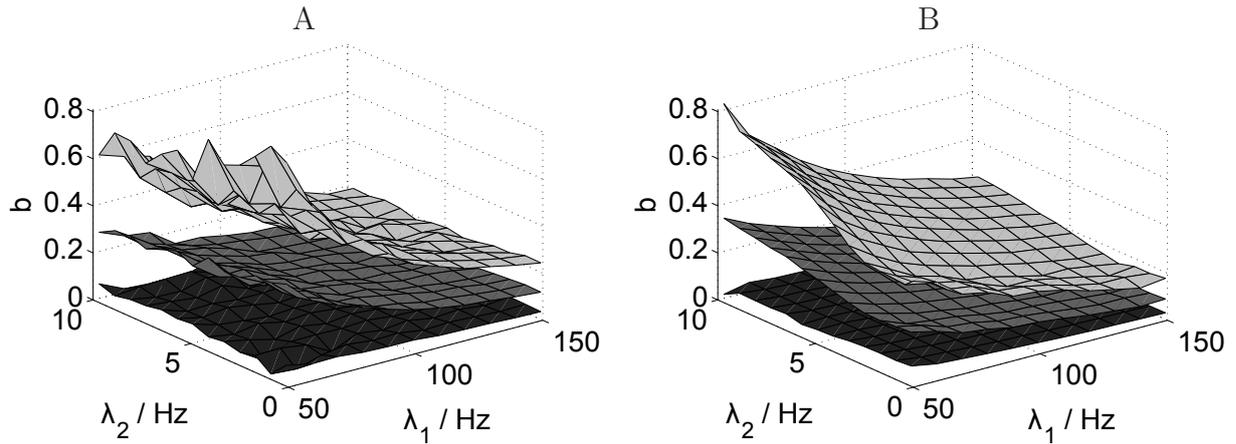

Abbildung 2.5: Optimale Modulationsfrequenz $f_{m,\text{opt}}$ als Funktion der Parameter der modulierten Pulsfolge. Es sind die drei Fälle $f_{m,\text{opt}}=1$ Hz (schwarze Fläche, unten), $f_{m,\text{opt}}=3$ Hz (dunkelgraue Fläche, Mitte) und $f_{m,\text{opt}}=6$ Hz (hellgraue Fläche, oben) dargestellt; (A) Simulation des originalen Quantalmodells, (B) analytische Berechnung basierend auf der nicht-iterativen Umschreibung. Parameter $\tau_{\text{facil}} = 50$ ms, $\tau_{\text{rec}} = 250$ ms, $U = 0.09$.

Hinsichtlich der Bevorzugung von burstenden/modulierten Pulsfolgen besteht wenig Abhängigkeit zwischen $f_{m,\text{opt}}$ und der Pulsrate zwischen den Bursts, d.h. $\lambda_2$. Dies könnte aus der Tatsache herrühren, dass eine Synapse während der Intervalle mit niedriger Pulsrate nur bis zu einem bestimmten Grad in den Ruhezustand zurückkehren kann. Während also die Ruhephase grundsätzlich wichtig ist, um ein hohes $\overline{uR}$ zu erreichen (wie im Folgenden bei Abbildung 2.6 ausgeführt), fällt die genaue Pulsrate während dieser Ruhephase für den Grad der Entspannung nicht ins Gewicht. Im Gegensatz dazu besteht eine klare Abhängigkeit zwischen dem Tastverhältnis $b$ und $f_{m,\text{opt}}$ für ein bestimmtes $\lambda_1, \lambda_2$ (siehe die für jeweils ein $f_{m,\text{opt}}$ von 1, 3 und 6 Hz eingezeichneten Flächen, die in regelmäßiger Progression übereinander liegen). Linear genähert lässt sich dies wie folgt für jeden Datenpunkt $(\lambda_1, \lambda_2)$ formulieren:

$$\frac{b}{f_m} = \text{const.} = b \cdot T = T_{\text{high}} , \qquad (2.9)$$

mit $T$ als Dauer einer Modulationsperiode (Abbildung 2.3) und $T_{\text{high}}$ als Dauer des Intervalls mit hoher Rate, d.h. der Länge des Bursts. Durch die lineare Abhängigkeit ergibt sich somit ein konstantes $T_{\text{high}}$, korrespondierend mit einer konstanten Anzahl an Pulsen während eines Bursts für eine gegebene Burstrate $\lambda_1$. Eine Erklärung dafür wäre, dass für einen gegebenen Parametersatz $U$, $\tau_{\text{facil}}$, $\tau_{\text{rec}}$ und ein gegebenes $\lambda_1$ ein optimales Burstpulsprofil existiert, welches $\overline{uR}$ maximiert. Wenn $b$ somit geändert wird, muss sich $f_m$ proportional mitändern um dieses optimale Profil, d.h. eine konstante Anzahl Pulse im Burst, zu erhalten. Gleichung 2.8 optimiert damit weniger die Modulationsfrequenz $f_m$, es wird eher ein optimales Burstpulsprofil gesucht.

Eine weitere interessante Eigenschaft der Ergebnisse in 2.5 ist die Abnahme des maximalen $b$, für das eine $f_{m,\text{opt}}$ gefunden werden kann, mit der Zunahme von $\lambda_1$. Mithin existiert eine inverse Beziehung zwischen dem maximalen Tastverhältnis, für das ein Burst mit höherem PSC übertragen wird als eine reguläre Pulsfolge, und der Pulsrate während des Bursts. Dies weist darauf hin, dass das optimale Pulsprofil eine (beinahe) invariante Anzahl an Pulsen während des Bursts beinhaltet, die auch bei Veränderung von $\lambda_1$ beibehalten wird. Nach Gleichung 2.9 kann die Anzahl an Pulsen während des Bursts zu $b/f_m \cdot \lambda_1$ berechnet werden, was eine mittlere Pulsanzahl von $N=4.3$ für $\lambda_1 = 150$ Hz und $N=3.6$ für $\lambda_1 = 50$ Hz ergibt. Die Ähnlichkeit dieser Werte liegt darin begründet, dass das optimale $\overline{uR}$ während des Bursts von der zeitlichen Entwicklung von $u$ und $R$ bestimmt wird. Diese wiederum hängen von den absoluten Zeitkonstanten ab, die mit $\lambda$ skalieren (siehe Gleichung 2.6); damit kompensieren sich die Veränderung der Zeitkonstanten und $\lambda$ zumindest teilweise gegenseitig, was trotz der Änderung von $\lambda_1$ sehr ähnliche optimale Burstprofile ergibt. Während also der Absolutwert von $\overline{uR}$ mit $\lambda_1$, $b$ und $f_m$ variiert, scheint das zugehörige qualitative Verhalten, d.h. das Burstprofil für maximales $\overline{uR}$, für einen gegebenen Synapsentyp (d.h. für einen bestimmten Parametersatz $U$, $\tau_{\text{facil}}$, $\tau_{\text{rec}}$) konstant zu sein.

Die Tatsache, dass oberhalb 8 Hz keine optimale Modulationsfrequenz mehr gefunden werden kann, d.h. dass in diesem Bereich eine reguläre Pulsfolge immer mehr PSC hervorruft als eine modulierte, liegt vermutlich darin begründet, dass bei derartigem $f_m$ ohnehin ein Übergang von modulierter zu regulärer Pulsfolge stattfindet. In diesem Bereich sind die Burst-/Hochfrequenzphasen zu kurz, um ein Gruppieren von Pulsen zu erlauben, während die Pausenintervalle mit niedriger Pulsrate zu kurz sind, um eine substantielle Erhohlung von $u$ und $R$ zuzulassen. Somit ruft eine bestimmte Anzahl Pulse ein höheres $\overline{uR}$ hervor, wenn sie über die betrachtete Zeitspanne gleichmässig verteilt sind.

Die bis jetzt angeführten Analysen haben die Verwendbarkeit der analytischen Lösung für deutlich im Aufwand reduzierte Parametersweeps gezeigt. Es wurden Aussagen zu den in einer Synapse bei einer modulierten Pulsfolge stattfindenden grundsätzlichen Mechanismen abgeleitet, basierend auf der optimalen Modulationsfrequenz. Im Folgenden soll darauf aufbauend in größerem Detail die Aussagekraft einzelner Kennzahlen der Herleitung, inbesondere der neuen absoluten Zeitkonstanten, für die Synapse untersucht werden.

Abbildung 2.6 zeigt die absoluten Zeitkonstanten von $u(t)$ und $R(t)$ in Abhängigkeit der Pulsfrequenz $\lambda$ für zwei unterschiedliche Parametersätze. Die zugehörigen optimalen Pulsfolgen aus (Natschläger and Maass, 2001, Abbildung 5B) zeigen einen auffälligen Übergang zwischen regulären Pulsfolgen mit $\lambda$=20 Hz (Abbildung 2.6A) und einem moduliertem Arbeitsbereich, mit einer Burstpulsfrequenz von ca. 100 Hz und einer Burstwiederholrate von 8 Hz (Abbildung 2.6B).

Es zeigt sich in Abbildung 2.6A, dass die absoluten Zeitkonstanten für u und R bei $\tau_{\text{rec}}=60$ ms für niedrige bis mittlere Pulsfrequenzen beinahe gleich sind. Wenn man zusätzlich in Betracht zieht, dass $u(t)$ und $R(t)$ genau entgegengesetztes Zeitverhalten haben (ansteigend/abfallend, siehe Abbildung 2.2), müssen folglich zur Maximierung des Produktes $\overline{uR}$ die Kurven von $u(t)$ und $R(t)$ in gleichmäßigen Intervallen mit Pulsen abgetastet werden. Für eine vorgegebene feste Pulszahl müssen damit diese Pulse gleichmässig über das Zeitintervall verteilt werden.

Im Unterschied hierzu ist für den höheren Wert von $\tau_{\text{rec}}$, 280 ms, die absolute Zeitkonstan-

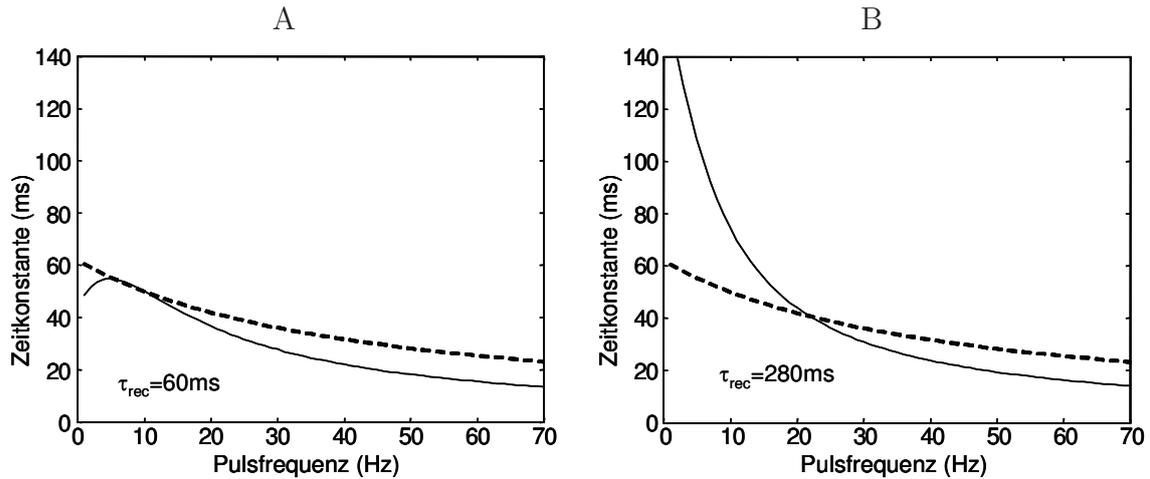

Abbildung 2.6: Zeitkonstanten $\tau_{u,\lambda}$ (gestrichelt) und $\tau_{R,\lambda}$ (durchgezogen) in Abhängigkeit der Pulsfrequenz $\lambda$ für zwei Parametersätze aus Abbildung 5B von (Natschläger and Maass, 2001), Synapsen bevorzugen (A) reguläre Pulsfolge, (B) burstende Pulsfolge; (restliche Parameter $U = 0.32$, $\tau_{\text{facil}} = 62$ ms).

te für $R(t)$, $\tau_{R,\lambda}$, bei niedrigen Pulsfrequenzen wesentlich höher als die Zeitkonstante für $u(t)$. Somit integriert $R(t)$ Ereignisse über einen längeren Zeitraum und würde durch eine reguläre Rate dauerhaft niedrig gehalten werden, während eine modulierte Pulsfolge eine Erholungszeit zwischen Bursts bereitstellt. Das Produkt $\overline{uR}$ hängt damit in den Intervallen niedriger Raten hauptsächlich von $R(t)$ ab, da $u(t)$ aufgrund der kürzeren Zeitkonstante sich bereits erholt hat. Während Bursts bewegen sich sowohl $R(t)$ als auch $u(t)$ im Mittel nahe bei ihren entspannten Werten aus der Erholungszeit, da die Burstdauer in derselben Größenordnung wie die entsprechenden absoluten Zeitkonstanten liegt. Es fällt also die Adaptation des Quantalmodells während hoher Pulsraten, insbesondere das Absinken von $R(t)$, für einen Burst von 3 oder 4 Pulsen nicht ins Gewicht, es wird ein hoher PSC produziert durch die Kombination aus hoher Rate und (immer noch) entspanntem $R(t)$ und $u(t)$. Der bevorzugte Arbeitsbereich einer derartigen Synapse ist damit eine Kombination aus kurzen Intervallen mit hoher Pulsrate und Erholungsphasen mit deutlich niedrigerer Pulsrate.

Die Entscheidung, ob von einer Synapse burstende (d.h. modulierte) oder reguläre Pulsfolgen bevorzugt werden, könnte somit beispielsweise auf einem Kriterium basieren, welches das Verhältnis zwischen der Konvergenzzeitkonstante für $R$ für niedrige und hohe Raten, $\tau_{R,\lambda_2}$ und $\tau_{R,\lambda_1}$, beinhaltet. Dieses Kriterium gibt die grundlegende Intuition wieder, dass $\tau_{R,\lambda_2}$ verglichen mit $\tau_{R,\lambda_1}$ relativ niedrig sein sollte, damit sich $R$ während der Intervalle mit niedriger Rate möglichst schnell auf einen hohen Wert erholen kann (Mayr et al, 2009a). Demgegenüber sollte $\tau_{R,\lambda_1}$ relativ zur Burstdauer möglichst groß sein, damit $R$ während der Bursts (Intervalle mit hoher Rate) nicht zu stark absinkt. Damit sollte ein Parametersatz, der in ein niedriges $\tau_{R,\lambda_2}$ relativ zu $\tau_{R,\lambda_1}$ resultiert, bevorzugt modulierte Pulsfolgen

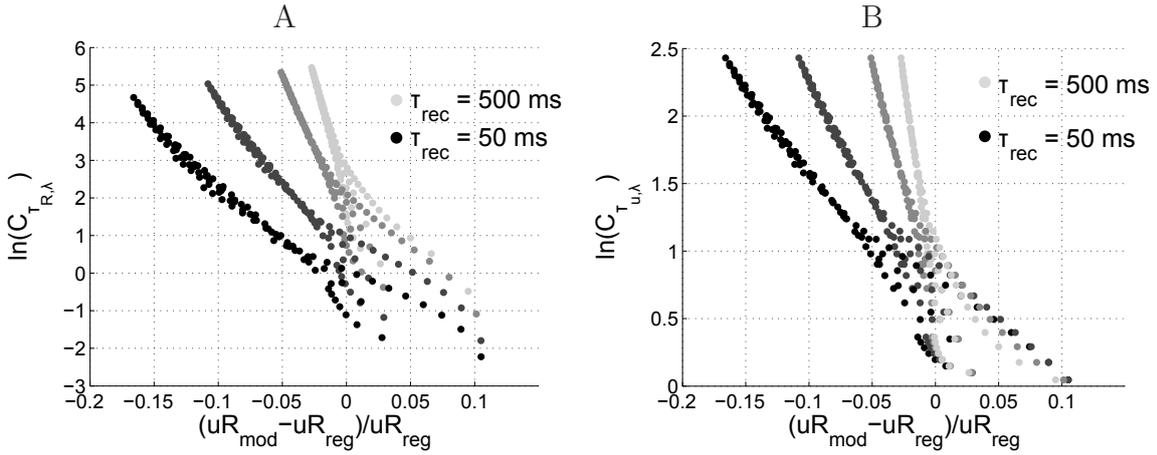

Abbildung 2.7: Korrelation zwischen dem Kriterium basierend auf den Konvergenzzeitkonstanten aus (Mayr et al, 2009a,b) (y-Achse) und dem normierten Unterschied zwischen der $\overline{uR}$-Antwort einer regulären und einer modulierten-Pulsfolge (x-Achse); (A) Kriterium aus den absoluten Zeitkonstanten $\tau_{R,\lambda_2}$ und $\tau_{R,\lambda_1}$, (B) Kriterium aus den absoluten Zeitkonstanten $\tau_{u,\lambda_2}$ und $\tau_{u,\lambda_1}$. Jeder der Punkte entspricht einem Parametersatz $(U, \tau_{\text{facil}}, \tau_{\text{rec}})$. Eine Zunahme von $\tau_{\text{rec}}$ wird als Übergang von schwarz zu hellgrau dargestellt (die einzelnen Werte für $\tau_{\text{rec}}$ sind 50 ms, 100 ms, 250 ms und 500 ms).

übertragen. Dies entspräche beispielsweise folgendem Diskriminanzkriterium:

$$\frac{\tau_{R,\lambda_2}}{\tau_{R,\lambda_1}} = C_{\tau_{R,\lambda}} = \frac{\frac{2}{3\tau_{\text{rec}} \cdot U\tau_{\text{facil}} \cdot \lambda_1} - \lambda_1^2 \cdot \tau_{\text{facil}} \cdot \ln(1-U) - \frac{2}{3}\lambda_1 \cdot \ln U}{\frac{2}{3\tau_{\text{rec}} \cdot U\tau_{\text{facil}} \cdot \lambda_2} - \lambda_2^2 \cdot \tau_{\text{facil}} \cdot \ln(1-U) - \frac{2}{3}\lambda_2 \cdot \ln U} \qquad (2.10)$$

Abbildung 2.7A zeigt eine Darstellung dieses Kriteriums, wobei jeder Punkt im Diagramm einem Datensatz $(U, \tau_{\text{facil}}, \tau_{\text{rec}})$ entspricht. Für die Beurteilung des bevorzugten Arbeitsbereichs der Synapse für den entsprechenden Datensatz wird auf der x-Achse der normalisierte Unterschied zwischen dem $\overline{uR}$ einer regulären Pulsfolge und dem $\overline{uR}$ einer burstenden Pulsfolge aufgetragen. Zur leichteren Übersicht ist auf der y-Achse statt des Quotienten $\tau_{R,\lambda_2}/\tau_{R,\lambda_1}$ selbst sein natürlicher Logarithmus aufgetragen. Es besteht eine klare Korrelation zwischen dem über die nicht-iterativen Zeitkonstanten definierten Kriterium und dem auf der x-Achse angetragenen Maß für die Bevorzugung einer modulierten Pulsfolge von der mittels des jeweiligen Parametersatzes definierten Synapse.

Interessanterweise scheint zusätzlich ein Parameter den Anstieg der Korrelationsgeraden zu beeinflussen. Aus der Grauwertdarstellung des $\tau_{\text{rec}}$-Verlaufs in Abbildung 2.7A wird ersichtlich, dass es sich dabei um die Zeitkonstante für $R_n$ der ursprünglichen iterativen Formulierung des Quantalmodells handelt, d.h. für hohe $\tau_{\text{rec}}$ wird bei gleichbleibendem $U$ und $\tau_{\text{facil}}$ das Regime modulierter Pulsfolgen früher erreicht (positiver Bereich der x-Achse). Für die Erklärung dieses Phänomens muss ausser den Zeitkonstanten noch die relative Höhe von $R_c$ für hohe und niedrige Raten hinzugezogen werden. Dies lässt sich damit begründen, dass für das Gesamt-$uR$ und damit die Bevorzugung einer burstenden oder regulären Rate sowohl das über die Zeitkonstanten vorgegebene dynamische Verhalten als auch der jeweils beim Einschwingen zu erreichende Konvergenzwert von Wichtigkeit sind (siehe Abbildung 2.3). Ein entsprechender Mechanismus für diesen von $\tau_{\text{rec}}$ abhängigen Kipppunkt zwischen burstender und regulärer Pulsfolge wird in (Mayr et al, 2009a) diskutiert.

Für die absoluten Zeitkonstanten $\tau_{u,\lambda_2}$ und $\tau_{u,\lambda_2}$ lässt sich, wie aus Abbildung 2.7B ersichtlich, eine ähnliche Gesamtüberlegung anstellen. Das zugehörige Kriterium lautet dann (es wird wieder der für die Darstellung eingeführte natürliche Logarithmus vernachlässigt):

$$\frac{\tau_{u,\lambda_2}}{\tau_{u,\lambda_1}} = C_{\tau_{u,\lambda}} = \frac{1 - \lambda_1 \cdot \tau_{\text{facil}} \cdot \ln(1-U)}{1 - \lambda_2 \cdot \tau_{\text{facil}} \cdot \ln(1-U)} \qquad (2.11)$$

Es ergibt sich, wie für das auf $\tau_{R,\lambda}$ basierende Kriterium, eine starke Diskriminanz bezüglich des bevorzugten Arbeitsbereiches der Synapse, und wiederum ein von $\tau_{\text{rec}}$ abhängiger stärkerer oder schwächerer Übergang zwischen den Arbeitsbereichen.

Mithin wurde in diesem Abschnitt gezeigt

- dass die nicht-iterative Formulierung das Verhalten des ursprünglichen Quantalmodells sehr gut wiedergibt (siehe Abbildung 2.5 und weitere Parametersweeps in (Mayr et al, 2009a))
- dass durch die nicht-iterative Formulierung grundsätzliche Wirkmechanismen an der Synapse aufgezeigt werden können, die in früheren Arbeiten nur fragmentarisch sichtbar waren (z.B. optimales Pulsprofil in Gleichung 2.9 im Vergleich zu der numerischen Optimierung aus (Natschläger and Maass, 2001))
- dass eine enge Verbindung zwischen den aus der nicht-iterativen Formulierung abgeleiteten absoluten Zeitkonstanten und dem Synapsenverhalten existiert, was die Charakterisierung des synaptischen Übertragungsverhaltens entlang eines Kontinuums ermöglicht

## 2.3 Relation des Quantalmodells zu Langzeitplastizität

Während die separate Betrachtung der präsynaptischen Kurzzeitplastizität im letzten Abschnitt sinnvoll zur Analyse von Übertragungsvorgängen auf entsprechenden Zeitskalen ist, kann damit noch keine Aussage über die Interaktion der verschiedenen zeitlichen Ausprägungen von Plastizität in einem Gesamtplastizitätsmodell getroffen werden. Speziell Kurzzeit- und Langzeitplastizität können jedoch nicht getrennt betrachtet werden, da kurzzeitige Adaptionseffekte maßgeblich die Induktion von Langzeitplastizität beeinflussen (Froemke and Dan, 2002; Froemke et al, 2006; Nevian and Sakmann, 2006; Sjöström et al, 2001; Wang et al, 2005). In gängigen aktuellen Plastizitätsmodellen wird dies in Form verschiedenster Filtervorgänge auf den prä- und/oder postsynaptischen Pulsfolgen berücksichtigt (Badoual et al, 2006; Clopath et al, 2010; Lu et al, 2007; Morrison et al, 2008; Pfister and Gerstner, 2006). Allerdings wurden in den erwähnten Modellen die Adaptionsvorgänge nur phenomenologisch eingeführt, d.h. zur Erklärung von Effekten bei der Langzeitplastizität. Somit fehlt für diese Varianten der Kurzzeitplastizität im Gegensatz zu dem explizit anhand Zellmechanismen eingeführten Quantalmodell (Markram et al, 1998) die biophysikalische Motivation sowohl hinsichtlich der Form der mathematischen Beschreibung als auch der durch die Beschreibung erfassten Wirkvorgänge.

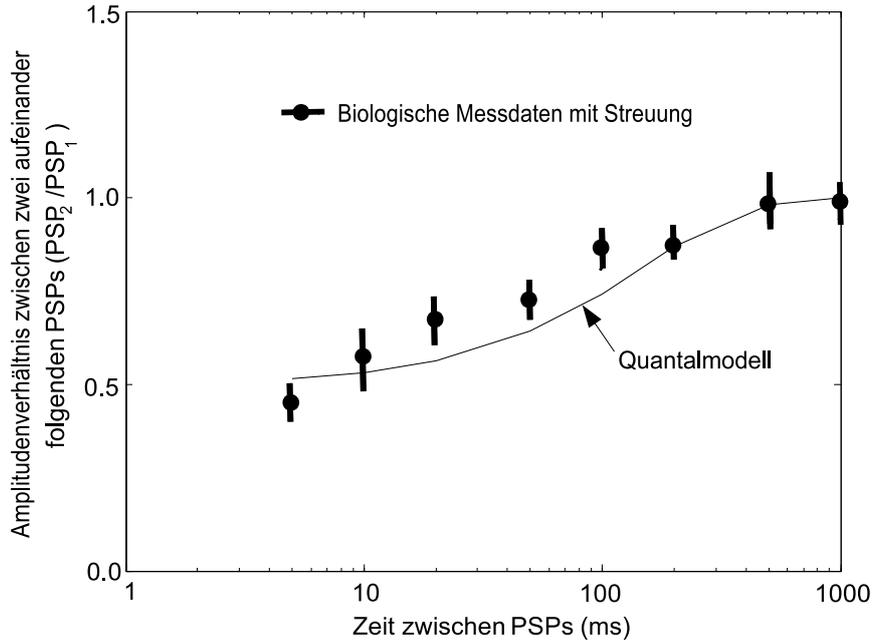

Abbildung 2.8: Verminderung des zweiten von zwei kurz aufeinander folgenden PSPs an einer Synapse (Paired Pulse Depression), als Funktion des zwischen den PSPs liegenden Zeitintervalls, Messdaten aus (Froemke et al, 2006). Parameter Quantalmodell $\tau_{\text{facil}} \to 0$, $\tau_{\text{rec}} = 150$ ms, $U = 0.5$

Für eine Verwendung innerhalb des im Kapitel 3 implementierten Gesamtplastizitätsmodells und der Hardwarerealisierung soll deshalb an Stelle einer der o.a. phenomenologischen präsynaptischen Formulierungen das Quantalmodell verwendet werden. Für diesen Verwendungszweck muss für das Quantalmodell jedoch ein zumindest indikativer Nachweis angetreten werden, dass diese Form der Kurzzeitplastizität kompatibel mit Messdaten und Mechanismen von Langzeitplastizität ist. Zu diesem Zweck soll versucht werden, mit dem Quantalmodell Messdaten der sogenannten Paired Pulse Depression (PPD) nachzustellen. PPD beschreibt das Phänomen der Amplitudenabschwächung des zweiten von zwei isolierten PSPs an einer Synapse (Koch, 1999). In (Froemke et al, 2006) wurde nachgewiesen, dass PPD an denselben Synapsen auftritt, die auch präsynaptische Kurzzeitadaption der Langzeitplastizität aufweisen. Ein numerischer Vergleich einer aus der PPD abgeleiteten Zeitkonstanten mit der zur Erklärung der Langzeitplastizitätseffekte nötigen präsynaptischen Zeitkonstanten zeigt ebenfalls eine gute Übereinstimmung (Froemke et al, 2006). Darüber hinaus wurde gezeigt, dass ein chemisches Unterdrücken von PPD die für die Langzeitplastizität gefundene präsynaptische Adaption ebenfalls zu großen Teilen aufhebt, d.h. dass PPD als Form von Kurzzeitplastizität maßgeblichen Einfluss auf die Ausprägung der Langzeitplastizität hat. Mithin ist ein Nachweis von PPD-Verhalten eine gute Indikation dafür, dass das Quantalmodell eine sinnvolle Ergänzung für ein Modell der Langzeitplastizität darstellt, da somit die bei Langzeitplastizität gefundene präsynaptische Adaption mit guter biophysikalischer Relevanz nachgestellt werden kann.

Da PPD eine reine Abschwächung der PSP-Amplitude darstellt, muss im Quantalmodell der Verstärkungsmechanismus abgeschaltet werden, d.h. $\tau_{\text{facil}} \to 0$, damit $u_{n+1} = u_n = U$.

Abbildung 2.8 zeigt die in (Froemke et al, 2006) aufgeführten Messdaten zur PPD sowie den Verlauf des Quantalmodells aus Gleichung 2.3. Das Quantalmodell stellt nicht den für PPD postulierten rein exponentiellen Zusammenhang zwischen der präsynaptischen Zeitdifferenz und dem Amplitudenverhältnis her (Froemke et al, 2006; Koch, 1999). Abbildung 2.8 zeigt aber, dass trotzdem für eine biologisch realistische Parametrisierung des Quantalmodells (Markram et al, 1998) eine gute Übereinstimmung mit den Messdaten hergestellt werden kann[5].

## 2.4 Überlegungen zur Schaltungsimplementierung des Quantalmodells

Gemäß der Schlussfolgerung des letzten Abschnitts soll der MAPLE (Kapitel 5) mit einer biologisch realistischen und für die Plastizitätsmodellierung relevanten Formulierung von Kurzzeitplastizität, d.h. mit einer Implementierung des Quantalmodells, ausgestattet werden. In der Schaltungsrealisierung lässt sich die in den Gleichungen 2.1 und 2.2 geforderte exponentielle Bewertung eines Zeitintervalls nicht direkt durchführen, d.h. die ursprüngliche iterative, auslesezeitpunktsbasierte Formulierung kann hier nicht verwendet werden. Üblicherweise wird deshalb z.B. in Schaltungsrealisierungen von STDP 'Trace'-basiert gearbeitet (Koickal et al, 2007), d.h. ein zeitkontinuierlicher exponentieller Abfall wird zu einem bestimmten Zeitpunkt gestartet und zu einem zweiten Zeitpunkt ausgelesen. Abbildung 2.9 zeigt den virtuellen Verlauf von $u_n$ und $R_n$ zwischen den jeweils bei den präsynaptischen Pulsen stattfindenden Aktualisierungen als zeitkontinuierliche Variablen $u(t)$ und $R(t)$, wobei $u(t_n) = u_n$ und $R(t_n) = R_n$. Wie aus Abbildung 2.9 ersichtlich, kann sowohl der Zeitverlauf von $u(t)$ als auch $R(t)$ als der Spannungsverlauf auf einem RC-Glied betrachtet werden, auf das zu jedem präsynaptischen Puls ein Ladungspaket addiert ($\Delta u_n$) oder subtrahiert ($\Delta R_n$) wird. Interessanterweise ist diese additive Formulierung damit in der schaltungstechnischen Realisierung sogar günstiger als beispielsweise die für Langzeitplastizität eingeführte vereinfachte Variante aus (Froemke et al, 2006), da dort die exponentiellen Zeitfenster aller vergangenen PSCs multiplikativ miteinander verknüpft werden, was eine große Anzahl an analogen Multiplizierern erfordern würde, um eine hinreichende Einbeziehung zurückliegender PSCs zu gewährleisten. Für eine 'Trace'-basierte Umschreibung von $u_n$ in $u(t)$ kann demnach folgende Gleichung abgeleitet werden:

$$u(t) = (u(t_n^{pre}) - U + \Delta u_n)\,\mathrm{e}^{-\frac{t-t_n^{pre}}{\tau_{\text{facil}}}} + U \ , \ t_n^{pre} < t \leq t_{n+1}^{pre} \tag{2.12}$$

d.h. wie aus Abbildung 2.9 ersichtlich besteht der Zeitverlauf von $u(t)$ zwischen präsynaptischen Pulsen aus einem Abklingen auf $U$ in Ruhephasen, während präsynaptische Pulse $u(t)$ um $\Delta u_n$ erhöhen, worauf wieder ein exponentieller Abfall mit $\tau_{\text{facil}}$ einsetzt. Gleichsetzen von Gleichung 2.12 und 2.1 bei $t = t_{n+1}^{pre}$ ergibt:

$$(u_n - U + \Delta u_n)\,\mathrm{e}^{-\frac{t_{n+1}^{pre}-t_n^{pre}}{\tau_{\text{facil}}}} + U = u_n \mathrm{e}^{-\frac{t_{n+1}^{pre}-t_n^{pre}}{\tau_{\text{facil}}}} + U \cdot \left(1 - u_n \mathrm{e}^{-\frac{t_{n+1}^{pre}-t_n^{pre}}{\tau_{\text{facil}}}}\right) \tag{2.13}$$

---

[5]Ein auf geringste Abweichung optimierter Parametersatz wäre nur unwesentlich besser als die gezeigte Variante, der qualitative Verlauf bzw. die Art der Abweichung bleibt gleich.

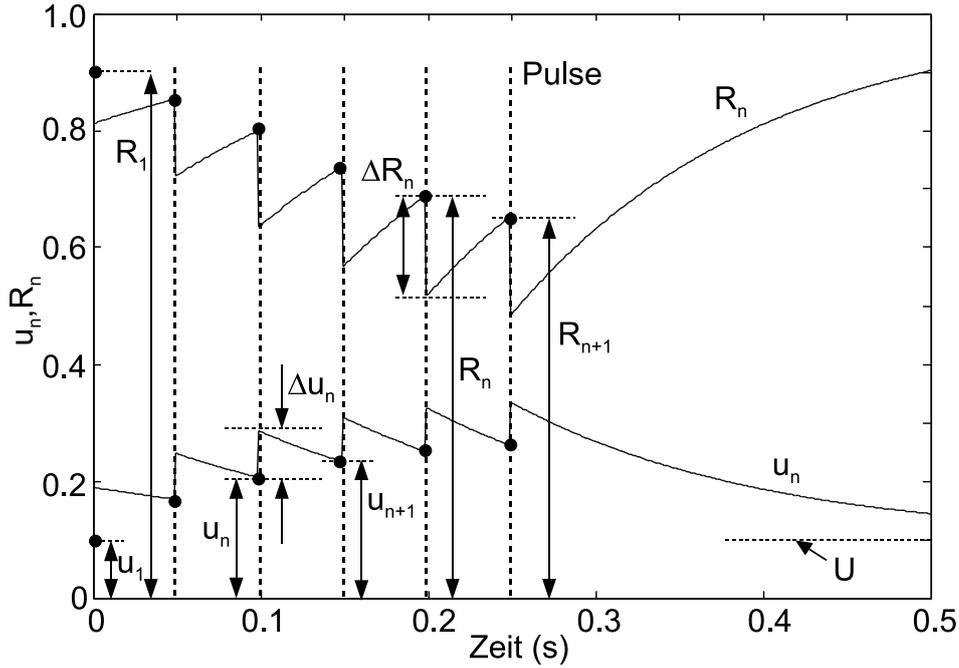

Abbildung 2.9: Detailverhalten des Quantalmodells gezeigt für eine reguläre Pulsfolge, 6 Pulse bei 20 s$^{-1}$, präsynaptische Eingangspulse, resultierende Utilization $u(t)$ sowie Recovery $R(t)$, Parameter $\tau_{\text{facil}} = 150$ ms, $\tau_{\text{rec}} = 150$ ms, $U = 0.1$. Die Kurven für $u(t)$ und $R(t)$ zeigen die jeweils zu den Pulszeitpunkten wirksamen Werte $u_n$ und $R_n$ (schwarze Punkte), die instantan nach dem Puls addierten $\Delta$'s sowie den exponentiell abklingenden Zeitverlauf bis zum nächsten Zeitpunkt eines präsynaptischen Pulses. Startwert für $u_1$ und $R_1$ sind 0.1 bzw. 0.9, d.h. es wird für den Anfang der Pulsfolge von einer Synapse in Ruhe ausgegangen.

Damit lässt sich $\Delta u_n$ wie folgt berechnen:

$$\Delta u_n = U \cdot (1 - u_n) = U \cdot (1 - u(t_n^{pre})) \tag{2.14}$$

An analogem Berechnungsaufwand für jeden neuen präsynaptischen Puls ist damit für $\Delta u_n$ nur eine feste (d.h. vorkonfigurierte) Skalierung von $u(t_n^{pre})$ nötig. Aus Gleichung 2.14 ist ausserdem ersichtlich, dass $u(t)$ bei 1 sättigt, d.h. dass dort der Zuwachs $\Delta u_n$ für jeden weiteren präsynaptischen Puls gegen 0 geht. Die vollständige additive Bildungsvorschrift für $u(t)$ ergibt sich zu (vergleiche Gleichung 2.12):

$$u(t) = (u(t_n^{pre}) - U + U \cdot (1 - u(t_n^{pre}))) \, e^{-\frac{t-t_n^{pre}}{\tau_{\text{facil}}}} + U \tag{2.15}$$

Eine ähnliche Herleitung kann für $R(t)$ durchgeführt werden:

$$R(t) = 1 - (1 - R(t_n^{pre}) + \Delta R_n) \, e^{-\frac{t-t_n^{pre}}{\tau_{\text{rec}}}} \, , \ t_n^{pre} < t \leq t_{n+1}^{pre} \tag{2.16}$$

d.h. der Zeitverlauf von $R(t)$ zwischen präsynaptischen Pulsen zeigt einen Anstieg auf 1

in Ruhephasen, während präsynaptische Pulse $R(t)$ um $\Delta R_n$ vermindern, worauf wieder ein exponentieller Anstieg mit $\tau_{\text{rec}}$ einsetzt. Über Gleichsetzen von Gleichung 2.16 und 2.2 lässt sich $\Delta R_n$ wie folgt berechnen:

$$\Delta R_n = R_n \cdot u_n = R(t_n^{pre}) \cdot u(t_n^{pre}) \tag{2.17}$$

Mithin sättigt $R(t)$ bei 0, d.h. das $\Delta R_n$ für jeden weiteren präsynaptischen Puls geht dort gegen 0. Die vollständige Bildungsvorschrift für $R(t)$ lautet:

$$R(t) = 1 - (1 - R(t_n^{pre}) + R(t_n^{pre}) \cdot u(t_n^{pre}))\, \mathrm{e}^{-\frac{t-t_n^{pre}}{\tau_{\text{rec}}}} \tag{2.18}$$

Für jeden neuen präsynaptischen Puls ist damit für $\Delta R_n$ eine analoge Multiplikation von $R(t_n^{pre})$ mit $u(t_n^{pre})$ nötig. Allerdings kann diese auch mit der ohnehin für die Bildung des PSC gemäß Gleichung 2.3 notwendigen Multiplikation kombiniert werden. $\Delta R_n$ und $\text{PSC}_n$ unterscheiden sich somit nur um eine Skalierung mit festem $A$. Der Gesamtaufwand für die Schaltungsumsetzung einer biologisch fundierten präsynaptischen Adaption gemäß des Quantalmodells beschränkt sich damit auf die Bereitstellung der RC-Glieder, Ladungsaddition/-subtraktion und eine einzige Multiplikation.

Um die obige additive Umschreibung des Quantalmodells in der Schaltungsumsetzung für ein gewünschtes Verhalten zu konfigurieren, kann nun wieder die nicht-iterative Umschreibung aus Abschnitt 2.1 verwendet werden. Wie in Abschnitt 2.2 gezeigt, eignet sich diese aufgrund des direkten Zusammenhanges zwischen z.B. den effektiven Zeitkonstanten $\tau_{u,\lambda}$ und $\tau_{R,\lambda}$ sehr gut zur Charakterisierung des Synapsenverhaltens. Es muss lediglich ein Weg von den Charakterisierungskenndaten zurück zu den für die Konfiguration der Schaltungsumsetzung nötigen originalen Parametern $U$, $\tau_{\text{facil}}$ und $\tau_{\text{rec}}$ gefunden werden.

Abbildung 2.7B zeigt, dass $\tau_{\text{rec}}$ je nach gewünschter 'Schärfe' des Übergangs zwischen den beiden Arbeitsbereichen gewählt werden kann. Danach wird ein absoluter Arbeitsbereich gemäß des auf der x-Achse angetragenen Bereichskriteriums $(uR_{\text{mod}} - uR_{\text{reg}})/uR_{\text{reg}}$ zugeordnet. Damit und mit dem gewählten $\tau_{\text{rec}}$ kann der zugehörige Wert für $C_{\tau_{u,\lambda}}$ (d.h. $\frac{\tau_{u,\lambda_2}}{\tau_{u,\lambda_1}}$) auf der y-Achse abgelesen werden. Aus Gleichung 2.11 lässt sich danach der für ein festes $C_{\tau_{u,\lambda}}$ geltende Zusammenhang zwischen $\tau_{\text{facil}}$ und $U$ herleiten:

$$\frac{1}{\tau_{\text{facil}}} = \frac{\lambda_1 - C_{\tau_{u,\lambda}} \cdot \lambda_2}{C_{\tau_{u,\lambda}} - 1} \cdot (-\ln(1-U)) = f^*_{\text{facil}} \cdot (-\ln(1-U)) \tag{2.19}$$

In der obigen Gleichung wurde zur leichteren Lesbarkeit eine Substitution mit $f^*_{\text{facil}}$ durchgeführt:

$$f^*_{\text{facil}} = \frac{\lambda_1 - C_{\tau_{u,\lambda}} \cdot \lambda_2}{C_{\tau_{u,\lambda}} - 1} \tag{2.20}$$

d.h. einer Frequenz, die noch mit $(-\ln(1-U))$ skaliert werden muss, um $1/\tau_{\text{facil}}$ zu ergeben. Wenn $\lambda_1$ und $\lambda_2$ z.B. anhand der Pulsprofildiskussion um Gleichung 2.9 und/oder einem zu erreichenden Absolutwert von $\overline{uR}$ ausgewählt werden, ist Gleichung 2.19 eindeutig bestimmt. Mit dem in Abbildung 2.7B verwendeten $(uR_{\text{mod}} - uR_{\text{reg}})/uR_{\text{reg}}$, dem vorab gewählten $\tau_{\text{rec}}$ und der Beziehung $\tau_{\text{facil}}$ zu $U$ aus Gleichung 2.19 kann nun aus Abbildung 8 in (Mayr et al, 2009a) ein eindeutiges $\tau_{\text{facil}}$ und $U$ ausgelesen werden. Somit ist der Parametersatz zur Konfiguration der Schaltungsimplementierung des Quantalmodells vollständig anhand der Bereichsanalyse des nicht-iterativen Modells bestimmbar.

Alternativ zu dem graphischen Auslesen aus Abbildung 8 in (Mayr et al, 2009a) können $\tau_{\text{facil}}$ und $U$ auch numerisch festgelegt werden. Dazu wird das auf $\tau_{R,\lambda}$ aufbauende Kriterium aus Abbildung 2.7A verwendet. Einsetzen von Gleichung 2.19 in Gleichung 2.10 ergibt:

$$\frac{\tau_{R,\lambda_2}}{\tau_{R,\lambda_1}} = C_{\tau_{R,\lambda}} = \frac{\frac{U \cdot \lambda_1^2}{f^*_{\text{facil}}} - \frac{2}{3}\lambda_1 \cdot U \cdot \ln(U) - \frac{2 \cdot f^*_{\text{facil}} \cdot \ln(1-U)}{3\tau_{\text{rec}} \cdot \lambda_1}}{\frac{U \cdot \lambda_2^2}{f^*_{\text{facil}}} - \frac{2}{3}\lambda_2 \cdot U \cdot \ln(U) - \frac{2 \cdot f^*_{\text{facil}} \cdot \ln(1-U)}{3\tau_{\text{rec}} \cdot \lambda_2}} \qquad (2.21)$$

Der Wert für $C_{\tau_{R,\lambda}}$ kann wiederum wie für Abbildung 2.7B anhand des verwendeten $(uR_{\text{mod}} - uR_{\text{reg}})/uR_{\text{reg}}$-Wertes und anhand von $\tau_{\text{rec}}$ bestimmt werden. Damit, mit $\lambda_1$ und $\lambda_2$, $C_{\tau_{u,\lambda}}$ und $\tau_{\text{rec}}$ kann mit Gleichung 2.21 ein eindeutiges $U$ bestimmt werden[6]. Über die rückwirkende Bestimmung von $\tau_{\text{facil}}$ sind damit ebenfalls alle Parameter definiert.

---

[6] Aufgrund von Ausdrücken wie $U \cdot \ln U$ in Gleichung 2.21 ist dies nur implizit möglich.

# 3 Langzeitplastizität

## 3.1 Bisherige Modelle für Langzeitplastizität

Wie in der Einleitung erwähnt, ist die Induktion von Langzeitplastizität an ein bestimmtes Zusammenspiel von prä- und postsynaptischen Pulsen und/oder weiteren Zustandsvariablen gebunden. Der aktuelle Stand der Wissenschaft lässt sich dabei grob in drei Bereiche einteilen:

- Induktion von Langzeitplastizität aufgrund von **prä- und/oder postsynaptisch anliegenden Pulsraten**. Die meistvertretene Ansicht ist, dass die postsynaptische Rate mit einem Schwellwert beaufschlagt über LTP bzw. LTD entscheidet, während die Stärke der Gewichtsänderung mit der präsynaptischen Rate skaliert (sogenannte BCM-Lernregel, siehe Abbildung 3.1A). Theoretische Untersuchungen derartigen Lernverhaltens finden sich in (Bienenstock et al, 1982), experimentelle Verifikation in (Dudek and Bear, 1992), Verwendung ratengesteuerter Plastizität in neuromorpher VLSI in (Fusi et al, 2000). BCM kann u.a. zur Erklärung von binokularem Wettbewerb und Dominanz (Bienenstock et al, 1982) sowie zum Nachstellen von pulsgebundener Informationsübertragung (Toyoizumi et al, 2005) herangezogen werden.

- Spannungsgesteuerte Induktion, wobei i.d.R. **präsynaptisch** ein sogenannter Tetanus verwendet wird, d.h. eine Abfolge von **100-1000 Pulsen** in kurzem Abstand, verbunden mit einem Setzen der **postsynaptischen Membran** auf einen bestimmten **Spannungswert**. Je nach Spannung kann so LTP oder LTD induziert werden, bzw. unterhalb einer bestimmten Spannung findet trotz Tetanus keine Gewichtsänderung statt (ABS-Lernregel, siehe auch Abbildung 3.1B). Theoretische Untersuchungen derartigen Lernverhaltens finden sich in (Clopath et al, 2010; Garagnani et al, 2009), experimentelle Verifikation in (Artola et al, 1990), Verwendung spannungsgesteuerter Plastizität in neuromorpher VLSI in (Fusi et al, 2000). Mit dieser Art von Plastizität kann beispielsweise das Entstehen von verarbeitungsspezifischen Unterpopulationen von Neuronen erklärt werden (Garagnani et al, 2009), oder verschiedene metaplastische Effekte nachgestellt werden (Ngezahayo et al, 2000).

- Induktion durch Kontrolle des **Zeitintervalls zwischen** Paaren (oder komplexeren Mustern) von **prä- und postsynaptischen Pulsen**. Bei Pulspaaren wird dabei i.d.R. LTP für prä-post Pulspaare ausgelöst, während LTD bei post-prä Paaren entsteht (sogenannte Spike Time Dependent Plasticity (STDP), siehe auch Abbildung 3.1C). Theoretische Untersuchungen derartigen Lernverhaltens finden sich in (Gerstner et al, 1996), experimentelle Verifikation in (Bi and Poo, 1998), Verwendung zeitintervallsgesteuerter Plastizität in neuromorpher VLSI in (Koickal et al, 2007). STDP wurde u.a. eingesetzt, um Reinforcement Learning (Izhikevich, 2007) oder die Entstehung von Small World Topologien (Siri et al, 2007) zu emulieren.

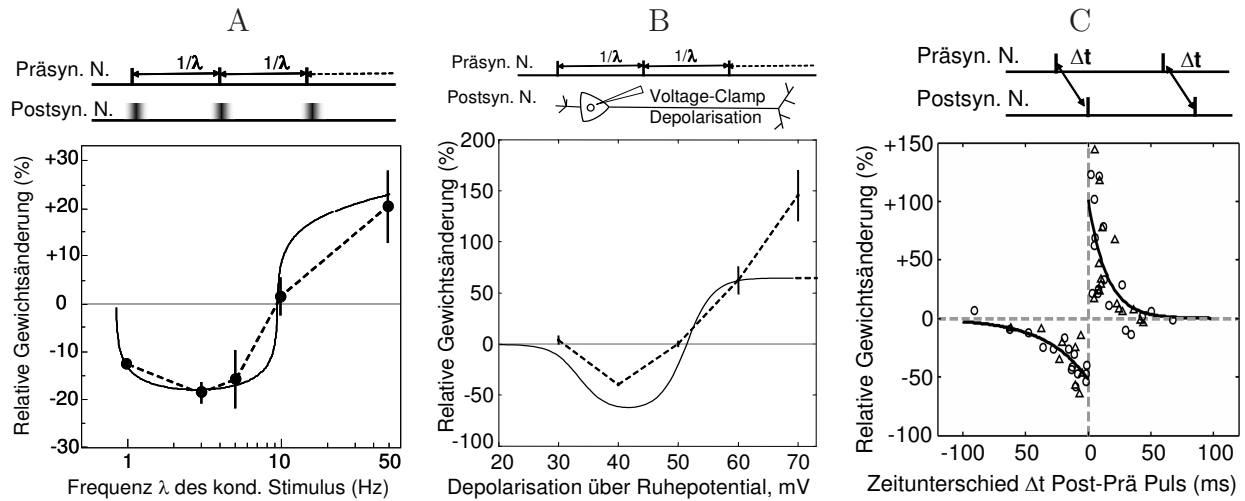

Abbildung 3.1:  Unterschiedliche Plastizitätsprotokolle, jeweils mit schematischer Darstellung des Experiments, von links nach rechts: (A) Ratenbasierte Plastizität, gestrichelte Linie Experiment (Dudek and Bear, 1992), durchgezogene Linie Theorie (Bienenstock et al, 1982) (Die 'verrauschte' Darstellung der postsynaptischen Pulse im Induktionsprotokoll versinnbildlicht die durch verschiedene Einflüsse nur partielle Korrelation zu den präsynaptischen Pulsen, wie im Text erwähnt); (B) Spannungsbasierte Plastizität, gestrichelte Linie Experiment (Ngezahayo et al, 2000), durchgezogene Linie Theorie (Artola and Singer, 1993) (C) Pulszeitpunktabhängige Plastizität, Punktewolke Experiment (Bi and Poo, 1998), durchgezogene Linie Theorie (Gerstner et al, 1996; Song et al, 2000).

Jeder dieser drei Bereiche hat diverse Familien von Modellen für das jeweilige Lernverhalten hervorgebracht, wobei in letzter Zeit in der Literatur verstärkt Anstrengungen unternommen werden, bereichsübergreifende Modelle für jeweils mehrere der o.a. Formen von Langzeitplastizität zu finden. Vor diesem Hintergrund wurde in (Mayr and Partzsch, 2010) der Versuch unternommen, eine gemeinsame Basis für einen Vergleich der verschiedenen aktuellen Plastizitätsmodelle zu schaffen. Die dortige tabellarische Auflistung des Vergleichs ist in verkürzter Weise in Tabelle 3.1 wiedergegeben.

Für den Vergleich wurde zuerst eine möglichst repräsentative und trotzdem überschaubare Menge an biologischen Experimenten ausgewählt, mit deren Nachstellung sich die Modelle bewerten lassen. Orientierungspunkt war dabei, welche Experimente zum einen die einzelnen Facetten der oben geschilderten Bandbreite möglichst gut wiederspiegeln. Zum anderen wurden naturgemäß Experimente gewählt, welche in der Literatur am häufigsten zum Test der Modelle eingesetzt werden (Froemke and Dan, 2002; Pfister and Gerstner, 2006; Senn, 2002; Shah et al, 2006). Jedes Experiment bzw. Plastizitätsprotokoll wird im Folgenden kurz vorgestellt und im weiteren Verlauf (beispielsweise in Tabelle 3.1) mit einer Experimentnummer in eckigen Klammern referenziert: [x]. Für eine weitergehende Diskussion der geschilderten Experimente wird auf (Mayr and Partzsch, 2010) verwiesen.

- Das erste für den Modellvergleich verwendete Experiment ist konventionelles STDP (Bi and Poo, 1998), d.h. die Langzeitplastizität, die durch mehrere Wiederholungen eines einzelnen Pulspaares aus präsynaptischem und postsynaptischem Puls mit verschiedenen positiven und negativen Zeitdifferenzen $\Delta t$ hervorgerufen wird [1] (siehe Abbildung 3.1C).

- In Erweiterung dieses Protokolls wurden in Sjöström et al (2001) Pulspaare bei festem $\Delta t$ in ihrer Wiederholungsfrequenz variiert [2]. Bei diesen Experimenten wurde eine mit dem Anstieg in der Wiederholungsfrequenz korrelierte Verstärkung des synaptischen Gewichts gefunden. Selbst für post-prä Pulspaare, die in einem STDP-Kontext (d.h. bei sehr langsamer Wiederholung der einzelnen Pulspaare) eine Abschwächung des Gewichtes verursachen, wurde für genügend hohe Wiederholungsfrequenzen eine Gewichtsverstärkung gefunden.

- Froemke and Dan (2002) haben in Experimenten mit Pulstripeln nachgewiesen, dass die durch einzelne STDP-Pulspaare ausgelösten Gewichtsänderungen nicht linear summiert werden können, d.h. dass beispielsweise ein prä-post-prä Triplet nicht als separate prä-post/post-prä Interaktion betrachtet werden kann [3].

- Ein weiteres häufiges Experimentprotokoll verwendet Pulsquadruplets in der Form post-prä und prä-post mit symmetrischer Zeitdifferenz $\Delta t$, separiert durch ein kurzes Zeitintervall $T$. Von Wang et al (2005) wurde eine Serie derartiger Experimente an Hippokampusneuronen mit $\Delta t = 5$ ms durchgeführt [4].

- Die Experimente in (Froemke et al, 2006) untersuchen Langzeitplastizität ausgelöst durch die relative Zeitdifferenz von prä- und postsynaptischen Pulsbursts [5].

- Mit den nächsten beiden Experimenten soll die Kompatibilität eines Modells mit dem einflußreichen BCM-Paradigma (Bienenstock et al, 1982) überprüft werden. Dieses Plastizitätsparadigma wurde bereits vielfältig experimentell nachgewiesen (Dudek and Bear, 1992; Mayford et al, 1995; Wang and Wagner, 1999). Wie oben angeführt, sind die Kernaussagen dieses Modells, dass eine niedrige postsynaptische Rate zu LTD führt, während eine hohe Rate LTP hervorruft, wobei die Höhe des jeweiligen LTP bzw. LTD mit der präsynaptischen Rate korreliert. Das erste in diesen Vergleich aufgenommene Experiment korrespondiert mit dem im ersten Anstrich bzw. in Abbildung 3.1A aufgeführten klassischen Tetanusexperiment (Dudek and Bear, 1992), d.h. es wird eine Sequenz von 900 APs im präsynaptischen Neuron ausgelöst [6]. Leider wurde in diesem Experiment die postsynaptische Seite nicht kontrolliert (Bear, 2008), so dass dort eine Rekonstruktion nötig ist, um Plastizitätsmodelle mit diesem Experiment testen zu können. Vermutlich kann für diese Rekonstruktion eine gewisse (verrauschte) Korrelation zwischen prä- und postsynaptischer Pulsfolge angenommen werden (Beggs, 2001), was in Abbildung 3.1A im Pulsprotokoll durch die Grauwertverläufe angedeutet ist. In Tabelle 3.1 werden zwei verschiedene Rekonstruktionen verwendet, die für die Randbedingungen des Versuchs am wahrscheinlichsten erscheinen (Bear (2008), siehe auch Mayr and Partzsch (2010)).

- In einem alternativen Ansatz für Ratenexperimente wurden in (Sjöström et al, 2001) prä-post Pulspaare mit zufälliger Zeitdifferenz verwendet, deren Wiederholungsfrequenz variiert wird [7].

- Die Vergleichsbasis an Experimenten weist eine Bevorzugung von pulsbasierten Zeitdifferenz- und Ratenexperimenten gegenüber spannungsbasierter Plastizität auf, da gegenwärtig nur wenige Modelle für Langzeitplastizität überhaupt einen Span-nungsterm beinhalten (Abarbanel et al, 2002; Clopath et al, 2010; Sjöström et al, 2001). Deshalb hat sich in der aktuellen Literatur auch noch kein Konsens bzgl. des wichtigsten experimentellen Aspekts von spannungsgesteuerter Langzeitplastizität eingestellt. Aus diesem Grund wird für Tabelle 3.1 kein einzelnes Experiment ausgewählt, ein Modell wird bereits als kompatibel mit dieser Art Plastizität eingestuft, wenn Reproduktion von mindestens einem der entsprechenden Experimente (Artola et al, 1990; Ngezahayo et al, 2000; Sjöström et al, 2001) nachgewiesen wird [8].

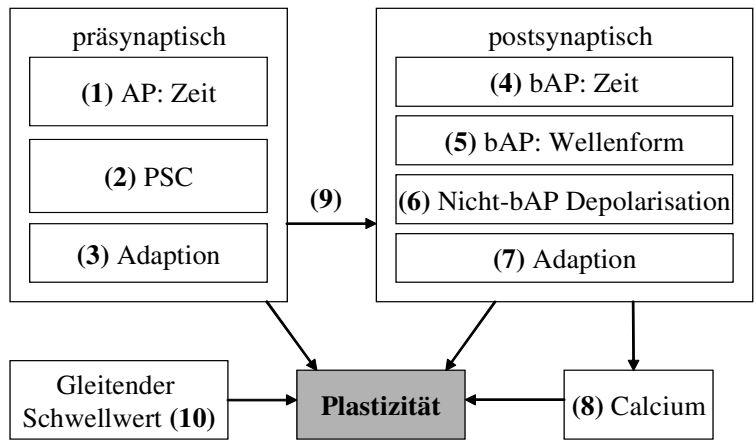

Abbildung 3.2: Übersicht über häufig in Plastizitätsmodellen verwendete Mechanismen. Die angeführte Zahlenkennzeichnung wird in der Klassifizierung der Modelle in Tabelle 3.1 verwendet. Entsprechend würde ein Standard-STDP-Modell (Abbott and Nelson, 2000; Froemke et al, 2006) mit einer Gewichtsänderung basierend auf der Zeitdifferenz von prä- and postsynaptischen Pulsen als (1,4) klassifiziert werden.

Für den in Tabelle 3.1 durchgeführten Vergleich empfiehlt es sich neben der reinen Experimentreproduktion zusätzlich eine Bewertung der Modelle nach den in ihnen enthaltenen Mechanismen auszuführen (siehe Abbildung 3.2). Die Motivation dabei ist, dass Mechanismen, die von mehreren in der Experimentreproduktion gut abschneidenden Modellen geteilt werden, ein Hinweis auf einen zugrundeliegenden fundamentalen biophysikalischen Prozess sein könnten. Solche Mechanismen sollten entsprechend sowohl in zukünftige Modelle aufgenommen werden als auch gezielt in der Biologie verifiziert werden.

Auf der präsynaptischen Seite wird als einfachster Mechanismus meist eine Registrierung der Ankunftszeit des präsynaptischen APs an der Synapse angenommen, korrespondierend mit dem Block 'präsynaptisches AP' in Abbildung 3.2. Die nächste Stufe an Detail wäre eine Nachbildung der analogen postsynaptischen Wellenform, die als Antwort auf einen präsynaptischen Puls entsteht, d.h. der PSC oder PSP. In diesem Fall hat dann nicht nur der Zeitpunkt, sondern auch die Wellenform Einfluß auf die Form der Plastizitätsfunktion (Abarbanel et al, 2002; Saudargiene et al, 2004; Shouval et al, 2002). Kurzzeitplastizität wie in Abschnitt 2.3 eingeführt, würde nach Abbildung 3.2 als Klassifikation eine '3' erhalten. In diesem Fall ist in Kontrast zu Kapitel 2 nicht der präsynaptische Filtervorgang per se interessant, sondern die Auswirkung der Interaktion zwischen den präsynaptischen Pulsen auf die gesamte Langzeitplastizität an der Synapse (Froemke et al, 2006; Lu et al, 2007;

Shah et al, 2006; Sjöström et al, 2001; Zou and Destexhe, 2007).

Das Gegenstück zur präsynaptischen Zeitpunktsregistrierung auf der postsynaptischen Seite ist das sogenannte 'Backpropagating Action Potential' (bAP). In den meisten Modellvorstellungen zu pulszeitpunktbasierter Plastizität wird angenommen, dass von der Soma neben dem Richtung Axon laufenden AP ein zweites Pulssignal erzeugt wird, eben das bAP, welches in den Dendriten zurückläuft (Koch, 1999) und damit an der Synapse den Zeitpunkt des postsynaptischen APs anzeigt. Ein klassisches STDP-Modell (Abbott and Nelson, 2000) wäre demnach anhand Abbildung 3.2 mit '1,4' zu klassifizieren, d.h. die Gewichtsänderung ist nur abhängig vom Zeitpunkt des prä- und postsynaptischen APs. Komplementär zum Einfluß der PSC-Wellenform auf die Plastizität gibt es ebenfalls Modelle, welche die Form des bAPs bzw. die verwandten Charakteristiken der postsynaptischen Membran in die Berechnung der Plastizität mit einbeziehen (Abarbanel et al, 2002; Badoual et al, 2006; Pfister et al, 2006; Saudargiene et al, 2004; Shah et al, 2006; Shouval et al, 2002). Ein Einfluß des präsynaptischen PSC auf das postsynaptische Potential wird durch eine Klassifikation von '9' angezeigt.

Falls die Einordnung eines Modells 'nicht-bAP Depolarisation (6)' enthält, wird davon ausgegangen, dass das Modell explizit einen spannungsgesteuerten Prozess beinhaltet, d.h. die Membranspannung überträgt nicht nur den Zeitpunkt oder die Wellenform eines AP, sondern kann zusätzlich über oder unter die Ruhespannung gezogen und damit die Plastizität gesteuert werden (Abarbanel et al, 2002; Clopath et al, 2010). Postsynaptische Kurzzeitplastizität '7' wird ebenfalls in einigen Modellen angenommen, meistens basierend auf der reduzierten Anregbarkeit des postsynaptischen Neurons direkt nach einem AP (Badoual et al, 2006; Froemke et al, 2006; Shah et al, 2006). Es gibt jedoch auch Modelle, die in diesem Fall eine gesteigerte Anregbarkeit annehmen (Pfister and Gerstner, 2006). Teilweise oder vollständig $Ca^{2+}$-basierte Modelle wie etwa (Kurashige and Sakai, 2006; Shah et al, 2006; Shouval et al, 2002) werden anhand 3.2 zu 'Calcium'-Mechanismen '8' zugeordnet.

Metaplastizität kann auf verschiedenen Organisationsebenen von einzelnen Synapsen bis hin zu kompletten Netzwerken stattfinden und ihren Ausdruck in verschiedensten Parametern der beteiligten Neuronen finden (Abraham, 2008). Allerdings verwendet der weitaus überwiegende Teil aktueller Lernregeln, welche Metaplastizität beinhalten, als Ausdruck der Metaplastizität nur den gleitenden Frequenzschwellwert der ursprünglichen BCM-Formulierung (Bienenstock et al, 1982). Drei mögliche Ausprägungen dieses gleitenden Schwellwertes lassen sich finden, etwa ein direkter Einfluß auf die Plastizitätsform (Benuskova and Abraham, 2007; Izhikevich and Desai, 2003), eine gleitende Anpassung einer der postsynaptischen Zustandsvariablen (Abarbanel et al, 2002), oder ein Schwellwert als eine Funktion des mittleren $Ca^{2+}$ Pegels (Brader et al, 2007; Kurashige and Sakai, 2006). Diese verschiedenen Ausprägungen werden einheitlich in Tabelle 3.1 als '10' klassifiziert, d.h. die zugehörige Lernregel enthält einen gleitenden Frequenzschwellwert im Sinne von Bienenstock et al (1982).

Tabelle 3.1: Vergleich von Modellen für LTP und LTD. Abkürzungen in der Tabelle: ZK(Zeitkonstante), SF(Skalierungsfaktor), prä(präsynaptisch), post(postsynaptisch). Die Nummern der Experimente korrespondieren mit den Nummern in der Experimentauflistung im Text (siehe auch die Experimenttabelle in (Mayr and Partzsch, 2010)). Die Experimentevaluierung der Modelle verwendet die folgenden Symbole: 'f' (full) Dieses Experiment wurde in der Literatur mit dem entsprechenden Modell nachgewiesen; 'h' (half) Nur eine Hälfte des Experimentes wurde in der Literatur mit dem entsprechenden Modell nachgewiesen; 'fc' (full compatible) Experiment wurde nicht nachgewiesen, aber ist vermutlich vollständig kompatibel basierend auf der nachgewiesenen Leistungsfähigkeit von ähnlichen Modellen; 'hc' (half compatible) nur eine Hälfte des Experimentes ist vermutlich kompatibel mit dem Modell; 'li' (likely incompatible) Dieses Experiment kann vermutlich nicht mit diesem Modell reproduziert werden; 'i' (incompatible) es wurde in der Literatur gezeigt, dass dieses Experiment nicht mit diesem Modell reproduziert werden kann; '?' Es gibt keine Daten bzw. Evaluierung des Modells für dieses biologische Experiment. Die letzte Spalte enthält eine Klassifizierung der im Modell enthaltenen Mechanismen aus Abbildung 3.2.

| Kurzbeschreibung und Referenz | Parameter | Experimentelle Protokolle | | | | | | | | Abb. 3.2 |
|---|---|---|---|---|---|---|---|---|---|---|
| | | [1] | [2] | [3] | [4] | [5] | [6] | [7] | [8] | |
| Abbott and Nelson (2000): Konventionelles Nächster-Nachbar STDP (Pfister and Gerstner, 2006) | 2 ZK, 2SF | f | i | h | i | i | i | i | i | 1,4,10 |
| Froemke et al (2006): STDP Modell mit zusätzlicher prä und post Abschwächung der Pulsauswirkung. Die Auswirkung/Effektivität eines prä APs ist abhängig von allen vorhergehenden APs, während das post AP nur vom Zeitintervall zum vorhergehenden AP abhängt | Standard-STDP: 2 ZK, 2 SF; Abschwächung: 2 ZK prä und post, 1 SF post Skalierung Abschwächung | fc | h | fc | fc | f | i | i | i | 1,3,4,7 |
| Pfister and Gerstner (2006): Reduziertes Triplet-Modell: Standard STDP-Modell, wobei die Amplitude des LTP mit der post Pulsfrequenz skaliert | konventionelles STDP: 2 ZK, 2 SF, zusätzlicher Mittelwert: 1 ZK, 1 SF | f | f | f | f | li | i | f | i | 1,4,7,10 |
| Benuskova and Abraham (2007): Die Parameter der konventionellen STDP-Gewichtsskalierung werden als Funktion eines post Mittelwertes angepasst (gleitender STDP Schwellwert) | konventionelles STDP: 2 ZK, 2 SF; Post Mittelwert: 1 ZK, 1 SF | fc | li | hc | li | li | ? | li | i | 1,3,10 |
| Senn (2002): STDP-Modellierung der Ausschüttungswahrscheinlichkeit der Neurotransmitter, basierend auf prä und post Mittelwerten, die an den korrespondierenden Pulsen abgetastet werden, zusätzlich prä- und post Adaptation | Rezeptoren: 1 ZK, 3 SF, Adaption: 1 ZK, 2 SF, Ausschüttungswahr.: 1 ZK, 2 SF | f | h | f | fc | fc | fc | fc | ? | 1,2,3,4, 7,10 |

Tabelle 3.1 – fortgesetzt von der letzten Seite

| Kurzbeschreibung und Referenz | Parameter | Experimentelle Protokolle | | | | | | | | Abb. 3.2 |
|---|---|---|---|---|---|---|---|---|---|---|
| | | [1] | [2] | [3] | [4] | [5] | [6] | [7] | [8] | |
| Abarbanel et al (2002): Beschreibung von prä Neurotransmitterausschüttung und post Spannung durch zwei DGLs, die zugehörigen Störfunktionen repräsentieren prä und post Aktionspotentiale. Gewichtsänderung als Mischung aus zeitbezogener Kooperation und Wettbewerb zwischen den Prozessen | prä&post DGLs: 2 ZK, 2 SF (reduziert sich auf 1 SF für $size(AP_{pre}) = size(AP_{post})$), Gewichtsänderung: 2 SF | f | h | hc | li | li | ? | ? | hc | 1,4,5,6,10 |
| Badoual et al (2006): Biophysikalisches Modell eines Kompartmentneurons und kinetische Gleichungen für separates LTP und LTD | LTP Kinetik: 2 ZK, LTD: 4 ZK, AMPA/NMDA Rezeptoren: 2 ZK, 3 SF, Calciumpumpe: 1 SF, 1 ZK, ca. 20 ZK/SF Neuronenmodell | f | f | h | li | ? | ? | ? | ? | 2,3,5,6,7,8,9 |
| Badoual et al (2006): Originales Abschwächungsmodell aus (Froemke and Dan, 2002) mit zusätzlichen weichen Gewichtsschranken | konventionelles STDP: 2 ZK; Abschwächung: 2 ZK pre and post suppression, weight scaling/bounds: 2 SF | f | i | f | f | f | li | li | i | 1,3,4,7 |
| Lu et al (2007): Zustandsbasiertes Modell, Übergänge durch Pulse gesteuert, drei Zustände: prä Ereignis, post Ereignis, Ruhezustand. Übergänge beinhalten Kurzzeitadaption | Übergänge: 2 ZK, Gewichtsberechnung: 3 SF | f | ? | f | f | fc | hc | ? | i | 1,3,4,7,10 |
| Pfister et al (2006): STDP Kurve basierend auf Neuronen/PSC-Characteristiken, abgeleitet von überwachtem Lernen von Pulsmustern | STDP: 2 ZK, 2SF; Randbedingung: 1 SF | f | li | hc | li | li | li | li | i | 2,5 |
| Shah et al (2006): Calziumbasiertes Modell mit Ausdrücken für bAPs und EPSP Einfluß auf die Calzium-Dynamiken, zusätzliche prä und post Plastizitätsabschwächung, Gewichtsänderung in Abhängigkeit von Calziumamplitude und -anstieg | $\Omega$: 5 SF; $\eta/\tau$: 2 ZK, 2 SF; Calzium: 1 ZK; prä&post Abschwächung 2 SF, 2 ZK | h | f | f | ? | fc | f | fc | fc | 1,2,3,4,5,7,8 |
| Clopath et al (2010): Triplet Modell von (Pfister and Gerstner, 2006) mit zwei zusätzlichen Spannungsschwellwerten | LTD: 1ZK, 2SF; LTP: 2ZK, 2 SF; Nachhyperpolarisation: 1 SF, 1 ZK | f | f | fc | fc | ? | fc | fc | f | 1,4,6,7,10 |
| Sjöström et al (2001): LTP Verhalten als sigmoide Abhängigkeit von der (gemessenen) Depolarisation, LTD als reine Skalierung in einem bestimmten Zeitfenster, post Nächster-Nachbar Interaction | LTP sigmoid: 3SF, 1ZK, LTD: 1 SF, 1 ZK, Frequenzabhängigkeit von LTP: 2 SF | fc | f | hc | fc | fc | fc | f | hc | 1,4,6,7 |

Tabelle 3.1 – fortgesetzt von der letzten Seite

| Kurzbeschreibung und Referenz | Parameter | Experimentelle Protokolle | | | | | | | | Abb. 3.2 |
|---|---|---|---|---|---|---|---|---|---|---|
| | | [1] | [2] | [3] | [4] | [5] | [6] | [7] | [8] | |
| Fusi et al (2000): Wahrscheinlichkeit von LTP und LTD abhängig von der post Membranspannung während des prä Pulses, Wahrscheinlichkeit entscheidet über Wechsel zwischen zwei Zuständen der Synapse | 8 SF, 4 ZK | f | f | hc | li | li | fc | fc | hc | 1,5,6,10 |
| LCP mit Puls-Antwort Neuron | 1 ZK PSC; 1 ZK, 2 SF post Pulsantwort; 1 SF post Kurzzeitplast. | f | h | f | f | i | i | f | h | 2,5,6,7, 10 |
| LCP mit Leaky-Integrate-and-Fire Neuron | wie oben, ein zusätzlicher SF für $U_{\text{PSP}}$ | f | f | f | f | h | f | f | h | 2,5,6,7, 9,10 |

Die Mehrheit der obigen Einschätzungen sind aus den angeführten Publikationen oder weiteren Modellanalysen der jeweiligen Autoren entnommen. Allerdings existiert gerade für den Test der Modelle mit typischen Ratenexperimenten (Experiment 6 und 7 in Tabelle 3.1) meist nur ein an (Izhikevich and Desai, 2003) angelehnter Nachweis, d.h. ratenabhängige Plastizität wird nur für ein Protokoll bestehend aus einer präsynaptischen festen Rate sowie einem postsynaptischen frequenzvariablen Poissonprozess gezeigt. Um diese Nachweislücke gegenüber konkret gemessener Ratenplastizität (Dudek and Bear, 1992; Sjöström et al, 2001) zu schliessen, wurden in (Mayr et al, 2010c) verschiedene pulszeitpunktsabhängige (STDP-)Modelle untersucht, für die von den jeweiligen Autoren eine generelle Kompatibilität mit Ratenplastizität beansprucht wird. Dies wird laut den Autoren entweder über einen zu STDP hinzugefügten ratenabhängigen Ausdruck (Froemke et al, 2006; Pfister and Gerstner, 2006) oder aufgrund eines speziellen STDP-Protokolls (Izhikevich and Desai, 2003) erreicht. In (Mayr et al, 2010c) wurde allerdings simulativ nachgewiesen, dass keine dieser Modifikationen beispielsweise für das Experiment aus (Dudek and Bear, 1992) mit einem rauschbehafteten LIAF-Neuron als postsynaptischer Rekonstruktion (Mayr and Partzsch, 2010) eine hinreichende Nachbildung der biologischen Messdaten bietet (vergleiche stark abweichende Kurvenverläufe von Abbildung 3.3A verglichen mit biologischen Messdaten in Abbildung 3.1A). Des weiteren wurde in (Mayr et al, 2010c) analytisch nachgewiesen, dass die meisten dieser Modellmodifikationen ebenfalls nicht hinreichend sind, die Ratenplastizität in (Sjöström et al, 2001) nachzubilden (siehe Abbildung 3.3B). Einzig das reduzierte Modell aus (Pfister and Gerstner, 2006) in der Nächster-Nachbar-Variante hat eine qualitativ mit BCM bzw. mit den Ergebnissen aus (Sjöström et al, 2001) kompatible Kurve, d.h. eine bei LTD beginnende Plastizitätskurve für niedrige Frequenzen, ein Wechsel von LTD zu LTP bei ca. 20-30 Hz, sowie einen weiteren Anstieg zu höherem LTP mit zunehmender Frequenz.

Mithin ist konträr zur in der Literatur vertretenen Ansicht nicht sichergestellt, dass mittels des häufig für einen BCM-Nachweis verwendeten (synthetischen) Protokolls aus (Izhikevich and Desai, 2003) eine echte Kompatibilität mit konkret biologisch gemessener Ratenplastizität (Dudek and Bear, 1992; Sjöström et al, 2001) nachgewiesen werden kann. Darüberhinausgehend gibt es auch Hinweise, dass der in (Izhikevich and Desai, 2003) verwendete Ansatz nicht nur für die zwei hier gezeigten biologischen Protokolle Defizite aufweist, son-dern dass dieser Ansatz sogar nur für einige wenige spezielle (synthetische) Protokolle funktioniert (Standage et al, 2007). Für das Zustandekommen der Einschätzung der Modelle in Tabelle 3.1 bzgl. der restlichen Experimente wird auf den Anhang von (Mayr and Partzsch, 2010) verwiesen. Die letzten beiden Zeilen der Vergleichstabelle sind dem in (Mayr and Partzsch, 2010; Mayr et al, 2008b) neu entwickelten Plastizitätsmodell Local Correlation Plasticity (LCP) gewidmet. Dieses Modell wird in zwei verschiedenen Ausprägungen in den nächsten Abschnitten vorgestellt und analysiert.

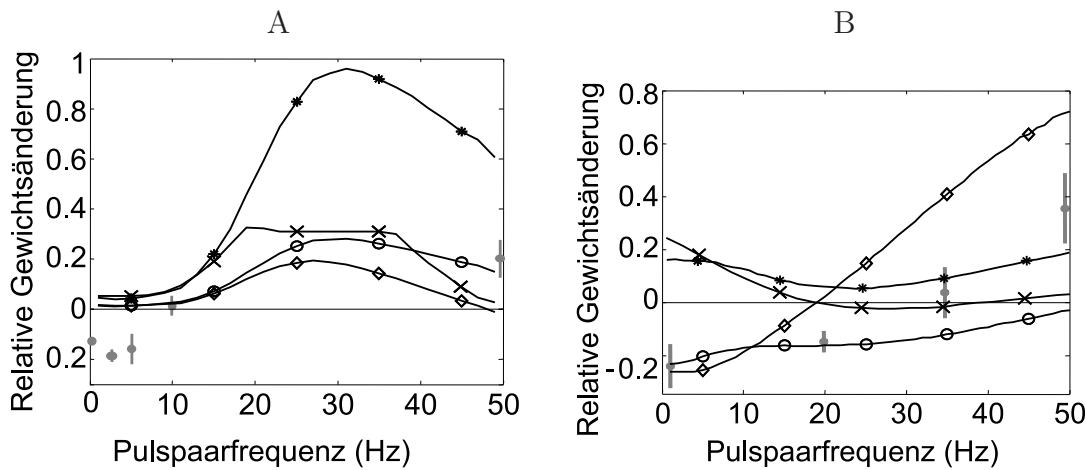

Abbildung 3.3: (A) Reproduktion des präsynaptischen Tetanus-Ratenexperiments aus (Dudek and Bear, 1992) für das Modell aus Pfister and Gerstner (2006) (Varianten: Nächster-Nachbar (Diamanten) und volle Interaktion (Kreise)), für das STDP-Modell aus Izhikevich and Desai (2003) (Sterne), sowie für das Tripletmodell aus Froemke et al (2006) (Kreuze); Es wurde ein verrauschtes LIAF Neuron zur Rekonstruktion der postsynaptischen Pulse dieses Experiments eingesetzt (Mayr and Partzsch, 2010; Beggs, 2001); (B) Zufälliges Ratenprotokoll aus Sjöström et al (2001), identische Zuordnung von Kurven zu Modellen. Die zugehörigen biologischen Messdaten sind jeweils in grau hinterlegt. Reproduktion des präsynaptischen Tetanus-Ratenexperiments (A) mit der LCP-Regel wurde in (Mayr and Partzsch, 2010) gezeigt, das zufällige Ratenprotokoll (B) unter Verwendung von LCP ist in Abbildung 3.6 bzw. Abbildung 3.10 zu sehen.

## 3.2 Local Correlation Plasticity (LCP) mit Puls-Antwort Neuron

Wie aus Tabelle 3.1 ersichtlich, haben die dort vorgestellten aktuellen Lernregeln trotz ihrer teilweise langjährigen Entwicklungsgeschichte (vergleiche (Brader et al, 2007; Fusi et al, 2000) und (Shah et al, 2006; Shouval et al, 2002)) deutliche Defizite bezüglich einer allgemeinen biologischen Relevanz. Ein Großteil der Modelle stellt nur einzelne Plastizitätseffekte nach, bzw. es werden in einer rein phänomenologischen Herangehensweise für jeden neuen Plastizitätseffekt Terme an bestehende Lernregeln angefügt (vergleiche (Pfister and Gerstner, 2006) mit (Clopath et al, 2010)), ohne die Lernregel grundsätzlich zu reformieren. Auf diese Weise entstehen jedoch Lernregeln, die zwar viele experimentelle Ergebnisse nachstellen können, jedoch sich aufgrund ihrer akkumulierten Komplexität und reinen Orientierung an Datennachbildung nicht zum Aufzeigen der zugrundeliegenden neurobiologischen Prozesse eignen (Wei, 1975). Die in diesem Abschnitt vorgestellte LCP-Regel baut demgegenüber ursprünglich nur auf der Fragestellung auf, wie eine Synapse mit den ihr zugänglichen lokalen Zustandsvariablen überhaupt Formen von Plastizität berechnen/aufweisen kann, da sich letztendlich alle Induktionsprotokolle aus Abbildung 3.1 in charakteristischen Spannungsverläufen z.B. des lokalen Membranpotentials an der Synapse wiederspiegeln. Phenomenologische Lernregeln nehmen teilweise unrealistische Mechanismen wie eine Messung der Zeitabstände von Pulsen an, wohingegen die LCP-Regel eher untersucht, wie beispielsweise diese Zeitfenster sich intrinsisch aus den Neuronen- und Synapseneigenschaften ergeben (Debanne et al, 1994). Es werden für die LCP-Regel folgende Mechanismen bzw. Voraussetzungen postuliert:

- **Prämisse 1:** Aus den Experimenten in (Artola et al, 1990; Holthoff et al, 2006; Kampa et al, 2007; Lisman and Spruston, 2005; Ngezahayo et al, 2000; Sjöström et al, 2008), kann geschlussfolgert werden, dass synapsenspezifische, postsynaptisch wirkende spannungsgesteuerte Prozesse einen signifikanten Anteil an synaptischer Plastizität haben.

- **Prämisse 2:** Nach (Aihara et al, 2007; Sjöström et al, 2008) produziert eine langsame Inaktivierung der Calziumkanäle nach einer niedrigen bis mittleren Erhebung des Calziumpegels LTD, während eine starke Erhöhung des intrazellulären Calzium gefolgt von einem schnellen Abklingen (Calziumpuls) LTP zur Folge hat.

- **Darauf aufbauende LCP-Hypothese:** Wenn ein präsynaptisch ausgelöster PSC während der postsynaptischen Refraktärszeit an der Synapse ankommt, verursacht dies eine kurzfristige, moderate Anhebung des Membranpotentials (wobei das Membranpotential typisch unterhalb des Ruhepotentials bleibt). Dies hat wiederum die erwähnte langsam abklingende mittlere Calziumerhöhung zur Folge (Koch, 1999), woraus LTD entsteht. Falls das postsynaptische Membranpotential allerdings zum Zeitpunkt des PSC über der Ruhespannung liegt, verursacht die in diesem Spannungsbereich gültige kürzere Zeitkonstante einen scharfen Anstieg des Calziumpegels gefolgt von einem ebenso schnellen Abklingen (siehe Spannungssteuerung des Calziumpegels und zugehörige Zeitkonstanten in Abbildung 9.3 aus Koch (1999)). Somit würde diese prä/postsynaptische Korrelation zu LTP führen (Aihara et al, 2007).

Diese Art von Gewichtsänderung in Abhängigkeit bestimmter Kombinationen von prä- und postsynaptischer Activität erinnert stark an die klassische BCM-Regel (siehe Abbildung 3.1A). In der BCM-Regel wird eine präsynaptische Aktivitätsvariable $d(t)$ mit einer verschobenen postsynaptischen Aktivitätsvariable $c(t)^1$ multipliziert und damit die Änderung des Gewichtes $w$ definiert:

$$\frac{\mathrm{d}w}{\mathrm{d}t} = \phi(c(t) - \Theta_M) \cdot d(t) \,, \tag{3.1}$$

---

[1]In der Regel werden diese Aktivitätsvariablen als verschiedene Ausprägungen von Tiefpassfilterungen der entsprechenden Pulsfolgen interpretiert

wobei $\Theta_M$ einen Aktivitätsschwellwert darstellt sowie mit $\phi(.)$ eine nichtlineare Funktion mit Nullpunkt im Ursprung zur Skalierung der postsynaptischen Aktivität zum Einsatz kommt.

In Einklang mit der oben angeführten Hypothese wird diese Regel nun neu interpretiert. Wie oben postuliert, stellt die postsynaptische Aktivität $c(t)$ im LCP-Kontext das Membranpotential dar, somit ist der Schwellwert $\Theta_M$ ein Spannungsschwellwert. Die Funktion $\phi(.)$ der postsynaptischen Aktivität wird als lineare Skalierungsfunktion mit der Proportionalitätskonstanten $B$ interpretiert. Ein Spannungsschwellwert zur Trennung von LTD und LTP in Abhängigkeit von einer postsynaptischen (spannungsbasierten) Zustandsvariable wird von den in (Artola et al, 1990; Fusi et al, 2000; Ngezahayo et al, 2000; Schreiter et al, 2002) vorgestellten Ergebnissen unterstützt.

Für die vollständige LCP-Regel muss nun noch die präsynaptische Aktivität $d(t)$ definiert werden. Da der Plastizitätsmechanismus direkten Zugang zum Membranpotential benötigt, muss das Auslesen der präsynaptischen Zustandsvariable ebenfalls postsynaptisch stattfinden, etwa über den durch einen präsynaptischen Puls ausgelösten PSC. Aus modellierungstechnischen Gründen wird untenstehend eine dem PSC äquivalente Größe verwendet, die Konduktanzänderung postsynaptischer Ionenkanäle an der Synapse. Diese wird von einer präsynaptischen Neurotransmitterausschüttung ausgelöst und verursacht in Folge den PSC. Mit dieser Konduktanzänderung liest sich die vollständige LCP-Regel wie folgt:

$$\frac{\mathrm{d}w}{\mathrm{d}t} = B \cdot (u(t) - \Theta_u) \cdot g(t) \ . \tag{3.2}$$

In dieser Gleichung ist $w(t)$ das Gewicht der Synapse, $g(t)$ die Konduktanz der präsynaptisch aktivierten Ionenkanäle, $u(t)$ das Membranpotential und $\Theta_u$ der Schwellwert zwischen Gewichtsverstärkung und -abschwächung. Diese Neuinterpretation der BCM-Regel erfüllt die in der obigen Auflistung eingeführte Hypothese sehr gut.

Es wird angenommen, dass eine Erhöhung des postsynaptischen Membranpotentials die Blockierung der NMDA-Rezeptoren aufhebt, wodurch $Ca^{2+}$ ausgeschüttet wird und in Folge das synaptische Gewicht sich ändert (Senn, 2002; Shouval et al, 2002). Somit sollte $B$ in Einheiten von $1/(As)$ definiert sein, was die $Ca^{2+}$-Ladung aufhebt. Das synaptische Gewicht ist damit dimensionslos, was für einem Vergleich mit der Mehrzahl der experimentellen Ergebnisse erforderlich ist (Bi and Poo, 1998; Froemke and Dan, 2002; Sjöström et al, 2001). Der postsynaptische Ausdruck $B \cdot (u(t) - \Theta_u)$ ergibt damit eine Einheit von $\Omega/s$, was als die Öffnungs/Schliessgeschwindigkeit der $Ca^{2+}$-Kanäle angesehen werden kann (d.h. ihre Widerstandsänderung). $\Theta_u$ kann als Umkehrpotential der $Ca^{2+}$-Kanäle interpretiert werden. Somit ergibt sich in Übereinstimmung mit neurobiologischen Erkenntnissen die Gewichtsänderung in der LCP-Regel aus der Rate der $Ca^{2+}$-Ausschüttung (Aihara et al, 2007).

Die LCP-Regel besteht damit im Kern nur aus der Subtraktion von $\Theta_u$, der Multiplikation von prä- und postsynaptischen Zustandsgrößen und der Integration des Gewichtes. Allerdings können natürlich die Eigenschaften der LCP-Regel erst mit Hilfe von Modellen für den Verlauf der prä- und postsynaptischen Zustandsgrößen untersucht werden, da diese starken Einfluß auf das Verhalten der Regel haben. Für die mathematische Analyse grundsätzlicher Mechanismen der LCP Lernregel in diesem und dem nächsten Abschnitt (3.3) werden für beide Zustandsgrößen vereinfachte Formulierungen angenommen.

Die synaptische Konduktanz $g(t)$ wird durch einen exponentiellen Abfall modelliert, aktiviert zum Zeitpunkt jedes präsynaptischen Pulses. Der Ausdruck entspricht sowohl einer Nachbildung des Zeitverlaufs des in Abschnitt 2 diskutierten PSCs als auch einer Vereinfachung des in (Gerstner and Kistler, 2002) beschriebenen Synapsenmodells:

$$g(t) = \hat{G} \cdot \mathrm{e}^{-\frac{t-t_j^{\mathrm{pre}}}{\tau_g}} \ , \ t_j^{\mathrm{pre}} \leq t < t_{j+1}^{\mathrm{pre}} \ , \tag{3.3}$$

mit $\tau_g$ als der Zeitkonstante des Abklingens und $\hat{G}$ als der Amplitude der synaptischen Pulsantwort.

Die postsynaptische Spannung wird über ein formalisiertes Modell eines LIAF-Neurons nachgebildet, das sogenannte Puls-Antwort-Neuron (Spike Response Model, SRM) (Gerstner and Kistler, 2002). Das SRM besteht aus einer Diracfunktion für das Aktionspotential und einer nachfolgenden exponentiellen Annäherung an das Ruhepotential, welche das Refraktärsverhalten nachstellt:

$$\begin{aligned} u(t) &= U_{\mathrm{p},n} \cdot \delta(t - t_n^{\mathrm{post}}) + U_{\mathrm{refr}} \cdot \mathrm{e}^{-\frac{t-t_n^{\mathrm{post}}}{\tau_{\mathrm{refr}}}} , \\ & t_n^{\mathrm{post}} \leq t < t_{n+1}^{\mathrm{post}} \end{aligned} \tag{3.4}$$

Wobei $t_n^{\mathrm{post}}$ den $n$-ten postsynaptischen Puls kennzeichnet. $U_{\mathrm{p},n}$ ist die Fläche unter dem Aktionspotential, $U_{\mathrm{refr}} < 0$ und $\tau_{\mathrm{refr}}$ sind die Amplitude und Zeitkonstante der Hyperpolarisation.

Zusätzlich zur Hyperpolarisation des Membranpotentials wurde eine Abschwächung der Pulsfläche während der Refraktärszeit angenommen, um die reduzierte Anregbarkeit des Neurons (Shah et al, 2006) bzw. die reduzierte Höhe des AP (Froemke et al, 2006) während dieser Zeit zu modellieren. Die Abschwächung wird realisiert als Skalierung der Fläche des $n$-ten postsynaptischen Pulses mit dem exponentiell gewichteten Zeitintervall zwischen dem $n$-ten und $(n-1)$-ten postsynaptischen Puls. Biologisch ist dies interpretierbar als Gewichtung mit der residualen Hyperpolarisation direkt vor dem Puls, d.h. die Tiefe der verbleibenden Refraktärsphase steuert die Pulsfläche (Mayr and Partzsch, 2010) (Froemke et al, 2006):

$$U_{\mathrm{p},n} = U_{\mathrm{p}} \cdot \left( 1 - \alpha_{\mathrm{att}} e^{-\frac{t_n^{\mathrm{post}} - t_{n-1}^{\mathrm{post}}}{\tau_{\mathrm{refr}}}} \right) \tag{3.5}$$

mit $U_{\mathrm{p}}$ als Parameter für die maximale Fläche des postsynaptischen Pulses und $\alpha_{\mathrm{att}}$ als Parameter, mit dem die maximale Stärke der Flächenreduktion des AP eingestellt werden kann. Im Unterschied zu der ähnlichen postsynaptischen Abschwächung aus (Froemke et al, 2006) wurde hier allerdings keine neue Zeitkonstante für diesen Sättigungseffekt angenommen, sondern wieder die der Refraktärszeit zugrundeliegende Membranzeitkonstante $\tau_{\mathrm{refr}}$ verwendet. Die Integration einer Abschwächung hat signifikanten Einfluß auf das Verhalten der Regel in Tripletexperimenten [3], wie in Abschnitt 3.3 gezeigt wird. Diese Variante der LCP-Regel wird im Weiteren LCP mit SRM genannt (vergleiche auch Lernregeldefinitionen in Mayr and Partzsch (2010)).

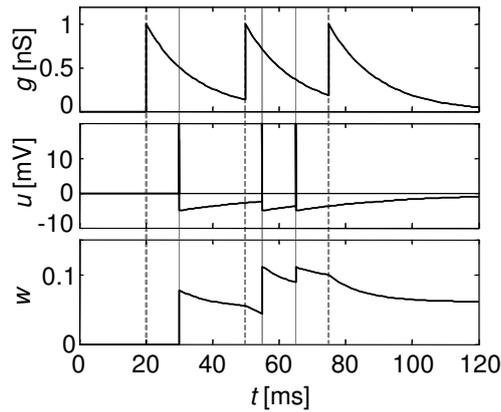

Abbildung 3.4: Zeitverlauf der Konduktanz $g$, des Membranpotentials $u$ und des synaptischen Gewichtes $w$ für eine Beispielpulsfolge. Die Einheiten und Skalierungskonstanten wurden in einen neurobiologisch realistischen Wertebereich gelegt: $\hat{G} = 1\text{nS}$, $B = 1/(1\text{pC})$ (Koch, 1999).

Natürlich öffnet man sich beim Hinzufügen zusätzlicher Effekte wie der in Gleichung 3.5 aufgeführten Abschwächung derselben Kritik wie die am Anfang dieses Abschnitts diskutierten phenomenologischen Modelle, d.h. wird dieser Effekt nur hinzugefügt, um das Modell mit weiteren Experimenten kompatibel zu machen? Konträr zu dieser Kritik lassen sich jedoch für die o.a. Form und Parametrisierung signifikante neurobiologische Argumente finden. Zum Beispiel verändert die experimentell gefundene postsynaptische Abschwächung ähnlich wie in Gleichung 3.5 hauptsächlich die Größe (Amplitude und Dauer) des postsynaptischen Aktionspotentials (Froemke et al, 2006; Tanaka et al, 1991), die Zeitkonstante stimmt quantitativ mit der Membranzeitkonstante überein (Shah et al, 2006) und die Abschwächung hat den postulierten Effekt auf die zustandekommende Plastizität (Froemke et al, 2006), siehe auch Abbildung 9 aus (Mayr and Partzsch, 2010).

## 3.3 Analyse der LCP-Regel mit Puls-Antwort Neuron

Abbildung 3.4 zeigt die prinzipielle Funktionsweise von LCP mit SRM. Im Gegensatz zu pulsbasierten Lernregeln findet eine kontinuierliche Gewichtsänderung statt, sobald das Membranpotential vom Spannungsschwellwert $\Theta_u$ abweicht und zeitgleich präsynaptische Aktivität vorhanden ist. Bei Verwendung des SRM-Neuronenmodells existiert für die LCP-Regel ein separater Mechanismus jeweils für LTP and LTD: LTP wird von einem postsynaptischen Puls getriggert. Da der Puls als Dirac-Stoß modelliert ist, wird die Konduktanzvariable $g(t)$ zu jedem postsynaptischen Pulszeitpunkt abgetastet. Dieser Mechanismus ist äquivalent zu einer iterativen Implementierung von LTP in pulspaarbasierten STDP-Modellen (Pfister and Gerstner, 2006).

LTD wiederum wird durch die kontinuierliche Integration der präsynaptischen Konduktanz während der Refraktärszeit des postsynaptischen Pulses verursacht. Damit wirkt der LTD-Mechanismus immer, wenn prä- und postsynaptische Pulse in enger zeitlicher Abfolge aufeinandertreffen. LTD ist ausserdem aktiv, wenn der Spannungsschwellwert zum

Zeitpunkt eines präsynaptischen Pulses über dem Ruhepotential ist, d.h. $\Theta_u > 0$.

Für die gesamte Gewichtsänderung ergibt sich damit, dass das relativ global wirkende LTD nur in Gewichtsverstärkung münden kann, wenn der präsynaptische Puls kurz vor dem postsynaptischen Puls auftritt (prä-post). In diesem Fall wird beinahe die gesamte präsynaptische Amplitude durch den postsynaptischen Puls abgetastet, d.h. es findet zuerst eine große Gewichtserhöhung statt, während in der darauffolgenden Refraktärszeit des postsynaptischen Pulses zwar wieder LTD produziert wird, dieses jedoch nicht die Wirkung des Aktionspotentials aufheben kann. Für die entgegengesetzte Pulsabfolge (post-prä) ergibt sich demgegenüber starkes LTD. Dies führt direkt zu einer zeitlichen Asymmetrie des Lernens vergleichbar mit pulspaarbasiertem STDP [1]. Dieser Zusammenhang lässt sich auch analytisch herstellen, indem die Gewichtsänderung für ein einzelnes, zeitlich isoliertes Pulspaar mit Zeitdifferenz $\Delta t = t^{\text{post}} - t^{\text{pre}}$ berechnet wird. Wenn für das prä-post Pulspaar die o.a. Gewichtsverstärkung und -abschwächung kombiniert werden, ergibt sich folgender Ausdruck (für die Herleitung, siehe Abschnitt 3.2.1 in (Mayr and Partzsch, 2010)):

$$
\begin{aligned}
\Delta w(\Delta t \geq 0) = {} & B\hat{G}\left(U_{\text{p}} + U_{\text{refr}}\tau_{\text{all}} - \Theta_u \tau_g\right) \mathrm{e}^{-\frac{|\Delta t|}{\tau_g}} \\
& - B\hat{G}U_{\text{refr}}\tau_{\text{all}} \cdot \mathrm{e}^{-\frac{t-t^{\text{pre}}}{\tau_g}} \cdot \mathrm{e}^{-\frac{t-t^{\text{post}}}{\tau_{\text{refr}}}} \\
& + B\hat{G}\Theta_u \tau_g \cdot \mathrm{e}^{-\frac{t-t^{\text{pre}}}{\tau_g}} \,,\ t > t^{\text{post}} \\
& \text{mit } \tau_{\text{all}} = \frac{1}{\frac{1}{\tau_g} + \frac{1}{\tau_{\text{refr}}}} \,.
\end{aligned}
\tag{3.6}
$$

Im Unterschied dazu führt ein post-prä Pulspaar, wie oben erklärt, nur zu Gewichtsabschwächung, weil zum Zeitpunkt des postsynaptischen Pulses noch keine präsynaptische Aktivität verzeichnet wird. Die Gewichtsabschwächung beginnt damit zum Zeitpunkt des präsynaptischen Pulses, d.h. $t^{\text{pre}}$. Die gesamte Gewichtsänderung kann wie folgt berechnet werden:

$$
\begin{aligned}
\Delta w(\Delta t < 0) = {} & B\hat{G}U_{\text{refr}}\tau_{\text{all}} \cdot \mathrm{e}^{-\frac{|\Delta t|}{\tau_{\text{refr}}}} \\
& - B\hat{G}U_{\text{refr}}\tau_{\text{all}} \cdot \mathrm{e}^{-\frac{t-t^{\text{pre}}}{\tau_g}} \cdot \mathrm{e}^{-\frac{t-t^{\text{post}}}{\tau_{\text{refr}}}} \\
& + B\hat{G}\Theta_u \tau_g \cdot \mathrm{e}^{-\frac{t-t^{\text{pre}}}{\tau_g}} - B\hat{G}\Theta_u \tau_g \,,\ t > t^{\text{pre}} \,.
\end{aligned}
\tag{3.7}
$$

Unter Verwendung von Gleichung 3.6 und 3.7 kann LCP mit SRM direkt auf pulspaarbasierte (Nächster-Nachbar) STDP-Regeln (Morrison et al, 2008) bezogen werden. Es wird $\Theta_u = 0$ gesetzt. Ein Pulspaar mit Zeitdifferenz $\Delta t = t^{\text{post}} - t^{\text{pre}}$ ergibt damit die folgende zeitkontinuierliche Gewichtsänderung:

$$
\begin{aligned}
\Delta w(t) = {} & \Delta w_\infty - B\hat{G}U_{\text{refr}}\tau_{\text{all}} \cdot \mathrm{e}^{-\frac{t-t^{\text{pre}}}{\tau_g}} \cdot \mathrm{e}^{-\frac{t-t^{\text{post}}}{\tau_{\text{refr}}}} \text{ mit} \\
\Delta w_\infty = {} & \begin{cases} B\hat{G}U_{\text{refr}}\tau_{\text{all}} \cdot \mathrm{e}^{-\frac{|\Delta t|}{\tau_{\text{refr}}}} & : \ \Delta t < 0 \\ B\hat{G}\left(U_{\text{p}} + U_{\text{refr}}\tau_{\text{all}}\right) \mathrm{e}^{-\frac{|\Delta t|}{\tau_g}} & : \ \Delta t \geq 0 \end{cases}
\end{aligned}
\tag{3.8}
$$

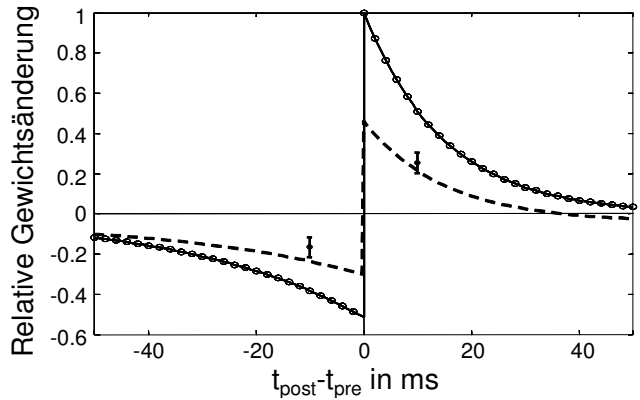

Abbildung 3.5: Normalisiertes STDP-Fenster [1] für LCP mit SRM: Analytischer Wert $\Delta w_\infty$ (durchgezogene Linie) und Simulationsergebnisse (Kreise, 60 Pulspaare mit 1 Hz Wiederholfrequenz, Protokoll aus (Bi and Poo, 1998; Froemke and Dan, 2002)). Gestrichelte Linie: Simulationsergebnisse für das leicht abweichende STDP aus (Wang et al, 2005), Fehlerschranken geben biologische Messdaten aus (Wang et al, 2005) wieder (Parameter für Simulation und analytische Kurve siehe Tabelle in (Mayr and Partzsch, 2010)).

Wie gezeigt kann die Lösung des Integrals aufgeteilt werden in einen zeitunabhängigen Term $\Delta w_\infty$, der von der Pulsreihenfolge abhängt, und einen zusätzlichen zeitabhängigen Term, der für $t \to \infty$ null wird. Für die in Standard-STDP-Experimenten (Bi and Poo, 1998; Froemke and Dan, 2002) verwendete niedrige Wiederholungsfrequenz kann die resultierende Gewichtsänderung in sehr guter Näherung durch den Term $\Delta w_\infty$ berechnet werden (siehe auch die Übereinstimmung zwischen Simulation mit biologischem Protokoll und analytischer Lösung in Abbildung 3.5). Dieser Ausdruck ist äquivalent zum exponentiellen Zeitfenster für Standard-STDP.

Aus dieser Korrespondenz können die Parameter von LCP mit SRM direkt aus den Parametern des exponentiellen STDP-Zeitfensters berechnet werden:

$$U_{\text{refr}} = \frac{1}{B\hat{G}} \cdot A_- \cdot \left(\frac{1}{\tau_g} + \frac{1}{\tau_{\text{refr}}}\right) \qquad \tau_{\text{refr}} = \tau_-$$

$$U_{\text{p}} = \frac{1}{B\hat{G}} \cdot (A_+ - A_-) \qquad \tau_g = \tau_+ \qquad (3.9)$$

mit $A_-(<0), \tau_-$ and $A_+, \tau_+$ als Amplitude und Zeitkonstante von LTD respektive LTP (Song et al, 2000). Um biologisch realistische Membranspannungswerte zu erhalten, wird der Verstärkungsfaktor $B\hat{G}$ so gesetzt, dass die Refraktärsamplitude $U_{\text{refr}}$ gleich -5 mV ist (Koch, 1999).

In diesem Zusammenhang ist auch interessant, dass sich mit der obigen Formulierung die nach phenomenologischen Daten generierten STDP-Parameter mit den biophysikalischen Parametern des SRM-Neurons verbinden lassen. Beispielsweise entspricht die STDP-Zeitkonstante für LTP (ca. 20 ms) der Zeitkonstante der synaptischen Konduktanz $\tau_g$ im LCP mit SRM-Modell, welche für NMDA-Synapsen in ähnlicher Größenordnung liegt (Ba-doual et al, 2006; Gerstner and Kistler, 2002). Im Unterschied zu (Pfister et al, 2006) ist diese Zeitkonstante bei LCP mit SRM nicht identisch mit $\tau_{mem}$, da diese Neuronencharakteristiken nicht zwingend korreliert sind (Koch, 1999; Senn, 2002). Die STDP-Zeitkonstante für LTD entspricht der Membranzeitkonstanten $\tau_{\text{refr}}$, welche in der Gegend von 10-40 ms liegt (Koch, 1999). Wie ebenfalls aus Gleichung 3.9 ersichtlich ist, legen die STDP-Parameter $A_+, \tau_+$ und $A_-, \tau_-$ das Verhältnis zwischen $U_{\text{p}}$ und $U_{\text{refr}}$ fest. Für die hier getroffene Annahme von $U_{\text{refr}} = -5$ mV ergeben die STDP-Parameter aus (Froemke and Dan, 2002) einen Flächeninhalt für das Aktionspotential von $U_{\text{p}}$=151 $\mu$Vs. Dies entspricht einem Rechteckpuls von 2 ms Länge und 75 mV Höhe, was wiederum einen neurobiologisch realistischen Wert darstellt (Koch, 1999).

Nachdem die grundsätzliche Kompatibilität mit pulszeitpunktsabhängigem Lernen durch die obige Herleitung und Simulation gezeigt wurde, soll nun ein Test bzgl. der Emulierung ratenbasierten Lernverhaltens nach Abbildung 3.1A durchgeführt werden. Dafür kommt das korrelierte Ratenprotokoll [7] zum Einsatz, da dies im Gegensatz zu konventionellen Tetanusexperimenten [6] eines der wenigen Experimente zu ratenabhängiger Plastizität ist, bei dem auch die postsynaptische Seite kontrolliert wurde. Dieses Protokoll führt also mithin zu eindeutig mit den experimentellen Resultaten vergleichbaren simulativen Ergebnissen, es besteht keine Mehrdeutigkeit in der Nachstellung des Experiments. Zudem stellt dieses Protokoll anscheinend generell einen härteren Testfall dar als z.B. [6]. Wie in (Mayr et al, 2010c) gezeigt wurde, kann dieses Experiment von mehreren aktuellen Lernregeln (Froemke et al, 2006; Izhikevich and Desai, 2003; Pfister and Gerstner, 2006) nicht nachgestellt werden, obwohl diese laut Literatur weitgehend kompatibel mit ratenbasierten Experimenten sind.

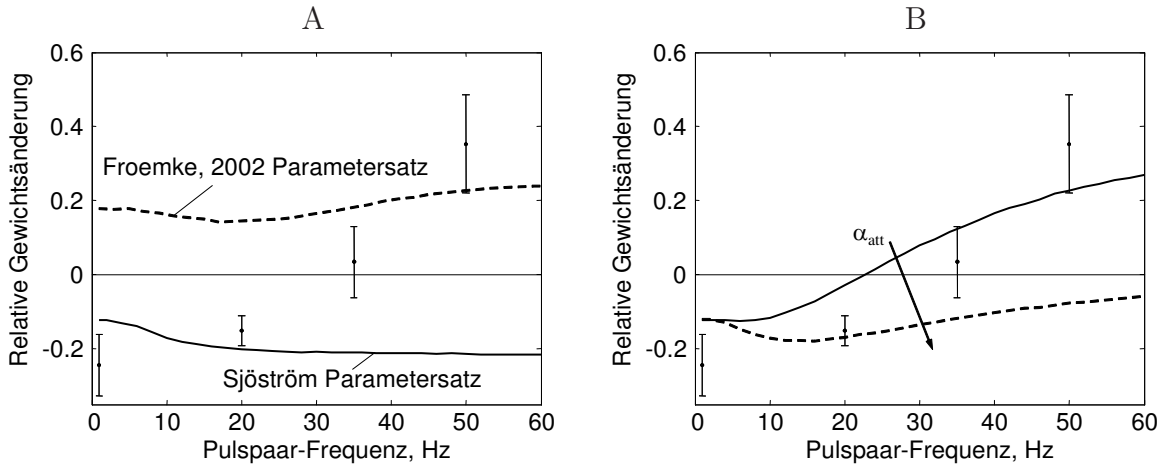

Abbildung 3.6: Korreliertes Ratenprotokoll [7], Mittelwert über 1000 Versuche; (A) Standard-Nächster-Nachbar-STDP; (B) LCP mit SRM ohne und mit postsynaptischer Abschwächung ($\alpha_{\text{att}} = 0$, $\alpha_{\text{att}} = 0.8$), restliche Parameter siehe Tabelle in (Mayr and Partzsch, 2010). Fehlerschranken entsprechen den experimentellen Ergebnissen aus (Sjöström et al, 2001).

Abbildung 3.6 zeigt die Gewichtsänderung für das korrelierte Ratenprotokoll für ein Standard-STDP- Modell (A) und für LCP mit SRM (B). Da für dieses Versuchsprotokoll Pulse zufällig generiert werden, tendiert die resultierende Gewichtsänderung zu einer

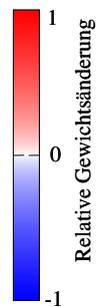

Abbildung 3.7: Triplet-Protokoll (Experiment [3]). Oberhalb der 45°-Geraden: 2 präsynaptische, 1 postsynaptischer Puls; Unterhalb der Geraden: 1 präsynaptischer, 2 postsynaptische Pulse, genaue Pulsabfolgen wie in Teilgrafik B dargestellt; (A) Experimentelle Daten aus (Froemke and Dan, 2002); (B) LCP mit SRM mit postsynaptischer Abschwächung ($\alpha_{att} = 0.8$), STDP-Parametersatz aus (Froemke and Dan, 2002); (C) Standard-STDP-Regel mit identischen Parametern zum Vergleich.

Als ein Beispiel für die nicht durch STDP erklärbare Interaktion von Pulsen in komplexeren Pulsprotokollen, wurden als weiteres Testprotokoll für LCP mit SRM sogenannte Pulstriplets [3] verwendet. Abbildung 3.7 zeigt die Datenpunkte der Experimente (A) und die mit LCP mit SRM simulierten Ergebnisse (B). Standard-STDP ist in (C) wieder zum Vergleich aufgeführt. LCP mit SRM (B) und STDP (C) zeigen in den meisten Quadranten ein miteinander und mit den Versuchsergebnissen gut übereinstimmendes Verhalten, dort muss also zur Erklärung der Versuchsdaten kein komplexeres Modell als STDP angenommen werden (Wei, 1975). Für die post-prä-post Triplets im unteren rechten Quadranten ergeben sich allerdings sichtbare qualitative Unterschiede: Während beide Modelle für kleine Zeitdifferenzen $t_1$ and $t_2$ eine Gewichtsverstärkung vorhersagen, zeigen die Messdaten eine Gewichtsabschwächung. Mit der Aktivierung der postsynaptischen Abschwächung, welche inhärent genau für diese dicht aufeinander folgenden postsynaptischen Pulsabfolgen greift, reduziert sich die Gewichtsverstärkung deutlich in diesem Quadranten. Mit einer Parametrisierung von ($\alpha_{att} = 0.8$) produziert LCP mit SRM ähnlich den experimentellen Ergebnissen für diese Triplets ebenfalls Gewichtsabschwächung, siehe Abbildung 3.7C. Somit kann LCP mit SRM die Tripletdaten nachstellen, was sonst i.d.R. nur mit deutlich aufwändigeren Modellen möglich ist (Badoual et al, 2006).

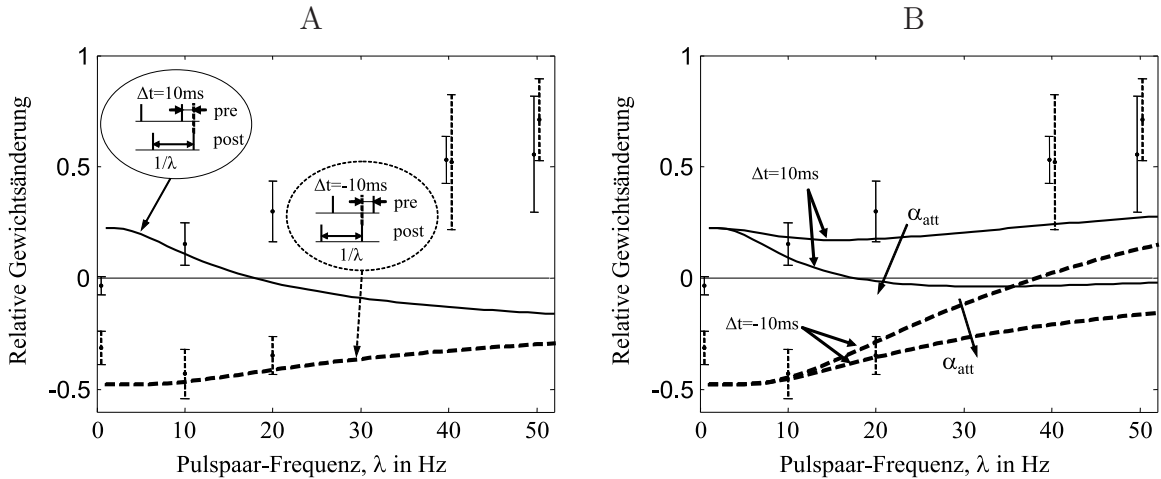

Abbildung 3.8: Frequenzabhängiges STDP-Experiment [2]; Fehlerschranken entsprechen den Messwerten aus (Sjöström et al, 2001). Gestrichelte Linien und Fehlerschranken: post-prä Pulspaare; durchgezogene Linien und Fehlerschranken: prä-post Pulspaare. Um die Lesbarkeit zu erhöhen, wurden die Fehlerschranken bei 40Hz und 50 Hz leicht auseinandergeführt. (A) Standard-STDP, (B) LCP mit SRM für $\alpha_{\text{att}} = 0$ und $\alpha_{\text{att}} = 0.8$.

In diesem Abschnitt wurde gezeigt, dass LCP mit SRM eine große Bandbreite an experimentellen Protokollen/Ergebnissen nachstellen kann (siehe auch die weiteren Experimente in (Mayr and Partzsch, 2010)). Allerdings ist dies nur für inkonsistente Einstellungen der postsynaptischen Abschwächung möglich: Während die Abschwächung zwingend erforderlich ist, um die Tripletexperimente von (Froemke and Dan, 2002) nachzustellen, führt die Abschwächung zu verschlechterter Reproduktion des korrelierten Ratenprotokolls [7]. Wie sieht es in diesem Zusammenhang mit der Reproduktion einer Kombination aus raten- und pulszeitpunktabhängigem Lernen aus? In Abbildung 3.8 wird LCP mit SRM mit dem Pulspaar/Wiederholungsfrequenz-Experiment [2] aus (Sjöström et al, 2001) getestet.

LCP mit SRM stimmt mit den experimentellen Ergebnissen deutlich besser überein als Standard-STDP, vergleiche Abbildung 3.8A. Allerdings fehlt bei LCP mit SRM die mit der Wiederholfrequenz ansteigende Gewichtsverstärkung für prä-post Pulspaare (Abbildung 3.8B). Zudem verschlechtert das Aktivieren der postsynaptischen Abschwächung wiederum die Nachbildung der experimentellen Ergebnisse. Im nächsten Abschnitt soll untersucht werden, ob diese Inkonsistenzen bzw. Fehler in der Experimentnachbildung durch ein realistischeres postsynaptischen Neurons als das in Gleichung 3.4 definierte aufgelöst werden können.

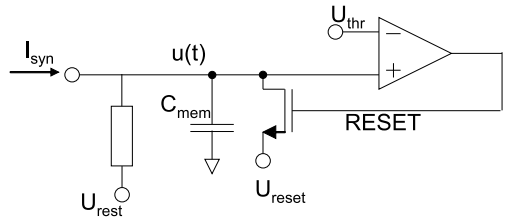

Abbildung 3.9: Standardschaltung eines LIAF-Neurons mit eingehendem PSC $I_\text{syn}$, Membrankapazität $C_\text{mem}$ und dem mit $R$ modellierten passiven Kanalleckstrom bezogen auf $U_\text{rest}$. Der Hystereseschalter erzeugt bei Erreichen eines Feuerschwellwertes $U_\text{thr}$ ein AP und setzt das Neuron danach auf $U_\text{reset}$ zurück.

## 3.4 LCP mit Leaky-Integrate-and-Fire Neuron

In Abschnitt 3.2 wurde ein stark vereinfachtes Neuron für die LCP-Regel eingeführt, das SRM-Neuron. Allerdings wurden dabei wichtige Mechanismen der neuronalen Signalübertragung vernachlässigt. Im Folgenden wird deshalb das Neuronenmodell um die postsynaptische Wirkung eines präsynaptischen Pulses erweitert. Dessen Einfluss kann durch den PSC-Stromfluss in der Zellmembran bestimmt werden, welcher durch eine präsynaptische Neurotransmitterausschüttung hervorgerufen wird. Um die Wirkung dieses synaptischen Stromes auf das Membranpotential zu berechnen, wird ein Leaky integrate-and-fire (LIAF) Neuron verwendet, definiert durch folgende Differentialgleichung:

$$C_\text{mem} \cdot \frac{\mathrm{d}u}{\mathrm{d}t} = -\frac{u}{R} + I_\text{syn}(t) \,, \tag{3.10}$$

wobei $C_\text{mem}$ und $R$ die Kapazität und den Widerstand der Neuronmembran darstellen, siehe Abbildung 3.9.

In Kombination ergeben $C_\text{mem}$ und $R$ wieder die Membranzeitkonstante $\tau_\text{refr} = C_\text{mem} \cdot R$, die auch bereits im SRM-Neuron in Gleichung 3.4 verwendet wurde. $I_\text{syn}$ repräsentiert den durch $g(t)$ hervorgerufenen PSC. Um das durch $I_\text{syn}$ entstehende postsynaptische Potential zu ermitteln, wird der PSC mit der LIAF-Neurongleichung integriert (Details in (Mayr and Partzsch, 2010)). Die resultierende explizite Formulierung für $u(t)$ zwischen Aktionspotentialen (d.h. während der Integration anliegender PSCs) ist dann[2]:

$$u(t) = u(0)\mathrm{e}^{-\frac{t}{\tau_\text{refr}}} + U_\text{PSP} \cdot W \cdot \left(\mathrm{e}^{-\frac{t}{\tau_g}} - \mathrm{e}^{-\frac{t}{\tau_\text{refr}}}\right)$$

$$\text{mit } W = \frac{\left(\frac{\tau_\text{refr}}{\tau_g}\right)^{\frac{\tau_\text{refr}}{\tau_\text{refr}-\tau_g}}}{1 - \frac{\tau_\text{refr}}{\tau_g}} \,. \tag{3.11}$$

$U_\text{PSP}$ stellt die Amplitude eines postsynaptischen Potentials dar, welches von einer beim postsynaptischen Ruhezustand ( $u(0) = 0$ ) präsynaptisch ausgelösten Strominjektion verursacht wird. Das übrige Verhalten des Neurons spiegelt das des SRM-Modells wieder (siehe Gleichung 3.4): Zu den vom Experimentprotokoll vorgegebenen postsynaptischen Pulszeitpunkten wird ein Dirac-Stoss mit der Fläche $U_p$ generiert. Danach wird das Membranpotential $u$ auf die Amplitude $U_{\text{refr}} = U_{\text{rest}} - U_{\text{reset}}$ der Refraktärszeit zurückgesetzt, von wo aus es sich gemäß Gleichung 3.11 weiterentwickelt. Für den Fall, dass keine präsynaptische Aktivität auftritt, ergeben die aufgeführten Verhaltensgleichungen einen mit dem SRM-Modell identischen Zeitverlauf.

---

[2]Ohne Beeinträchtigung der allgemeinen Gültigkeit wird für diese Herleitung angenommen, dass $U_\text{rest}$ auf den Nullpunkt von $u(t)$ bezogen ist.

LCP with SRM konnte das korrelierte Ratenexperiment bei aktivierter postsynaptischer Abschwächung nicht nachstellen. Die Verwendung der postsynaptischen Potentiale und der präsynaptischen alle-zu-allen Interaktion bei LCP mit LIAF kompensiert diese negativen Auswirkungen, wie in Abbildung 3.10B gezeigt wird. Die postsynaptischen Potentiale führen bereits bei kleinen Wiederholfrequenzen zu verstärktem LTP. Für hohe Wiederholfrequenzen führt die alle-zu-allen Interaktion zu erhöhtem postsynaptischen Potential, welches wiederum die Hyperpolarisation nach postsynaptischen Pulsen schneller beenden und damit indirekt die bei dieser hohen Rate stark wirkende postsynaptische Verkleinerung der Aktionspotentialfläche kompensieren.

Aufgrund der postsynaptischen Potentiale und der speziellen STDP-Parameter aus (Sjöström et al, 2001) wird bei LCP mit LIAF die LTP-Hälfte des STDP-Lernfenster verzerrt, wie in Abbildung 3.10A dargestellt ist. Interessanterweise ist dieses zusätzliche LTD-Fenster in der LTP-Hälfte auch in den Experimentdaten von Abbildung 2D aus Sjöström et al (2001) vorhanden.

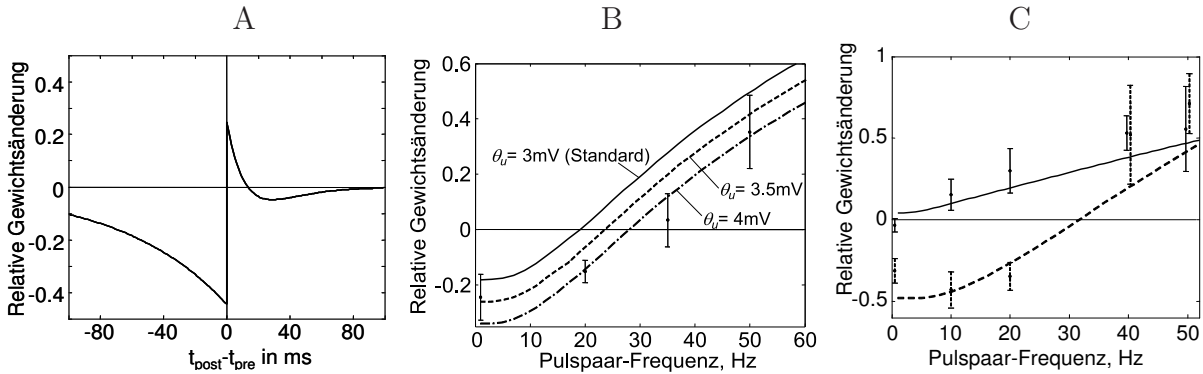

Abbildung 3.10: Experimente aus (Sjöström et al, 2001) unter Verwendung von LCP mit LIAF (Parameter siehe Tabelle in (Mayr and Partzsch, 2010)). A: auf (Sjöström et al, 2001) optimiertes STDP-Fenster (Experiment [1]); B: korreliertes Ratenprotokoll [7]; C: Frequenzabhängiges STDP [2], obere Kurve prä-post Pulspaare, untere Kurve post-prä (vergleiche Abbildung 3.8).

Die erhöhte Gewichtsverstärkung durch die akkumulierte präsynaptische Konduktanz (alle-zu-allen Interaktion) und der Einfluss der korrespondierenden postsynaptischen Potentiale auf die Membranspannung führt trotz der postsynaptischen Abschwächung zu mit der Wiederholfrequenz ansteigendem LTP für pre-post Pulspaare, vergleiche Abbildung 3.10C. LCP mit LIAF ist damit im Gegensatz zu LCP mit SRM sowohl für prä-post als auch für post-prä Pulspaare mit den experimentellen Daten kompatibel (siehe Abbildung 3.8B).

Mithin konnte für die Experimente [1], [2], [3] und [7] gezeigt werden, dass LCP mit LIAF eine deutlich gesteigerte Fähigkeit zur Nachbildung experimenteller Ergebnisse besitzt und Inkonsistenzen von LCP mit SRM auflöst. Dieser Trend setzt sich auch für die restlichen in Abschnitt 3.1 aufgelisteten Bechmark-Experimente fort. Für eine ausführliche Analyse und weitere Ergebnisse von sowohl LCP mit SRM als auch LCP mit LIAF, siehe (Mayr and Partzsch, 2010).

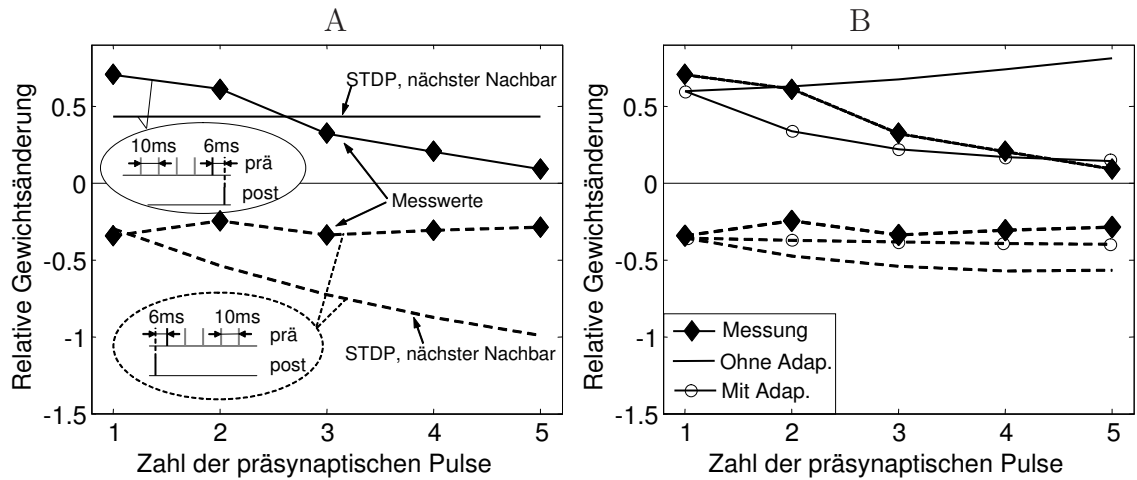

Abbildung 3.11: Präsynaptisches Burstexperiment aus (Froemke et al, 2006), A: Nächster-Nachbar-STDP; B: LCP mit LIAF in der in (Mayr and Partzsch, 2010) eingeführten Version (ohne präsyn. Adaption) und mit der präsynaptischen Adaption des Quantalmodells; Parameter: STDP-Parametersatz Froemke 2 aus (Mayr and Partzsch, 2010), zusätzliche LCP-Parameter $U_{\text{PSP}} = 0.5$ mV, $\Theta_U = 0$ mV, Parameter Quantalmodell wie in Abbildung 2.8.

Aus diesen erweiterten Analysen soll zum Abschluss dieses Kapitels nur ein Beispiel herausgegriffen werden, welches auch mit dem bisher diskutierten LCP mit LIAF Plastizitätsmodell nicht zufriedenstellend nachgebildet werden kann. Dieses Beispiel (Experiment [5]) untersucht Langzeitplastizität ausgelöst durch präsynaptische Bursts (siehe Abbildung 4 in (Froemke et al, 2006)). Aus den zugehörigen Messdaten in Abbildung 3.11A&B ist ersichtlich, dass für den post-prä-burst Fall (untere Kurven) die gesamte Gewichtsveränderung durch das erste post-prä Zusammentreffen verursacht wird, weitere präsynaptische Pulse haben keinen signifikanten Einfluss. Abbildung 3.11A zeigt, dass dieser Fall durch konventionelles STDP nicht zufriedenstellend nachgebildet wird, während LCP mit LIAF (in seiner bisherigen Form, d.h. ohne präsynaptische Adaption) die Messdaten gut emuliert (Abbildung 3.11B). Dieses Experiment erfordert allerdings auch keine präsynaptische Adaption.

Im prä-burst-post Fall ist dagegen mit zunehmender Anzahl an präsynaptischen Pulsen eine Verminderung des LTP festzustellen, d.h. das erste prä-post Pulspaar fällt anscheinend stärker ins Gewicht als folgende, was auf eine zunehmende Abschwächung der präsynaptischen Pulse mit der Pulsanzahl hindeutet. Weder STDP noch LCP mit LIAF in der bisherigen Form können diese Messdaten nachbilden. Wenn das LCP mit LIAF Modell hingegen mit der präsynaptischen Adaption des Quantalmodells entsprechend Abschnitt 2.3 kombiniert wird, verbessert sich die Experimentreproduktion deutlich. Dies ist ein weiteres starkes Indiz dafür, dass das biophysikalische Quantalmodell bei entsprechender Parametrisierung sowohl Kurzzeitplastizität wie etwa PPD nachstellen kann als auch in Modellen für Langzeitplastizität sinnvoll einsetzbar ist. Mithin kann das Quantalmodell sehr gut in ein Gesamtmodell für zeitskalenübergreifende Plastizität integriert werden und wird deshalb gemäß der Umschreibung in Abschnitt 2.4 in die in Kapitel 5 geschilderte Hardwarerealisierung des MAPLE inkludiert.

# 4 Metaplastizität

## 4.1 Metaplastizität in der LCP Regel

Wie in Abschnitt 3.1 erwähnt, hat Metaplastizität, d.h. die Änderung von Plastizität auf sehr langen Zeitskalen, verschiedenste synapsen- und neuronenspezifische Ausdrucksformen. Allerdings gibt es für Lernregeln auf Einzel-Synapsen-Niveau bisher nur einen experimentell nachgewiesenen metaplastischen Mechanismus, den gleitenden Schwellwert der BCM-Formulierung in seinen verschiedenen Ausprägungen (Mayford et al, 1995; Ngezahayo et al, 2000). Um Metaplastizität im Kontext der LCP-Regel zu diskutieren, muss demnach ein Bezug zu diesem gleitenden Schwellwert hergestellt werden. Spezifisch sollte dabei eine Äquivalenz zwischen dem Spannungsschwellwert $\Theta_u$ der LCP-Regel und dem Frequenzschwellwert $\Theta_M$ der BCM-Regel nachgewiesen werden. Dies hätte den zusätzlichen Vorteil, die Kompatibilität von LCP sowohl mit experimentell nachgewiesenen metaplastischen Frequenz- als auch mit Spannungsschwellwerten aufzuzeigen. Diese beiden unterschiedlichen Schwellwerte werden zwar teilweise im selben Kontext diskutiert (Ngezahayo et al, 2000), jedoch gibt es aktuell keine Lernregel, welche explizit beide vereinen würde.

Ausserdem würde ein derartiger Nachweis einen willkommenen analytischen Rückbezug zur BCM-Regel darstellen. Einerseits wurde zwar die math. Formulierung der LCP Regel von BCM abgeleitet (Gleichung 3.2) und in Abschnitt 3.3 simulativ eine mit den BCM-Vorgaben konforme ratenabhängige Plastizität gezeigt. Andererseits ist die Umdefinition der Zustandsgrößen von BCM zu LCP doch so disjunkt, dass nicht automatisch von einer Kongruenz der beiden Formulierungen ausgegangen werden kann.

Um einen analytischen BCM-LCP Bezug herzustellen, wird eine Herangehensweise ähnlich zu (Izhikevich and Desai, 2003) verwendet. Begonnen wird mit dem allgemeinen Ausdruck für eine Gewichtsänderung für ein einzelnes prä-/postsynaptisches Pulspaar aus den Gleichungen 3.6 und 3.7 in Abschnitt 3.3. Es wird wiederum niedrige Wiederholfrequenz der Pulspaare angenommen, somit können von $t$ abhängige Terme vernachlässigt werden. Im Gegensatz zu der STDP-Herleitung wird allerdings $\Theta_u$ nicht Null gesetzt, sondern als Parameter mitgeführt. Der resultierende Ausdruck für LTP ist dann:

$$\Delta w(\Delta t) = B\hat{G}\left(U_{\mathrm{p}} + U_{\mathrm{refr}}\tau_{\mathrm{all}} - \Theta_u \tau_g\right) \mathrm{e}^{-\frac{|\Delta t|}{\tau_g}} \qquad (4.1)$$

und der Ausdruck für LTD:

$$\Delta w(\Delta t) = B\hat{G}U_{\mathrm{refr}}\tau_{\mathrm{all}} \cdot \mathrm{e}^{-\frac{|\Delta t|}{\tau_{\mathrm{refr}}}} - B\hat{G}\Theta_u \tau_g \qquad (4.2)$$

Unter der Annahme einer Poissonverteilung $p(\Delta t) = \lambda \cdot exp(-\lambda \cdot \Delta t)$ für die postsynaptischen Pulszeitpunkte kann der Erwartungswert der Gewichtsänderung $\Delta w(\lambda)$ für einen einzelnen präsynaptischen Puls und eine postsynaptische Pulsrate $\lambda$ explizit berechnet werden. Es wird das Integral des Produktes aus der Gewichtsänderung als Funktion der Zeit

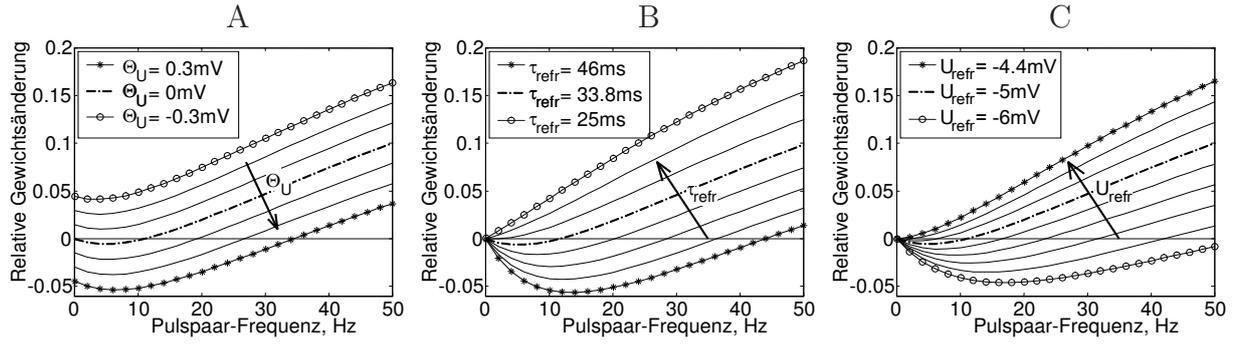

Abbildung 4.1: Parametrische Kurven des analytischen BCM-Ausdrucks aus Gleichung 4.4. Der Standardparametersatz ($\Theta_u = 0$ mV, $U_\mathrm{p}=151$ mV·ms, $U_\mathrm{refr} = -5$ mV, $\tau_\mathrm{refr} = 33.4$ ms, $\tau_g = 16.8$ ms) wird immer durch die gestrichelte Linie angegeben. (A) Durchlauf von $\Theta_u$ als Parameter; (B) Durchlauf von $\tau_\mathrm{refr}$; (C) Durchlauf von $U_\mathrm{refr}$

$\Delta t$ und der Poissonverteilung dieses $\Delta t$ als Funktion der postsynaptischen Rate separat für LTP und LTD aufgestellt. Eine Addition beider Integrale ergibt die gesamte Gewichtsänderung (Izhikevich and Desai, 2003):

$$\Delta w(\lambda) = \int_0^\infty B\hat{G}\left(U_\mathrm{p} + U_\mathrm{refr}\tau_\mathrm{all} - \Theta_u\tau_g\right) \mathrm{e}^{-\frac{\Delta t}{\tau_g}} \cdot \lambda \mathrm{e}^{-\lambda\Delta t}\mathrm{d}\Delta t$$
$$+ \int_{-\infty}^0 B\hat{G}\left(U_\mathrm{refr}\tau_\mathrm{all}\mathrm{e}^{\frac{\Delta t}{\tau_\mathrm{refr}}} - \Theta_u\tau_g\right) \cdot \lambda \mathrm{e}^{\lambda\Delta t}\mathrm{d}\Delta t \qquad (4.3)$$

Eine Berechnung dieses Integrals führt zu folgender Gleichung:

$$\Delta w(\lambda) = \lambda B\hat{G}\left(\frac{U_\mathrm{p} + U_\mathrm{refr}\tau_\mathrm{all}}{\frac{1}{\tau_g} + \lambda} + \frac{U_\mathrm{refr}\tau_\mathrm{all}}{\frac{1}{\tau_\mathrm{refr}} + \lambda}\right) - B\hat{G}\Theta_u\tau_g\left(\frac{\lambda}{\frac{1}{\tau_g} + \lambda} + 1\right) \qquad (4.4)$$

Der erste Teil des obigen Ausdrucks ist eine mit (Izhikevich and Desai, 2003) identische STDP-BCM-Übersetzung, wobei die Standard-STDP-Parameter entsprechend Gleichung 3.9 substituiert wurden. Der zweite Teil des Ausdrucks verschiebt die gesamte Kurve der Gewichtsänderung über die Wiederholfrequenz in vertikaler Richtung in Abhängigkeit von $\Theta_u$. Die zusätzliche Abhängigkeit dieser Verschiebung von $\lambda$ kann für kleine $\lambda$ vernachlässigt werden, während aus großen $\lambda$ ($> 1/\tau_g$) eine Verstärkung des Anstiegs der Kurve resultiert, welche den Übergangspunkt von LTD to LTP leicht zu niedrigeren Frequenzen verschiebt. Mithin ist, wie aus Gleichung 4.4 und den Kurven in Abbildung 4.1A ersichtlich, der neu eingeführte Spannungsschwellwert äquivalent zu dem gleitenden Frequenzschwellwert, welcher im Kontext von ratenbasierten Plastizitätsprotokollen gefunden wurde (Abraham et al, 2001; Mayford et al, 1995; Wang and Wagner, 1999). Aus Abbildung 4.1A kann ausserdem gefolgert werden, dass $\Theta_u$ mit dem in (Ngezahayo et al, 2000) experimentell gefundenen positiven (d.h. LTD/LTP) Spannungsschwellwert korrespondiert.

Eine interessante Beobachtung in Abbildung 4.1A ist, dass für niedrige Frequenzen, in denen die meisten BCM-Formulierungen keine Gewichtsänderung annehmen, die Plastizitäts-kurve der LCP-Regel in Abhängigkeit von $\Theta_u$ einen Offset zu sowohl LTD als auch LTP aufweisen kann. Diese Art von Offset zeigt große Übereinstimmung mit den in (Sjöström et al, 2001) vorgestellten Experimenten, bei denen eine Depolarisation für niedrige Wiederholungsfrequenzen einen Offset Richtung LTP verursacht, während eine Hyperpolarisation für niedrige Frequenzen zu verstärktem LTD führt. Dass eine vertikale Verschiebung der Plastizitätskurve eine horizontale Verschiebung des Frequenzschwellwertes zur Folge hat, stimmt ausserdem gut mit der in (Beggs, 2001) postulierten Plastizitätstheorie überein.

Wie aus einem Vergleich der analytischen Darstellung in Abbildung 4.1A und der Simulation in Abbildung 3.10B hervorgeht, verursacht ein $\Theta_u \neq 0$ eine mit der analytischen Lösung qualitativ übereinstimmende Verschiebung der Kurve. Allerdings unterscheiden sich die für eine bestimmte Verschiebung nötigen absoluten Werte von $\Theta_u$ deutlich. Die Ursache hierfür ist, dass $\Theta_u$ vor allem als Offset für den Refraktärsverlauf wirkt, von dem in der analytischen Herleitung eine unendliche Ausdehnung angenommen wird. Mithin ist die integrierte Gesamtfläche und damit der Effekt von $\Theta_u$ deutlich größer im analytischen Ausdruck als in der Simulation, da dort der Refraktärsverlauf vom nächsten Puls abgeschnitten wird. Folglich muß $\Theta_u$ in der Simulation höhere absolute Werte annehmen, um denselben Effekt zu erreichen.

Im Gegensatz zu $\Theta_u$ beeinflußt ein Variieren der Amplitude $U_{\text{refr}}$ und Zeitdauer $\tau_{\text{refr}}$ des Refraktärsverlaufs nur den Schwellwert von LTD zu LTP, während der Nullpunkt der Kurve bzw. das Verhalten für niedrige Widerholfrequenzen unverändert bleibt (Abbildung 4.1B,C). Veränderungen beider Parameter haben neben der Hebung/Senkung des Schwellwertes den Effekt, die Plastizitätskurve vom klassischen U-förmigen Verlauf bis zu einer beinahe linearen Charakteristik zu verändern. Die Auswirkung beider Parameter auf die Kurve könnte mit der metaplastischen Veränderung der Nachhyperpolarisation (post-spike after-hyperpolarization, AHP) in Verbindung stehen (Abraham, 2008; Zelcer et al, 2006), da sich z.B. eine Vertiefung des AHP in der LCP Lernregel als Vergrößerung der Amplitude $U_{\text{refr}}$ bzw. als eine größere Zeitkonstante $\tau_{\text{refr}}$ darstellen lässt.

## 4.2 Einbettung in den makroskopischen biologischen Kontext

Unser Modell könnte mithin durch die metaplastische Modulation weiterer Parameter (neben dem gleitenden Schwellwert) zusätzliche in Experimenten gefundene Metaplastizitätsmechanismen emulieren (Lebel et al, 2001; Zelcer et al, 2006). Beispielsweise scheint reduzierte AHP in Versuchstieren erhöhte Lernfähigkeiten zu verursachen (Disterhoft and Oh, 2006; Zelcer et al, 2006). Wie oben bereits ausgeführt, würde dies in der LCP-Regel einer verkürzten Zeitkonstante $\tau_{\text{refr}}$ oder einer Reduzierung der Refraktärsamplitude $U_{\text{refr}}$ entsprechen, was mit einer Erhöhung der LTP-Fläche bzw. einer Verschiebung des gleitenden Schwellwertes hin zu niedrigeren Frequenzen und damit einer erhöhten Lernbereitschaft einhergeht (siehe Abbildung 4.1B&C). Nachdem in den entsprechenden biologischen Experimenten allerdings das Training eingestellt wird, ist nicht nur eine Reduzierung des AHP zurück zum Ruhezustand zu verzeichnen (Disterhoft and Oh, 2006), was in Folge LTD und LTP wieder ausgleichen würde, sondern es kann zusätzlich noch ein 'Überschwingen' hin zu verstärktem LTD beobachtet werden (Lebel et al, 2001).

Eine Hypothese zur Erklärung dieses Effektes könnte ein zweiter Mechanismus sein, der beispielsweise $\Theta_u$ homeostatisch dem reduzierten AHP nachführt, aber in seinen Auswirkungen andauert, obwohl das AHP bereits wieder im Ruhezustand angekommen ist. Unterstützen die zugehörigen Zeitkonstanten die postulierte Abfolge der Ereignisse? Aus (Lebel et al, 2001; Zelcer et al, 2006) kann gefolgert werden, dass die Zeitkonstante für die Anpassung des AHP ungefähr einen Tag beträgt. Wenn angenommen wird, dass $\Theta_u$ der spannungsbasierten Metaplastizität in (Ngezahayo et al, 2000) entspricht, scheint hier ein Widerspruch vorzuliegen, da diese metaplastische Anpassung deutlich schneller induziert werden kann als die AHP-Änderung und damit vermutlich auch wesentlich schneller als diese wieder abklingt, anstatt sie zu überdauern. Allerdings dürfte diese schnelle Induktion der $\Theta_u$-Änderung 'in vitro' nicht die 'in vivo' auftretende 'echte' Zeitkonstante widerspiegeln. In dieser Arbeit wurde bereits ein Zusammenhang zwischen den gleitenden Spannungs- und Frequenzschwellwerten hergestellt (siehe Gleichung 4.4). Damit kann in diesem Kontext angenommen werden, dass die spannungsbasierten (Ngezahayo et al, 2000) und frequenzbasierten (Abraham et al, 2001; Wang and Wagner, 1999) metaplastischen Effekte auf ähnlichen Mechanismen beruhen. Beide können mit starken Stimuli ('in vitro') in kurzen Zeiträumen hervorgerufen werden (Ngezahayo et al, 2000; Wang and Wagner, 1999), ähnlich der 'one-shot' Plastizitätsinduktion in (Holthoff et al, 2006), welche allerdings nicht den tatsächlichen Zeitverlauf der Plastizitätsinduktion 'in-vivo' widerspiegelt (Lisman and Spruston, 2005). Korrespondierend mit den Beobachtungen in (Lisman and Spruston, 2005) könnte die der Anpassung von $\Theta_u$ zugrundeliegende tatsächliche Zeitkonstante eher aus der Relaxationszeit abgeleitet werden, d.h. der Zeit, in der $\Theta_u$ wieder zum Ruhewert zurückkehrt, d.h. ungefähr zehn Tagen (Abraham et al, 2001). Dies würde die obige Hypothese unterstützen, d.h. dass zuerst von einer dem Versuchstier gestellten Aufgabe mittels AHP-Reduktion eine verstärkte Lernbereitschaft ausgelöst wird, gefolgt von einem langsameren metaplastischen bzw. homeostatischen Nachführen des LTD/LTP-Schwellwertes, welcher initial versucht, diese AHP-Reduktion wieder auszugleichen, dann allerdings diese nach dem Training überdauert. Wie oben angeführt, könnten die AHP-Änderungen in der LCP-Regel mit einem sekundären Prozess modelliert werden. Dieser würde in Zeiten globaler starker Gewichtsänderungen, d.h. in einem mikroskopischen Zustand, der dem makroskopischen gesteigertem Lernen des Versuchstieres entspricht, $\tau_{\text{refr}}$ und/oder $U_{\text{refr}}$ mit einer Zeitkonstante von ca. einem Tag anpassen. Im Hintergrund würde ein weiterer Prozess laufen, der ähnlich wie in der originalen BCM-Formulierung (Bienenstock et al, 1982) über einen Mittelwert der postsynaptischen Aktivität mit einer Zeitkonstante von ca. zehn Tagen $\Theta_u$ nachführt (siehe Abbildung 4.1) und damit das Flächenverhältnis LTP/LTD wieder ausgleicht.

# 5 Schaltungsimplementierung: Multiscale Plasticity Experiment - MAPLE

## 5.1 Systemaufbau und analoges Testkonzept

Die Umsetzung synaptischer Plastizität in CMOS VLSI, d.h. die Schaltungsentwicklung und Chiprealisierung von Modellen aus Tabelle 3.1 und ähnlichen Ansätzen, stellt ein junges, jedoch äusserst dynamisches Forschungsgebiet dar. Die resultierenden Schaltkreise finden Einsatz als Interface zu biologischem neuronalen Gewebe (Bontorin et al, 2009), als Forschungswerkzeug zur Entwicklung biologieinspirierter Informationsverarbeitung (Schemmel et al, 2007), in Systemen zur Nachbildung olfaktorischer Klassifikation (Koickal et al, 2007), zur Kompensation von Bauelementestreuungen in integrierten Schaltungen (Cameron et al, 2005), in technischer Bildverarbeitung (Mitra et al, 2009) und in vielen weiteren Feldern, in denen pulsbasierte adaptive Verarbeitung eingesetzt wird. Eine wichtige Zielsetzung ist in vielen dieser Ansätze, die Gesamtverarbeitung des neuronalen Netzes möglichst detailgetreu vom biologischen Vorbild zu übernehmen. Naturgemäß wird dabei mit einer biologienahen Implementierung von synaptischer Lang- und Kurzzeitplastizität begonnen, da die 'mikroskopische' synaptische Verarbeitung entscheidend die 'makroskopischen' Eigenschaften des Gesamtnetzwerkes beeinflusst (Del Giudice and Mattia, 2001).

Als ein Beitrag zu diesem Forschungsgebiet wird in diesem Kapitel ein Testschaltkreis vorgestellt, der die in den Kapiteln 2, 3 und 4 entwickelten Modelle für synaptische Plastizität auf verschiedenen Zeitskalen in Silizium umsetzt. Dieser Testschaltkreis, genannt 'Multiscale Plasticity Experiment (MAPLE)', stellt in mehrfacher Hinsicht einen wegweisenden Beitrag zur Abbildung synaptischer Plastizität in CMOS dar:

- Umsetzung der ersten kombinierten, ineinander greifenden Implementierung der drei hauptsächlichen synaptischen Plastizitätsmechanismen (Kurzzeit-, Langzeit- und Metaplastizität)
- Biologische Begründung jedes einzelnen Mechanismus (siehe Kapitel 2 bis 4)
- Signifikant bessere Nachbildung von biologischen Versuchsdaten (hier insbesondere die Langzeitplastizität in Form der LCP-Lernregel: Versuchsemulierung Spannungs-, Raten-, und Pulszeitpunktsplastizität bis Stand ca. 2007, bisherige Schaltungsrealisierungen bis Stand ca. 1998, d.h. nur STDP.)
- Trotzdem deutlich geringere schaltungstechnische Komplexität als gängige biologienahe Implementierungen, bedingt durch Vereinfachungen der vorhergehenden Kapitel und die im folgenden geschilderte komplexitätsoptimierte Topologie

**Optimierte neuromorphe Topologie**

In einer typischen Schaltungsimplementierung eines pulsenden neuronalen Netzes (Schemmel et al, 2007) werden deutlich mehr Synapsen als Neuronen realisiert, konform mit der gemessenen großen Ausfächerung biologischer Neuronen (Binzegger et al, 2004). Somit wird der Platzbedarf und die Komplexität der Gesamtschaltung wesentlich durch die Synapsenrealisierung beeinflusst. Aktuelle Implementierungen verwenden in der Regel zum Aufbau von Lernzeitfenstern für z.B. STDP (vergleiche Abschnitt 3.1) in der Synapse Variablen, welche tiefpassgefilterte Versionen der prä- und postsynaptischen Pulsfolgen darstellen und für den Gesamtlernvorgang mit ihrem jeweiligen Pendant (prä-Variable mit post-Puls und post-Variable mit prä-Puls) kombiniert werden (Arthur and Boahen, 2006; Koickal et al, 2007; Saeki et al, 2008). Das individuelle Bereitstellen der Tiefpassfilterung und der Zustandsvariablen in jeder Synapse stellt naturgemäß einen großen Teil der Synapsenschaltung dar. Im Unterschied hierzu wird in der in Kapitel 3 beschriebenen LCP-Lernregel der postsynaptische Spannungsverlauf, der bereits durch die Membranzeitkonstante eine tiefpassgefilterte Variante der postsynaptischen Pulsfolge enthält, direkt als Zustandsvariable verwendet. Mithin verwenden alle am Dendriten desselben Neurons liegenden Synapsen dieselbe postsynaptische Zustandsvariable, wodurch sich die Komplexität der einzelnen Synapsenschaltung verglichen mit z.B. (Koickal et al, 2007) signifikant reduziert. In einem weiteren Schritt kann auch die PSC-Generierung von Synapsen, die mit demselben präsynaptischen Neuron verbunden sind, über einen zentralen Baustein erfolgen, da diese Synapsen einen identischen Pulseingang erhalten und damit auch identische PSC-Folgen generieren. Diese Herangehensweise wurde z.B. in (Schemmel et al, 2007) verwendet. Damit enthält die Synapse effektiv nur noch die individuelle Berechnung der Überlagerung von prä- und postsynaptischer Zustandsvariable (d.h. die Subtraktion von $\Theta_u$ und die Multiplikation in Gleichung 3.2), während alle rein prä- oder postsynaptischen Zustandsvariablen ausgelagert sind. Es kann damit auch der Einfluss zusätzlicher (rein prä- bzw. postsynaptischer) Kurzzeitadaption auf die Langzeitplastizität effektiv in das Modell eingebunden werden, ohne die Komplexität der Synapsenschaltung zu erhöhen. Beispielsweise wird die in Abschnitt 2.3 diskutierte präsynaptische Adaption in die PSC-Schaltung integriert, während die postsynaptische Adaption der Fläche des APs aus Gleichung 3.5 ins Neuron eingebaut werden kann.

Die Auslagerung besonders der PSC-Generierung hat aber auch negative Aspekte, da dadurch evtl. die Konfigurierbarkeit der Netztopologie durch die restriktive Zuordnung der Synapsen eingeschränkt wird. In den meisten Chipimplementierungen werden die Synapsen in einer Matrix angeordnet (Giulioni, 2008; Schemmel et al, 2007), wobei in diesen Beispielen wie auch beim MAPLE angenommen wird, dass alle mit demselben postsynaptischen Neuron verbundenen Synapsen sich in einer Reihe befinden (ersichtlich in Abbildung 5.1). In (Schemmel et al, 2007) wird zusätzlich angenommen, dass jeweils eine PSC-Schaltung alle Synapsen einer Spalte versorgt. Damit sind alle Synapsen einer Spalte mit derselben Pulsquelle verbunden oder abgeschaltet, was die Anzahl an möglichen Eingangssignalen auf die Anzahl an Spalten beschränkt. Im Kontrast hierzu kann beispielsweise im FLANN-System (Giulioni, 2008) jede Synapse von einer unterschiedlichen Pulsquelle stimuliert werden, was eine größtmögliche topologische Flexibilität gewährleistet, aber auch einen teilweise redundanten topologischen Overhead erzeugt, da alle Synapsen eine eigene PSC-Schaltung benötigen.

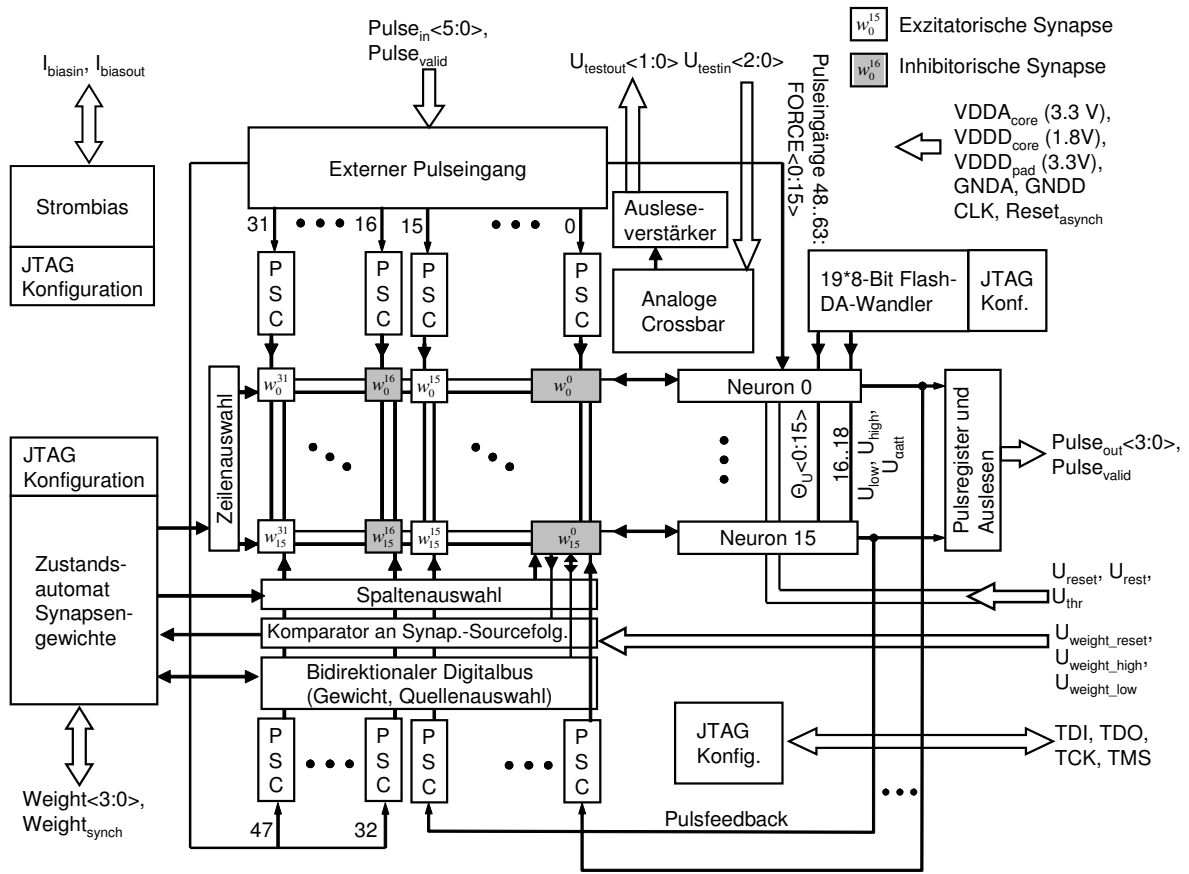

Abbildung 5.1: Gesamtübersicht des MAPLE: neuromorphe Architektur mit optimierter Topologie, periphären Baugruppen und Aussenanschlüssen (=Signale an Blockpfeilen).

Ein Kompromiss zwischen beiden Herangehensweisen wird in (Noack et al, 2010) vorgestellt: Jede Synapse einer Spalte kann zwischen zwei Pulsquellen auswählen, einer oberen PSC-Zeile für die externen Pulseingänge sowie einer unteren Zeile, die den Feedback von den auf dem MAPLE befindlichen Neuronen und weitere externe Pulseingänge realisiert (siehe Abbildung 5.1 bzw. auch Abbildung 5.10). Das Gesamtsystem besteht aus 16 Neuronen und einer Synapsenmatrix mit 16x32 Synapsen, daraus resultierend 64 PSC-Schaltungen[1]. Es wird somit nur eine um den Faktor zwei höhere Anzahl an PSC-Schaltungen als in (Schemmel et al, 2007) verwendet, wodurch aber die topologische Flexibilität deutlich verbessert wird. In Abbildung 5.2 wurde ein entsprechender Vergleich der Konfigurierbarkeit vorgenommen für diese drei unterschiedlichen Topologien:

- Der präsynaptische PSC Eingang ist jeweils für eine Spalte derselbe
- Die Synapse kann zwischen zwei PSCs auswählen

---

[1] Somit wird ein Faktor 8 weniger PSC-Schaltungen als in einer vergleichbaren FLANN-Architektur verwendet. Das Skalierungsverhalten für größere Matrizen ist noch deutlich vorteilhafter, da in der diskutierten Architektur (ähnlich wie bei (Schemmel et al, 2007)) die Anzahl der PSC-Schaltungen mit $2 \cdot \sqrt{N_{Synapsen}}$ zunimmt, während für die FLANN-Architektur eine Anzahl $N_{Synapsen}$ von PSC-Schaltungen benötigt wird.

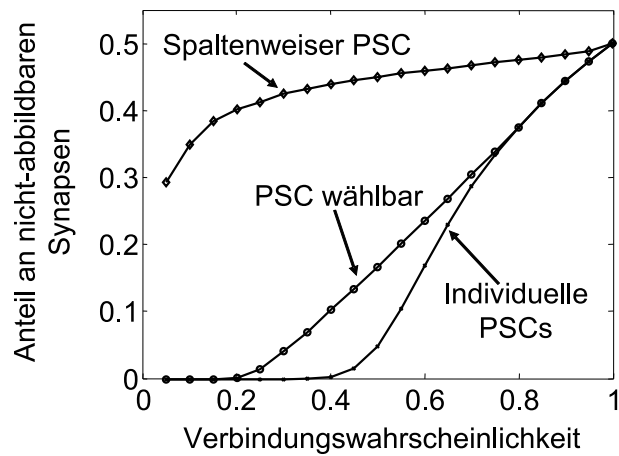

Abbildung 5.2: Vergleich verschiedener Systemarchitekturen (spaltenweiser PSC (Schemmel et al, 2007), PSC wählbar individueller PSC (Giulioni, 2008)) für zufällig verbundene neuronale Netze mit 64 Neuronen. Jede der möglichen synaptischen Verbindungen des Netzwerkes $64 \cdot 64 = 4096$ wird mit der auf der x-Achse gegebenen Wahrscheinlichkeit erzeugt.

- Jede Synapse kann über einen individuellen PSC angesprochen werden

Zur besseren Vergleichbarkeit wurde die Größe eines einzelnen ICs in der jeweiligen Topologie identisch mit dem MAPLE angenommen, d.h. 16 Neuronen und $32 \cdot 16$ Synapsen, und ein Gesamtnetzwerk aus 4 Chips untersucht, d.h. 64 Neuronen mit $64^2 = 4096$ möglichen Synapsen. Es wurde jeweils der Anteil an nicht abbildbaren Synapsen für verschiedene Verbindungswahrscheinlichkeiten der Einzelverbindungen (=Synapsen) bestimmt[2].

Für eine Verbindungswahrscheinlichkeit von 1 werden alle 4096 möglichen Synapsen tatsächlich ausgebildet. Da unabhängig von der PSC-Architektur nur $64 \cdot 32 = 2048$ Synapsen auf den Chips physisch implementiert sind, können 50% der Synapsen des Netzwerkes nicht auf die Chiptopologie abgebildet werden. Im Umkehrschluss kann anhand der Anzahl vorhandener Synapsen erwartet werden, dass der Chip für eine Verbindungswahrscheinlichkeit unter $p = 2048/4096 = 0.5$ genügend Synapsen für eine Abbildung des Netzwerkes auf den Chip bereitstellt. Die FLANN-Architektur repräsentiert beinahe dieses Optimum, sichtbar aus der 'Individuelle PSCs'-Kurve. Das Optimum wird nicht ganz erreicht, da beim stochastischen Aufbau des Netzwerks mit $p = 0.5$ zwar im Mittel 32 Synapsen pro Neuron benötigt werden, jedoch auf einzelne Neuronen auch mehr als 32 synaptische Eingänge entfallen können. Hier macht sich bemerkbar, dass die Synapsen im FLANN zwar individuell präsynaptisch adressiert werden können, jedoch jeweils 32 an dasselbe postsynaptische Neuron gebunden sind und mithin nicht beispielsweise Neuronen mit weniger als 32 synaptischen Eingängen ihre Synapsen den Neuronen mit mehr als 32 synaptischen Eingängen zur Verfügung stellen können.

---

[2]Da in der Hardware aufgrund der ressourcensparenden Implementierung nicht genug Synapsen für eine vollbesetzte Verbindungsmatrix vorhanden sind und zusätzliche Einschränkungen hinsichtlich der logischen Zuordnung der Synapsen zu Neuronen je nach den oben angeführten Topologien bestehen, gibt es für eine vorgegebene Netzwerkstruktur jeweils mehrere verschiedene Arten, diese auf die Hardwaretopologie abzubilden, mit unterschiedlichen Anteilen an nicht abbildbaren Synapsen. Für die dargestellten Beispiele wurde bei der Umsetzung der Netzstruktur auf die jeweilige Hardwaretopologie die im Sinne des Synapsenverlustes bestmögliche Variante gewählt. Die Kurven in Abbildung 5.2 repräsentieren einen Mittelwert über jeweils 1000 zufällige Netzrealisierungen.

Für die stark beschränkte Architektur aus (Schemmel et al, 2007) ist der Anteil der nicht-abbildbaren Synapsen über alle möglichen Verbindungswahrscheinlichkeiten besonders hoch, da nur 32 der möglichen 64 Pulsquellen in die Synapsenmatrix eingespeist werden können. Die in (Noack et al, 2010) vorgeschlagene Architektur erreicht im Unterschied dazu beinahe die Flexibilität der vollen Konfigurierbarkeit der FLANN-Architektur (Giulioni, 2008). Zusätzlich zur deutlich verminderten PSC-Anzahl[1] ist dieses Ergebnis auch dahingehend bemerkenswert, als in der Synapse nur ein zusätzliches Konfigurationsbit integriert wurde, mit dem in der jeweiligen Spalte die Pulsquelle umgeschaltet werden kann.

**Gesamtkonzept des MAPLE**

Abgesehen von den eigentlichen neuromorphen Bestandteilen, d.h. PSCs, Neuronen und Synapsen, enthält der Chip noch weitere analoge und digitale Baugruppen zur Ansteuerung der neuromorphen Funktionalität bzw. zum Test. Da in den Synapsen nur die Gewichtsberechnung analog stattfindet, jedoch die Gewichtsspeicherung über SRAM-Zellen realisiert ist (siehe Abschnitt 5.4), existiert ein Zustandsautomat, der das entsprechende zyklische Auslesen und Schreiben der Synapse ähnlich wie bei den Pixelmatrizen in (Mayr et al, 2008a; Henker et al, 2007) übernimmt. Der Systemtakt für alle synchronen Digitalbaugruppen ist 100 MHz. Die Gewichte werden zur Überwachung des Lernverhaltens während des Lesens auch extern ausgegeben. Die 4 Bit des digitalen Gewichtsspeichers in jeder Synapse skalieren über binär gewichtete Stromspiegel den von den PSC-Schaltungen bereitgestellten präsynaptischen Stromverlauf und geben ihn auf eine Stromsammelschiene Richtung Neuronenmembran weiter. In Übereinstimmung mit biologischen Erkenntnissen werden nicht nur exzitatorische Synapsen (d.h. Membran-aufladende), sondern auch inhibitorische (d.h. Membran-entladende) bereitgestellt (d.h. die Stromausgänge der 0. und 16. Synapsenspalte sind an den Neuronenmembranen nach Masse geschaltet, während die anderen Synapsenspalten im Fall eines präsynaptischen Pulses einen aufladenden PSC von VDD bereitstellen). Synapsen und PSC-Schaltungen der Spalten 0 und 16 sind ansonsten identisch mit den übrigen, beispielsweise ist also auch das Lernverhalten dieser inhibitorischen Synapsen gleich dem der exzitatorischen. Ein Interface zur externen Eingabe von Pulsen steuert die PSC-Schaltungen an (siehe Abschnitt 5.3) und kann ebenfalls dazu verwendet werden, die Neuronen des MAPLE von extern zu triggern, d.h. der Puls eines Neurons ist in diesem Fall keine Antwort auf ein über den Feuerschwellwert aufgeladenes Membranpotential, sondern kann gezielt ausgelöst werden. Dies dient dazu, verschiedene Plastizitätsexperimente wie etwa STDP (1) mit dem MAPLE nachzustellen. Die Pulszeitpunkte der Neuronen werden als Digitalwerte über ein weiteres Interface ausgegeben (siehe Abschnitt 5.2). Über einen 19-fach ausgeführten 8-Bit Digital-Analog-Wandler (DAC) werden die 16 $\Theta_U$ der Neuronen sowie die Parameter der postsynaptischen Adaption bereitgestellt (siehe Abschnitt 5.5).

Mit einer analogen Auswahl- und Verschaltmatrix ähnlich der im (Mayr et al, 2010d) oder (Mayr et al, 2008a) verwendeten können bis zu 3 Analogspannungen an verschiedene Stellen des MAPLE zu Testzwecken eingespeist werden bzw. bis zu 2 Analogspannungen über

Pufferverstärker ausgelesen werden[3]. Es ist damit beispielsweise möglich, die Adaption der PSC-Schaltungen, die Funktion der Neuronen oder das Lernverhalten der Synapsen zu überwachen bzw. durch gezielte Ansteuerung zu testen. Eine Strombank zur Bereitstellung der Biassignale der verschiedenen Verstärker bzw. zur Einstellung der strombasierten Parameter der neuromorphen Baugruppen vervollständigt die analogen Peripheriekomponenten.

Ein konventionelles Konfigurations- und Testinterface für ICs (JTAG, siehe (IEEE, 2001)) wird zur Ansteuerung der verschiedenen analogen und digitalen Konfigurationsparameter verwendet. Damit ist es beispielsweise möglich, die verschiedenen Biasströme digital einzustellen, in der Synapsenmatrix die initialen Synapsengewichte und die PSC-Auswahl individuell zu setzen, das Lernverhalten der Synapsen im Zustandsautomat synapsenindividuell zu aktivieren oder zu deaktivieren, die DACs anzusteuern oder die analoge Verschaltmatrix zu konfigurieren. Das JTAG-Interface kann mit maximal 10 MHz betrieben werden.

Detaillierte Simulationsergebnisse des MAPLE-Entwurfes, d.h. zu Einzelbaugruppen und Gesamtverhalten insbesondere des neuromorphen Teils sind in Abschnitt 5.6 dokumentiert. Wie bereits erwähnt, arbeitet MAPLE auf einer gegenüber biologischer Echtzeit um den Faktor $10^4$ beschleunigten Zeitbasis, d.h. 1 $\mu s$ im MAPLE-Kontext entspricht 10 ms in der Biologie. Bei den Vergleichen mit biologischen Messdaten in Abschnitt 5.6 wurde die MAPLE-Zeitbasis auf die biologische umgerechnet, bei einzeln stehenden Simulationskurven (insbesondere in den Abschnitten zu den Einzelbaugruppen) wurde die reale (beschleunigte) Zeitbasis beibehalten. Die Abschnitte 5.7 und 5.8 enthalten zukünftige Erweiterungen des MAPLE zur Steigerung der Einsetzbarkeit in Kortex-adaptierten Verarbeitungsfunktionen.

## 5.2 Schaltung des adaptiven Neurons

Die Neuronenschaltung, welche den $u(t)$-Verlauf in Abbildung 3.4 konform mit der LCP-Regel generiert, wird in (Mayr et al, 2010a) vorgestellt. Es handelt sich dabei um ein erweitertes LIAF-Neuron mit einer Schaltung gemäß Abbildung 5.3. Alle in das Neuron fließenden Ströme werden auf der Kapazität $C_{mem}$ integriert, in Folge wird das Membranpotential $u(t)$ des Neurons gebildet, hier in der Schaltungsrealisierung als $U_{mem}$ bezeichnet.

Das Modul 'Leckstrom' in Abbildung 5.3 besteht aus einem OTA, bei dem sowohl der negative Eingang als auch der Ausgang mit der Membrankapazität verbunden sind (Mayr et al, 2010a)(Koickal et al, 2007). Dadurch wird ein über den Biasstrom $I_{refr}$ des OTA steuerbarer Widerstand $R$ wirksam, mithin wird ein LIAF-Neuron mit einer einstellbaren Zeitkonstanten $\tau_{refr} = R \cdot C_{mem}$ realisiert. Das Ruhepotential kann über die Spannung $U_{rest}$ am nicht-invertierenden Eingang des OTA eingestellt werden. Bei Erreichen eines Schwellwertes von $U_{thr}$ oder Aktivierung des externen Eingangs zum Pulsauslösen (FORCE) wird ein RS-Flipflop gesetzt, dessen Ausgang ein AP signalisiert und das Membranpotential über M1 auf VDDA führt. Dies ist einer der Hauptunterschiede der hier vorgestellten Im-plementierung zu konventionellen LIAF-Neuronen, bei denen das Membranpotential und das AP zwei unterschiedliche Wellenformen darstellen und das Membranpotential während eines AP letzlich nicht definiert ist (Koickal et al, 2007). In dem hier vorgestellten Neuron wird zwar wie in anderen LIAF-Realisierungen das AP separat am Ausgang gemeldet, aber auch das Membranpotential während eines APs auf VDDA gesetzt, so dass der über einen Pufferverstärker Richtung Synapse gegebene Membranspannungsverlauf wie in Abbildung 3.4 für die LCP-Regel gefordert sowohl die Dynamiken während der Aufladung als auch während des APs beschreibt. Das für die LCP-Synapse ausserdem benötigte $\Theta_U$ wird aus dem zum Neuron gehörigen DA-Wandler Richtung Synapse durchgeschleift. Während des APs ist das Digitalsignal $SPK$ gesetzt, ein Reset findet über die Schaltung zur Pulsgenerierung (Abbildung 5.4) statt. Nach dem AP wird das Membranpotential während des kurzen Überlappungsintervalls zwischen $\overline{SPK}$ und $SPK_{delayed}$ über M2/3 auf eine Spannung $V_{reset}$ unter das Ruhepotential zurückgesetzt und durchläuft danach eine Hyperpolarisationskurve, wodurch sich ein Spannungsprofil entsprechend Gleichung 3.11 bzw. Gleichung 3.4 ergibt. Wie bereits in der Systembeschreibung erwähnt, hat die Membran jeweils einen über Stromspiegel gepufferten Eingang für eine inhibitorische und eine exzitatorische PSC-Sammelschiene aus der Synapsenmatrix[4].

---

[3]Momentan werden für die Auslese-/Pufferverstärker (der Verschaltmatrix, aber auch der Puffer der Membranspannung Richtung Synapsenmatrix) Klasse A Verstärker verwendet. Zukünftige Versionen des MAPLE werden hier die adaptive Bias aus (Ellguth et al, 2009) verwenden, da damit mit minimalem Modifizieraufwand (jetzige Verstärker bleiben gleich) eine energetisch günstige Umrüstung auf AB-Betrieb möglich ist

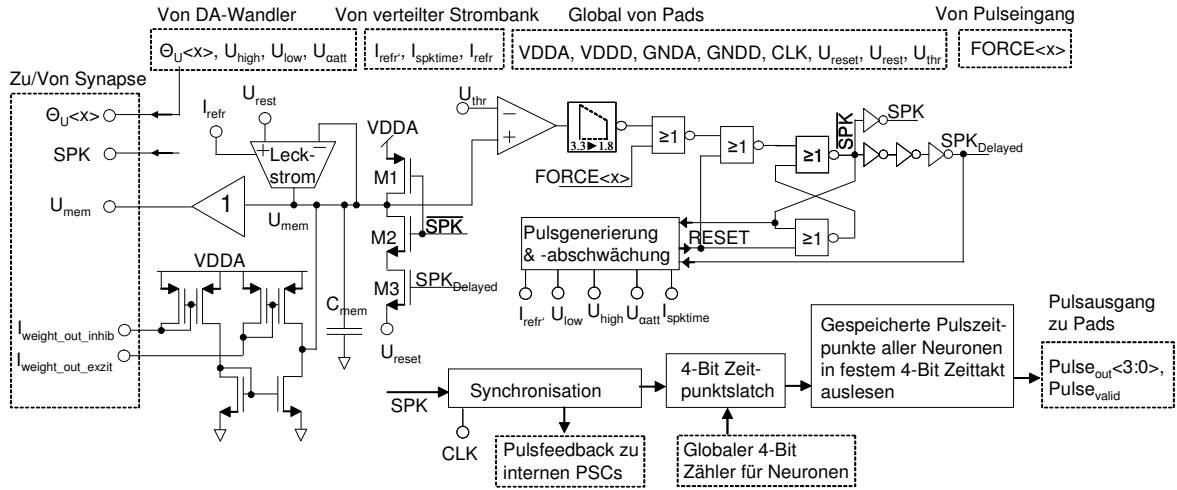

Abbildung 5.3: Schaltung des LIAF-Neurons mit postsynaptischer Adaption für die LCP-Lernregel, mit gruppierten Schnittstellen zu anderen Schaltungsgruppen und den Baublöcken, die das synchrone Pulsauslesen realisieren. Schnittstellen zwischen analogen und digitalen Baugruppen enthalten, wie nach dem Schwellwertschalter des Neurons angedeutet, entsprechende Pegelwandler. Diese wurden im restlichen Teil der Abbildung (etwa zwischen der Pulsgenerierung und dem RS-Flipflop) zum leichteren Verständnis weggelassen.

Der im asynchronen RS-Flipflop registrierte Puls wird zur weiteren Verarbeitung mit dem Systemtakt synchronisiert. Er wird im Folgenden einerseits zu den für die interne Pulsrückkopplung zuständigen PSC-Schaltungen weitergeleitet, andererseits wird zum Zeitpunkt des Pulses auch ein 4-Bit Register mit dem aktuellen Wert eines fortlaufenden 4-Bit Zählers belegt, d.h. der Pulszeitpunkt wird registriert. Diese Registerbank mit den Pulszeitpunk-ten aller Neuronen wird in fortlaufendem Zyklus (alle 16 Takte dasselbe Neuron) ausgelesen, so dass die Pulszeitpunkte der Neuronen mit der Auflösung des Systemtaktes (d.h. 1/100 MHz=10 ns bzw. 0.1 ms biologisch) und mit einem maximalen Durchsatz pro Neuron von 100 MHz/16 (d.h. $6.25 \cdot 10^6$ Pulse/s oder 625 Pulse/s biologisch), nach aussen gemeldet werden können[5].

---
[4]Die Sammelschienen sind aus Synapsensicht definiert und deshalb im Subskript als '...out' tituliert

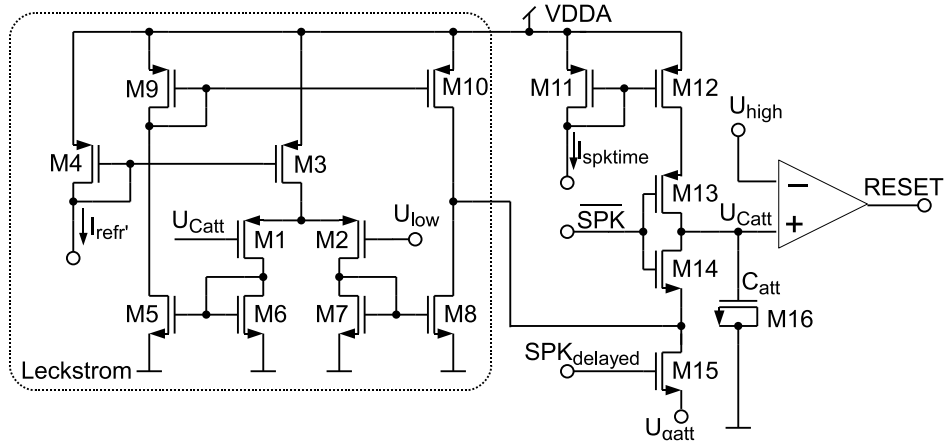

Abbildung 5.4: Schaltung zur Pulsgenerierung und -abschwächung.

Der Baublock 'Pulsgenerierung und Abschwächung' erzeugt eine postsynaptische Adaption der Fläche des AP gemäß Gleichung 3.5. Unter der Annahme einer rechteckigen Fläche für das AP kann der Flächeninhalt über die Höhe oder die Zeitdauer variiert werden. Allerdings ist zur Ausnutzung eines möglichst großen Dynamikbereichs für $U_{mem}$ der Schwellwert $U_{thr}$ sehr hoch gesetzt. Da wie oben bereits erwähnt die LCP-Regel darauf basiert, dass eine einzige Zustandsvariable sowohl den Membranspannungsverlauf unterhalb des Feuerschwellwertes als auch das AP selbst abbildet (siehe Gleichung 3.4), könnte die Höhe des AP nur in dem übrigen Bereich zwischen $U_{thr}$ und VDDA variiert werden. Eine Änderung der AP-Fläche über die Pulsdauer ist damit eine deutlich günstigere Alternative. Kern der Pulsgenerierung und -abschwächung ist die Kapazität $C_{att}$, die über einen exponentiellen Abfall von $U_{C_{att}}$ die Anpassung der Pulsdauer bestimmt:

Im Ruhezustand ist $\overline{SPK}$ high, $SPK_{delayed}$ low und $C_{att}$ ist auf $U_{low}$ geladen. Damit ist $RESET$ ebenfalls low. Wenn das Membranpotential die Schwellspannung $U_{thr}$ erreicht, wird $\overline{SPK}$ low, was den über M14 aus $C_{att}$ abfliessenden Strom abschaltet. Anschliessend steigt die Spannung über $C_{att}$ durch Ladung mit dem Konstantstrom $I_{spktime}$ (über M11-13) linear an. Bei Erreichen der Schwellspannung $U_{high}$ des nachfolgenden Schwellwertschalters wird $RESET$ high, was über das RS-Flipflop in Abbildung 5.3 den Stromfluss in $C_{att}$ beendet und die Membranspannung zurücksetzt. Die Zeitdauer, die zum Laden von $C_{att}$ nötig ist, bestimmt die Dauer $t_{spk,n}$ des Pulses/APs:

$$t_{spk,n} = C_{att} \frac{U_{high} - U_{C_{att}}(t_n^{post})}{I_{spktime}} \tag{5.1}$$

---

[5]Das sequentielle Auslesen wurde hier gegenüber asynchronen, arbitrierten AER-Auslesemethoden bevorzugt, da es bei kleinen Neuronenzahlen wie im MAPLE mit deutlich weniger Schaltungsaufwand umgesetzt werden kann (Culurciello and Andreou, 2003)

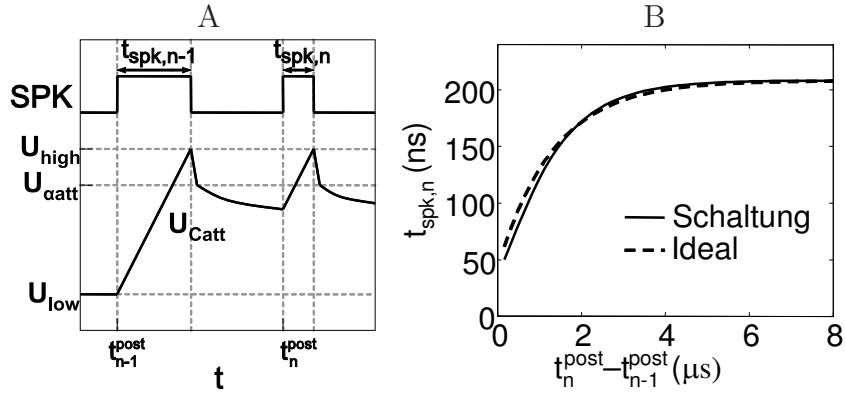

Abbildung 5.5: (A) Schema des Zeitverlaufs von $U_{C_{att}}$ und $SPK$ (B) AP-Dauer als Funktion des zwischen den APs liegenden Zeitintervalls

Nachdem $\overline{SPK}$ wieder high ist, lässt eine Leckstromschaltung (M1-10) ähnlich wie in Abbildung 5.3 über eine einstellbare Zeitkonstante $\tau_{refr'}$[6] $U_{C_{att}}$ wieder exponentiell auf $U_{low}$ abfallen. Wenn kurz nach einem AP ein zweites AP stattfindet, wird die Pulsdauer verkürzt, da die Zeit zum Laden von $C_{att}$ auf $U_{high}$ kürzer ist (siehe Abbildung 5.5A). Die Stärke der Adaptation kann über $U_{\alpha att}$ konfiguriert werden, auf das $C_{att}$ während der Zeitdifferenz zwischen $\overline{SPK}$ und $SPK_{delayed}$ über M14/15 entladen wird. Da diese Zeit klein ist gegenüber $\tau_{refr'}$, kann $U_{\alpha att}$ als neuer Startwert für den exponentiellen Abfall angenommen werden:

$$U_{C_{att}}(t) = U_{low} + (U_{\alpha att} - U_{low})e^{-\frac{t-t_{n-1}^{post}}{\tau_{refr'}}} \tag{5.2}$$

Einsetzen von Gleichung (5.2) in Gleichung (5.1) unter der zusätzlichen Annahme, dass $t_{spk,n-1}$ klein gegenüber der Zeitspanne $t_n^{post} - t_{n-1}^{post}$ ist, ergibt:

$$t_{spk,n} = \frac{C_{att}(U_{high} - U_{low})}{I_{spktime}} \left(1 - \frac{U_{\alpha att} - U_{low}}{U_{high} - U_{low}} \cdot e^{-\frac{t_n^{post} - t_{n-1}^{post}}{\tau_{refr'}}}\right) \tag{5.3}$$

Eine gedankliche Substitution von $\alpha_{att} = \frac{U_{\alpha att} - U_{low}}{U_{high} - U_{low}}$ und $U_p = t_{spk0} * VDDA$, bei $t_{spk0} = \frac{C_{att}(U_{high} - U_{low})}{I_{spktime}}$ in Gleichung 5.3 macht deutlich, dass die Schaltung aus Abbildung 5.4 im Rahmen der getroffenen Näherungen die Pulsabschwächung aus Gleichung 3.5 exakt nachbildet[7]. Abbildung 5.5B zeigt einen Vergleich zwischen einer Simulation der Schaltung und dem idealem Verhalten laut Gleichung 3.5.

---

[6]$\tau_{refr'}$ ist über $I_{refr'}$ separat von $\tau_{refr}$ einstellbar, wird jedoch üblicherweise identisch zu $\tau_{refr}$ gewählt, da die postsyn. Adaption im idealen LCP-Modell ebenfalls mit $\tau_{refr}$ operiert.

[7]Leichte Abweichungen entstehen wie erwähnt durch die getroffenen Näherungen zu den Zeitdauern und durch die nicht ideale Linearität des Leckstrom-OTA.

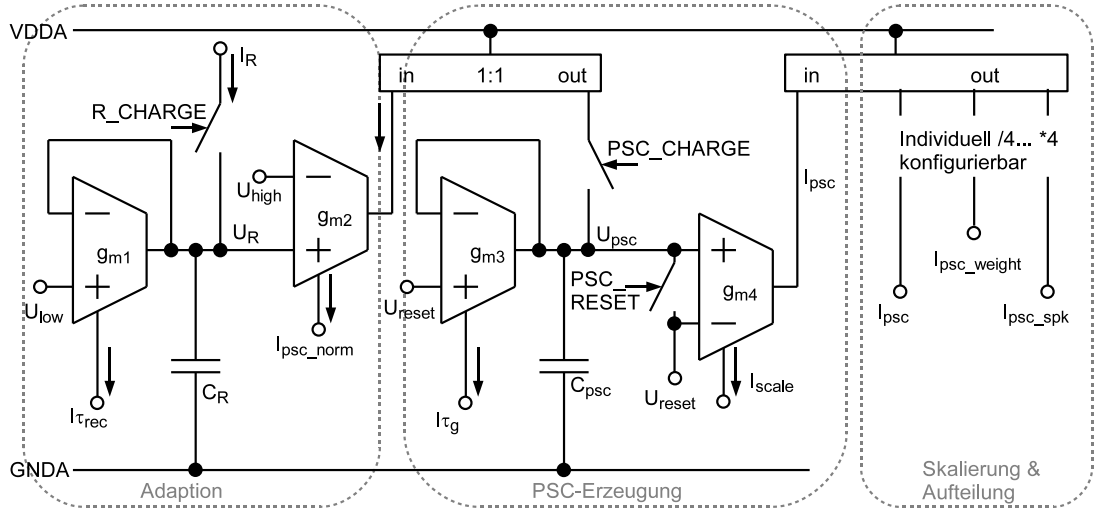

Abbildung 5.6: PSC-Schaltung aus präsynaptischer Adaption (links) und PSC-Generierung (rechts)[8]. Die eigentliche Adaptionsschaltung $g_{m1}$ und $C_R$ erzeugt den Laufzeitwert $U_R$ der Adaption, der über $g_{m2}$ in einen Spannungshub der PSC-Variable $U_{psc}$ umgesetzt wird; im weiteren Verlauf erfolgt ein exponentielles Abklingen der PSC-Variable über $C_{psc}$ und $g_{m3}$ und Umsetzung des Zeitverlaufs in einen PSC mit $g_{m4}$. Der PSC wird in drei unterschiedlich skalierten Versionen für die Synapsen bereitgestellt. Die OTA-Schaltungen $g_{m1}$ bis $g_{m4}$ sind aus Koickal et al (2007) adaptiert, mit zusätzlichen Kaskodestromspiegeln zur Reduzierung des Offset und einer Source-Gegenkopplung zur Erweiterung des linearen Arbeitsbereiches. Für weitere Schaltungsdetails sei auf (Noack et al, 2010) verwiesen.

## 5.3 Präsynaptische Rekonstruktion und Kurzzeitplastizität

In (Noack et al, 2010) werden Schaltungen vorgestellt zur Generierung des PSC-Verlaufs gemäß Gleichung 3.3 sowie einer analog zur PPD in Abschnitt 2.3 implementierten Kurzzeitplastizität des PSC. Die hier vorgestellte Schaltung stellt eine Weiterentwicklung dieses Konzeptes dar, wobei insbesondere auf die exakte Umsetzung des Quantalmodells gemäß der Umschreibung in Abschnitt 2.4 Wert gelegt wurde. Wie aus Abbildung 5.6 ersichtlich, wird der exponentielle PSC-Verlauf (als Spannung) über das aus $C_{psc}$ und $g_{m3}$ gebildete RC-Glied mit einer Zeitkonstanten $\tau_g$ erzeugt. Der OTA $g_{m4}$ wandelt über seine mit $I_{scale}$ einstellbare Transkonduktanz die Spannung über $C_{psc}$ in einen Ausgangsstrom $I_{psc}$, d.h. den eigentlichen PSC. Der so erzeugte PSC wird über individuell konfigurierbare Stromspiegel in drei unterschiedlichen Amplituden Richtung Synapsenmatrix gegeben (zur Verwendung dieser drei Varianten siehe Abschnitt 5.4).

Die Amplitude des PSC ist bei fester Zeitdauer, die PSC_CHARGE aktiv ist (siehe Abbildung 5.7B), und bei Vernachlässigung des Stroms von $g_{m3}$ während der Ladephase, linear vom Ausgangsstrom von $g_{m2}$ und damit vom Momentanwert der Adaptionsvariablen $U_R(t)$ (äquivalent zu $R(t)$) abhängig. Über das Signal PSC_RESET, das vor der Aufladung von $C_{psc}$ gesetzt werden kann, sind zwei verschiedene Betriebsarten der PSC-Schaltung möglich. Wenn PSC_RESET gesetzt ist, befindet sich die Schaltung in der 'Nächster-Nachbar'-

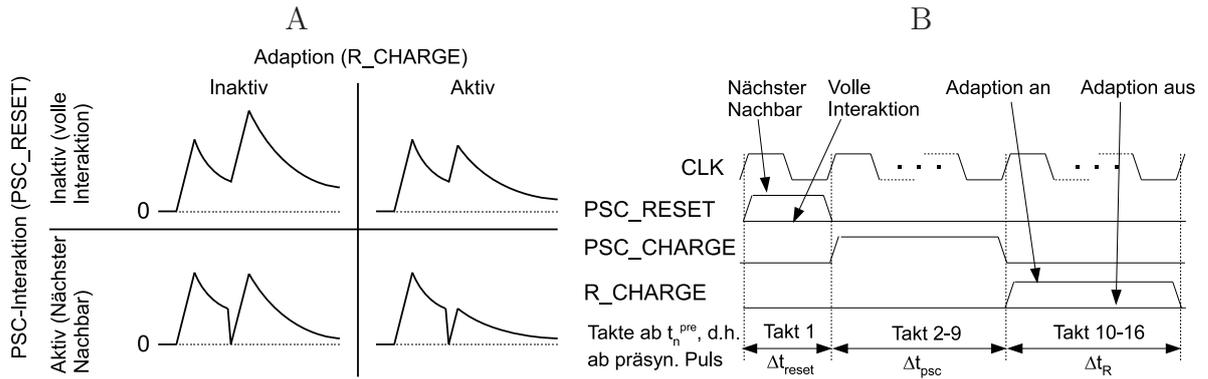

Abbildung 5.7: (A) Arbeitsmodi der PSC-Schaltung. Die Adaption kann an- oder abgeschaltet werden. Unabhängig davon kann die Interaktion zwischen den einzelnen PSCs auf 'voll' (Überlagerung, verstärkender Effekt) oder auf 'Nächster Nachbar' (keine Verstärkung, unabhängige PSC-Amplituden) gesetzt werden. (B) Signaldiagramm der PSC-Schaltung: Ein eingehender präsynaptischer Puls startet einen 4-Bit Zähler. Im ersten Takt wird bei aktiviertem Nächster-Nachbar-Modus $C_{psc}$ über RESET_PSC zurückgesetzt. In Takt 2-9 findet die Aufladung von $C_{psc}$ zum Spitzenwert des PSC statt. In Takt 10-16 wird bei aktivierter Adaption $U_R$ erhöht.

Betriebsart. Vor der erneuten Aufladung von $C_{psc}$ durch einen PSC-signalisierenden Strompuls wird die Spannung $U_{psc}$ durch einen kurzen Puls von PSC_RESET auf $U_{reset}$[8] zurückgesetzt. Damit beeinflusst nur der aktuelle präsynaptische Puls die Höhe des PSC. Wenn PSC_RESET dauerhaft Low ist, überlagern sich aufeinanderfolgende PSC, d.h. der jeweils während PSC_CHARGE auf $C_{psc}$ geladene neue Spannungshub wird auf den Momentanwert von $U_{psc}$ addiert (siehe Abbildung 5.7A).

Die Schaltung der Adaptionsvariablen $U_R(t)$ in Abbildung 5.6 versucht, wie oben erwähnt, möglichst genau das Verhalten von $R(t)$ aus Gleichung 2.18 nachzustellen, d.h. eine mit zunehmender präsynaptischer Pulsfrequenz abnehmende PSC-Amplitude zu realisieren. Diese Äquivalenz wird im Folgenden hergeleitet.

Für die Herleitung ist zu beachten, dass ein verstärkender Effekt gemäß Gleichung 2.15 nicht Teil dieser Realisierung ist, es wird wie in Abbildung 2.8 angenommen, dass $\tau_{\text{facil}} \to 0$, d.h. $u_n = u_{n+1} = U$. Der einzige 'verstärkende' Effekt ist der oben angeführte 'Volle Interaktion'-Modus, der jedoch inhärent mit der gleichen Zeitkonstante $\tau_g$ wie das PSC selbst stattfindet, d.h. viel kleiner als die üblicherweise angenommene Größenordnung für $\tau_{facil}$ ist. Dieser Modus ist unabhängig von der Zeitskala auch im Verhalten nicht direkt kompatibel mit der Herleitung in Abschnitt 2.4, weshalb im Folgenden der 'Nächste-Nachbar'-Modus angenommen wird. Die Amplitude von $U_{psc}$ für den n-ten präsynaptischen Puls $\hat{U}_{psc,n}$ kann

---

[8] Aufgrund der begrenzten Anzahl an verfügbaren Biasspannungen wurden in der obigen Schaltung unkritische Spannungen wiederverwendet. So sind etwa $U_{low}$ und $U_{high}$ identisch mit den für die postsynaptische Adaption verwendeten Spannungen, $U_{reset}$ ist die Rücksetzspannung des Neurons. Die unabhängige Einstellbarkeit beispielsweise der prä- und postsynaptischen Adaption ist durch die jeweils dort vorhandenen zusätzlichen Freiheitsgrade trotzdem gesichert.

nach durch PSC_CHARGE erfolgter Aufladung wie folgt beschrieben werden:

$$\hat{U}_{psc,n} = U_{reset} + \frac{(U_{high} - U_{\mathrm{R}}(t_n^{pre})) \cdot g_{m2} \cdot \Delta t_{\mathrm{psc}}}{C_{psc}} \ , \ t = \Delta t_{\mathrm{reset}} + \Delta t_{\mathrm{psc}} + t_n^{pre} \qquad (5.4)$$

d.h. die maximale Spannung wird auf $C_{psc}$ nach $t_n^{pre} + \Delta t_{\mathrm{reset}} + \Delta t_{\mathrm{psc}}$ erreicht. Diese wird linear in einen Strom umgewandelt:

$$\hat{I}_{psc,n} = (\hat{U}_{psc,n} - U_{reset}) \cdot g_{m2} \ , \ t = \Delta t_{\mathrm{reset}} + \Delta t_{\mathrm{psc}} + t_n^{pre} \qquad (5.5)$$

Zu beachten ist, dass das Signal R_CHARGE erst nach den obigen Gleichungen aktiv wird (siehe Abbildung 5.7B), d.h. es wird in Übereinstimmung mit dem Ablauf aus Abbildung 2.9 nach dem Reset von $U_{psc}$ zuerst auf Basis des alten Wertes von $U_R$ der PSC für den aktuellen Puls erzeugt. Danach erfolgt die Aufladung von $U_R$. Wenn angenommen wird, dass $g_{m1}$ und $C_R$ gemeinsam eine Zeitkonstante $\tau_{rec}$ realisieren, kann der exponentielle Abfall von $U_{\mathrm{R}}(t)$ nach dieser Aufladung, d.h. nach $\Delta t_{\mathrm{R}}$, wie folgt beschrieben werden:

$$U_{\mathrm{R}}(t) = U_{low} + \left( U_{\mathrm{R}}(t_n^{pre}) + \frac{\Delta t_{\mathrm{R}} \cdot I_{\mathrm{R}}}{C_{\mathrm{R}}} - U_{low} \right) \mathrm{e}^{-\frac{t-t_n^{pre}}{\tau_{rec}}} \ , \ t_n^{pre} + \Delta t_{\mathrm{reset}} + \Delta t_{\mathrm{psc}} + \Delta t_{\mathrm{R}} < t \leq t_{n+1}^{pre}$$
$$(5.6)$$

Gleichung 5.6 ist zwar erst ab dem Zeitpunkt $t_n^{pre} + \Delta t_{\mathrm{reset}} + \Delta t_{\mathrm{psc}} + \Delta t_{\mathrm{R}}$ gültig, es wird jedoch im Folgenden angenommen, dass $\Delta t_{\mathrm{reset}} + \Delta t_{\mathrm{psc}} + \Delta t_{\mathrm{R}} << t_{n+1}^{pre} - t_n^{pre}$. Diese Annahme ist typischerweise erfüllt. Der Zeitverlauf des PSC ist deutlich länger ($\tau_g \approx 3\mu s$ bzw. 30 ms biologisch) als die 16 für $\Delta t_{\mathrm{reset}} + \Delta t_{\mathrm{psc}} + \Delta t_{\mathrm{R}}$ benötigten Systemtakte (äquivalent zu 160 ns bzw. 1.6 ms biologisch). Auf Grundlage dieser Näherung wird beispielsweise im Exponenten der e-Funktion in Gleichung 5.6 nur $t_n^{pre}$ aufgeführt. Zu beachten ist, dass $U_{\mathrm{R}}(t)$ eine zu $R(t)$ umgekehrte Polarität besitzt (vergleiche Abbildung 5.8 und Abbildung 2.9). Die vorzeichenrichtige Entsprechung von $R(t)$ in der Schaltungsrealisierung kann wie folgt dargestellt werden:

$$R(t_n^{pre}) = \frac{U_{high} - U_{\mathrm{R}}(t_n^{pre})}{U_{high} - U_{low}} \ \mathrm{bzw.} \ R(t_{n+1}^{pre}) = \frac{U_{high} - U_{\mathrm{R}}(t_{n+1}^{pre})}{U_{high} - U_{low}} \qquad (5.7)$$

Einsetzen von Gleichung 5.6 in Gleichung 5.7 und Umformen ergibt:

$$R(t) = \frac{U_{high} - U_{low} - \left( U_{\mathrm{R}}(t_n^{pre}) + \frac{\Delta t_{\mathrm{R}} \cdot I_{\mathrm{R}}}{C_{\mathrm{R}}} - U_{low} \right) \mathrm{e}^{-\frac{t-t_n^{pre}}{\tau_{rec}}}}{U_{high} - U_{low}} \qquad (5.8)$$

$$= 1 - \left( 1 - R(t_n^{pre}) + \frac{\Delta t_{\mathrm{R}} \cdot I_{\mathrm{R}}}{C_{\mathrm{R}} \cdot (U_{high} - U_{low})} \right) \mathrm{e}^{-\frac{t-t_n^{pre}}{\tau_{rec}}} \qquad (5.9)$$

Ein Vergleich von Gleichung 5.9 mit Gleichung 2.18 macht deutlich, dass $\Delta R_n = \frac{\Delta t_{\mathrm{R}} \cdot I_{\mathrm{R}}}{C_{\mathrm{R}} \cdot (U_{high} - U_{low})}$. Wie aus dieser Herleitung ersichtlich, kennt $R(t)$ in der Schaltungsimplementierung keine multiplikative (d.h. weiche) Begrenzung, $U_{\mathrm{R}}(t)$ kann über $U_{high}$ steigen. Die Adaption wird allerdings durch die PMOS-Stromspiegel (in Abbildung 5.6 durch den

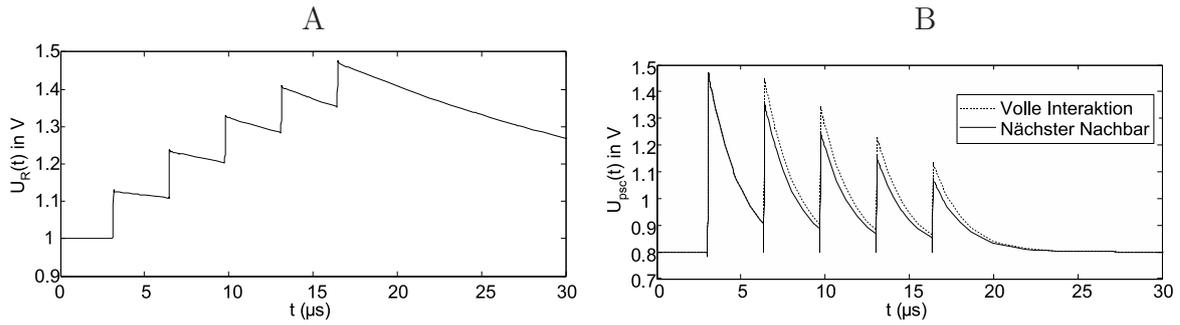

Abbildung 5.8: Beispielkurven für $U_R$ (Vergleiche Abbildung 2.9) und $U_{\text{psc}}$ mit $\tau_{rec} = 24\,\mu s$ (240 ms biologisch), $\tau_g = 1.8\,\mu s$ (18 ms), $U_{high} = 1.6\,V$, $U_{low} = 1.0\,V$ und einem Spannungshub pro präsynaptischem Puls von $\Delta R_n \cdot (U_{high} - U_{low}) = \frac{\Delta t_R \cdot I_R}{C_R} = 0.12\,V$

Stromspiegel ab VDDA angedeutet) hart begrenzt, d.h. wenn bei vielen dicht aufeinander folgenden präsynaptischen Pulsen $U_R(t)$ über $U_{high}$ steigt, ist die PSC-Amplitude 0. Bei entsprechender Parametrisierung ist der Dynamikbereich für $U_R(t)$ leicht groß genug, um eine biologisch relevante Anzahl an präsynaptischen Pulsen zu berücksichtigen, ohne zu begrenzen (siehe Abbildung 5.8).

Für die Bildung des Gesamt-PSC aus Gleichung 2.3, d.h. $\hat{I}_{psc}$ in der Schaltungsrealisierung, wird Gleichung 5.4 in Gleichung 5.5 eingesetzt und gemäß Gleichung 5.7 ein Rückbezug zu $R(t_n^{pre})$ hergestellt:

$$\hat{I}_{psc,n} = \frac{(U_{high} - U_R(t_n^{pre})) \cdot g_{m2} \cdot \Delta t_{\text{psc}} \cdot g_{m4}}{C_{psc}} \tag{5.10}$$

$$= R(t_n^{pre}) \cdot \frac{(U_{high} - U_{low}) \cdot g_{m2} \cdot \Delta t_{\text{psc}} \cdot g_{m4}}{C_{psc}} \tag{5.11}$$

Wie oben erwähnt wird angenommen, dass Gleichung 2.3 bei $\tau_{\text{facil}} \to 0$ in der Form $\text{PSC}_n = A \cdot R_n \cdot U$ Anwendung findet. Es ergibt sich damit aus einem Koeffizientenvergleich mit Gleichung 5.11, dass der neben $R(t_n^{pre})$ in Gleichung 2.3 auftretende konstante Term $A \cdot U$ gleich $\frac{(U_{high} - U_{low}) \cdot g_{m2} \cdot \Delta t_{\text{psc}} \cdot g_{m4}}{C_{psc}}$ ist.

Mithin wurde eine frei parametrisierbare Schaltungsnachbildung einer Hälfte des Quantalmodells implementiert, welche neben der sehr guten Nachbildung biologischer Lern- und Adaptionsexperimente (siehe Abbildung 5.13 bzw. Abbildung 3.11) ausserdem die Umsetzung komplexer dynamischer Verarbeitungsfunktionen analog zu (Mejias and Torres, 2009; Sussillo et al, 2007) im MAPLE erlaubt. Mit ungefähr dem doppelten Schaltungsaufwand wäre gemäß der Herleitung in Abschnitt 2.4 auch eine Implementierung des vollständigen Quantalmodells möglich. Durch die am Anfang dieses Kapitels diskutierte optimierte Topologie mit aus der Synapsenmatrix ausgelagerten PSC-Schaltungen wäre dieser Schaltungsaufwand für zukünftige Versionen des MAPLE durchaus vertretbar. Mit der Gesamtschaltung generierte PSC-Wellenformen sind in Abschnitt 5.6 enthalten.

## 5.4 Synapse und Management der synaptischen Gewichte

Der Kern der synaptischen Gewichtsberechnung in der LCP-Regel, d.h. des Lernverhaltens, besteht gemäß Gleichung 3.2 aus der Subtraktion des Schwellwertes $\Theta_U$ von der postsynaptischen Membranspannung und aus der Multiplikation dieser Differenz mit der präsynaptischen Aktivität. Eine Multiplikation liesse sich in CMOS beispielsweise im Subthresholdbereich der Transistoren mit dem dort gültigen exponentiellen Zusammenhang zwischen Gate-Source-Spannung und Drainstrom mittels einer translinearen Masche abbilden (Gilbert, 1975). Allerdings wirken sich die fertigungstechnischen Schwankungen der CMOS-Transistoren in diesem Bereich besonders stark aus (Bartolozzi and Indiveri, 2007). Zudem ist die Kombination einer derartigen Multiplikation mit der ausserdem nötigen Subtraktion nicht offensichtlich. Die synaptische Gewichtsberechnung in Abbildung 5.9 wurde daher statt dessen ähnlich wie in (Schreiter et al, 2002) als ein differentielles Paar (M9/10) ausgeführt, wobei man sich zunutze macht, dass die Stromdifferenz zwischen beiden Zweigen zum einen proportional zur Differenz $U_{mem} - \Theta_U$ ist (Subtraktion des Schwellwertes), zum anderen aber auch proportional zum Tailstrom (=Zuführung der PSC-Wellenform), der dem differentiellen Paar über M2 vorgegeben wird. Die Gewichtsänderung $\frac{dw}{dt}$ in Gleichung 3.2 liegt damit im differentiellen Paar als Stromdifferenz vor, die über die Pfadstromspiegel M11/13 und M12/14 sowie über den Kaskodestromspiegel M15-18 gebildet wird. Das absolute Gewicht wird durch Aufintegration der Stromdifferenz auf $C_{weight}$ gebildet. Der LCP-kompatible Verlauf von $U_{mem}$ gemäß Gleichung 3.4 wird gepuffert von dem jeweiligen Neuron der Synapsenzeile bereitgestellt (siehe Abbildung 5.1 bzw. Abbildung 5.3). M7/8 bilden eine Source-Gegenkopplung zur Herabsetzung der Transkonduktanz und Erhöhung des linearen Aussteuerbereichs (Sanchez-Sinencio and Silva-Martinez, 2000). Da die Stromspiegel M11-14 den durch das differentielle Paar fliessenden Strom um einen (relativ großen) Faktor 9 herabspiegeln um den Spannungshub pro Puls auf $C_{weight}$ zu reduzieren, wurde für besseres Matching und zur Flächenreduktion eine 1/3-zu-3 seriell-parallel Struktur verwendet (angedeutet mit den gestrichelten Transistoren, nach Arnaud and Galup-Montoro (2003)) statt einer einfachen 1:9-Stromspiegelung.

Es besteht bei $U_{mem}$ in dieser Schaltungsrealisierung ein ähnliches Problem wie bei der postsynaptischen Adaption in Abschnitt 5.2. Der zwischen $V_{thr}$ und VDDA liegende Dynamikbereich ist für eine biologisch realistische Abbildung der Wellenformen zu klein. Wie im Text zu Gleichung 3.9 ausgeführt, sind für realistische STDP-Fenster die Flächen der Hyperpolarisation und des Aktionspotentials ungefähr gleich. Um dies zu erreichen, müsste das Verhältnis zwischen der Refraktärsamplitude (in der Schaltungsrealisierung $U_{rest} - U_{reset}$) und der Höhe des APs (in der Schaltungsrealisierung $VDDA - U_{thr}$) umgekehrt proportional zum Verhältnis zwischen der Pulsdauer (ca. 200 ns, 2 ms biologisch) und der Zeitkonstante $\tau_{refr}$ der Membran ((ca. 3 $\mu s$, 30 ms biologisch) sein. Damit müsste der Puls einen Faktor 15 höher sein als die Membrandynamiken unterhalb des Schwellwertes, was die Membrandynamiken auf einen zu kleinen Spannungsbereich einschränken würde. Der hier gewählte Lösungsweg ist vielmehr, während eines APs einen zusätzlichen, entsprechend skalierten $I_{psc\_spk}$ über M3 und den Schalter M5 in das differentielle Paar einzuspeisen, wodurch die Amplitudenverhältnisse wiederhergestellt werden. Abbildung 5.15 illustriert die korrekte Abbildung des idealen LCP-Modells in die vorliegende Schaltung.

Die Schaltung in Abbildung 5.9 dient lediglich der Gewichtsberechnung in Übereinstim-

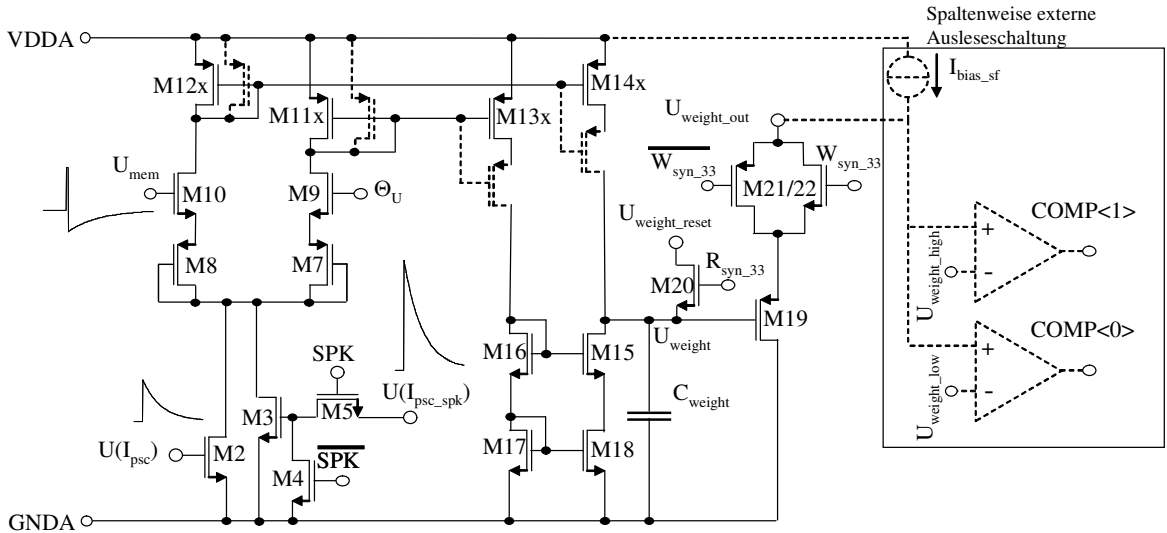

Abbildung 5.9: Analoger Teil der Synapsenschaltung, mit Gewichtsberechnung gemäß Gleichung 3.2, Reset und Auslese-Sourcefolger. Die Wellenformen des PSC und der postsynaptischen Membranspannung werden von den in Abschnitt 5.3 und Abschnitt 5.2 beschriebenen Schaltungen generiert.

mung mit der LCP-Regel, eine Gewichtswirkung basierend auf $U_{weight}$ findet nicht direkt statt. Das auf $U_{weight}$ gespeicherte Gewicht ist aufgrund verschiedener Analogeffekte nur flüchtig gespeichert (Zeitfenster ca. 5 ms, d.h. 50 s biologisch), was den von der LCP-Regel beschriebenen LTP/LTD-Vorgängen widerspricht, welche Gewichtsspeicherung auf biologischen Zeitskalen von Stunden bis Tagen benötigen (mithin im MAPLE für Zeiträume $> 0.5\ s$). Die Herangehensweise im MAPLE ist deshalb ähnlich gewählt wie in (Mitra et al, 2009; Schemmel et al, 2006), d.h. das analoge Zeitfenster wird nur für die Gewichtsberechnung verwendet, wohingegen die Gewichtsspeicherung und damit auch die Gewichtswirkung Richtung Neuron mit einem in der Synapse liegenden 4-Bit SRAM digital ausgeführt werden (siehe Abbildung 5.10). Ähnlich wie die Diodenspannung bei einem Pixelsensor (Mayr et al, 2008a) wird mit dem über M21/22 schaltbaren Sourcefolger in Abbildung 5.9 in regelmäßigen Abständen durch den ausserhalb der Synapsenmatrix liegenden Zustandsautomat (FSM) das synapsenindividuelle Gewicht ausgelesen und in einer für jede Spalte ausgeführten Komparatorschaltung mit zwei Schwellwerten bewertet. Wenn $U_{weight}$ zwischen beiden Schranken liegt, findet keine Aktion statt, die Synapse führt ihr bisheriges Lernverhalten fort (siehe auch Abbildung 5.11). Falls $U_{weight}$ kleiner als die untere Schranke ist, wird das ebenfalls ausgelesene 4-Bit Gewicht dekrementiert, zurück in die Synapse geschrieben und $U_{weight}$ für erneutes Lernen auf eine zwischen den beiden Schranken liegende Spannung $U_{weight\_reset}$ zurückgesetzt. Falls $U_{weight}$ oberhalb der oberen Schranke ist, wird das Gewicht inkrementiert, zurückgeschrieben und ebenfalls $U_{weight}$ zurückgesetzt[9].

---

[9]Wie bereits in Abschnitt 5.1 erwähnt, kann in der FSM als Alternative zu der analogen Plastizitätsberechnung das digitale Gewicht auch statisch gehalten werden. Die FSM enthält einen 512 Bit Speicher, mit dem individuell für jede Synapse die Auswertung der Komparatorsignale aktiviert oder deaktiviert werden kann. Im deaktivierten Fall bleibt die Synapse auf dem festen Gewicht, das bei der Initialisierung geschrieben wurde, oder kann über JTAG auch von aussen zur Laufzeit geschrieben werden

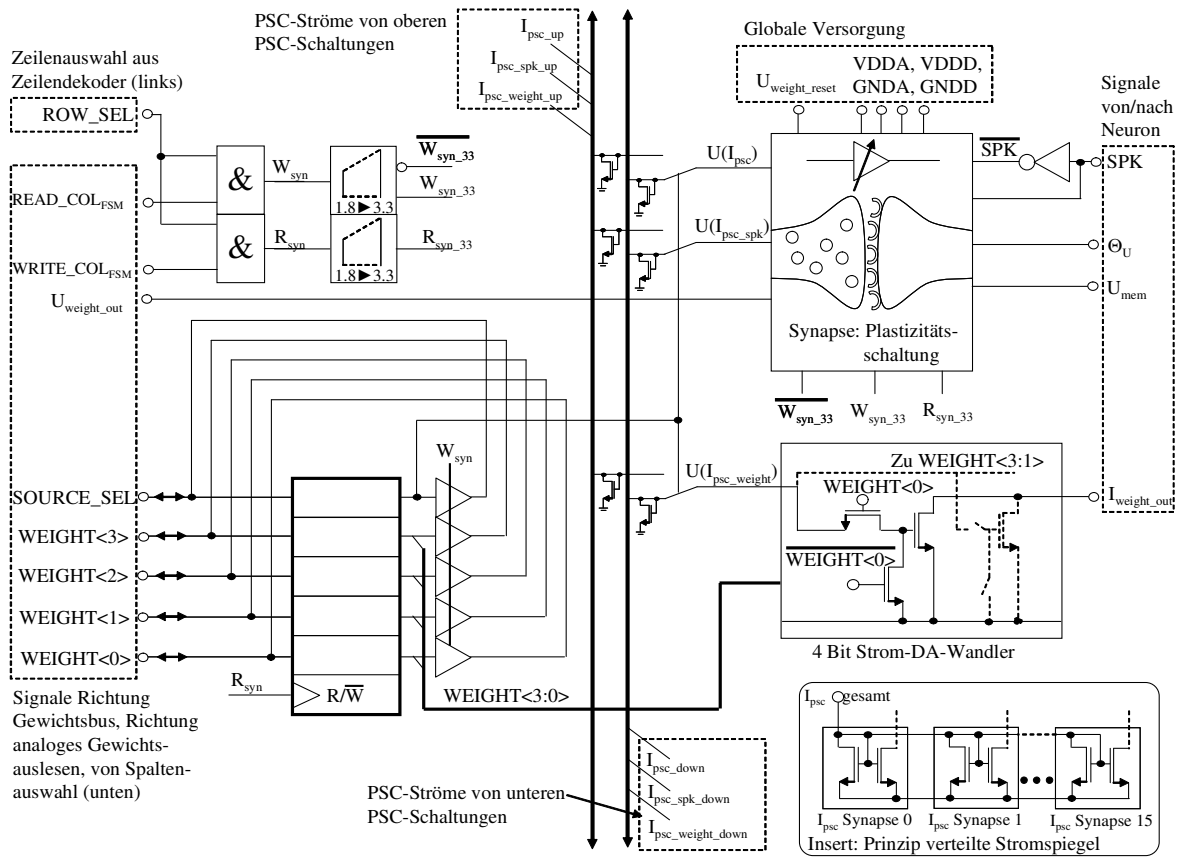

Abbildung 5.10: Blockschaltbild der Synapse mit interner Logik, DA-Stromwandler für die Gewichtswirkung, Auswahl der PSC-Ströme, Darstellung des Prinzips der verteilten Stromspiegel und der Plastizitätsschaltung aus Abbildung 5.9.

Abbildung 5.10 zeigt das komplette Blockschaltbild aller Baugruppen der Synapse. Über SOURCE_SEL werden entweder die drei Ströme der oberen oder unteren PSC-Schaltung für die Gewichtsberechnung ($I_{psc}$, $I_{psc\_spk}$) und die Gewichtswirkung ($I_{psc\_weight}$) ausgewählt. Die $I_{psc}$-, $I_{psc\_weight}$-, und $I_{psc\_spk}$-Verteilung von einer PSC-Schaltung auf alle Synapsen einer Spalte findet wie in Abbildung 5.10 dargestellt über einen direkt von der PSC-Schaltung kommenden Strom statt. Es wird nicht, wie sonst bei einer Stromverteilung üblich, eine Gatespannung eines Dioden-verbundenen Transistors an das Gate vieler Ausgangstransistoren kopiert. In jeder Synapse ist damit ein kompletter Stromspiegel ausgeführt, der von der PSC-Schaltung aus gesehen in Summe aller Synapsen wie ein Spiegel mit der 16-fachen Weite/Länge (bei 16 Synapsen pro Spalte) des jeweils einzelnen Synapsenspiegels gesehen wird. Das Prinzip ist in der rechten unteren Ecke von Abbildung 5.10 dargestellt. Der Vorteil dieser Variante gegenüber einer konventionellen Strombank ist zum einen, dass der aus der PSC-Schaltung kommende Strom für jede Synapse durch den Faktor N geteilt wird, wodurch sich der Spannungshub auf $C_{weight}$ weiter reduziert und damit die Integration von mehr Lernereignissen auf $C_{weight}$ ermöglicht. Zum anderen ergibt sich eine geringere Kennlinienstreuung durch die enge räumliche Kopplung von Eingangs- und Ausgangstransistor. Um ein festes Spiegelverhältnis und eine möglichst wenig variierende Last für die PSC-Schaltungen zu gewährleisten, sind die Eingangstransistoren der Stromspiegel fest mit allen sechs Stromleitungen verbunden, es werden nur die Ausgangstransistoren über SOURCE_SEL geschalten.

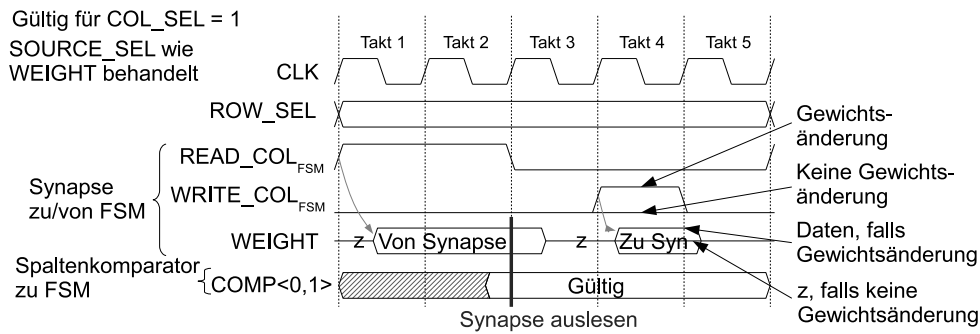

Abbildung 5.11: Taktdiagramm der Synapse sowie der Kommunikation zwischen Synapse und Zustandsautomat (FSM), inklusive der Signale der unter jeder Synapsenspalte liegenden Komparatoren.

Der in Abbildung 5.11 dargestellte Zyklus für das Lesen und optionale Schreiben der Synapse benötigt fünf 100 MHz-Takte. Zu Anfang wird mit dem über den Zeilendekoder generierten ROW_SEL die Reihe angesprochen und mit READ_COL$_{FSM}$ die Gewichtsbits WEIGHT$<x>$ und das SOURCE_SEL auf den bidirektionalen Bus Richtung FSM gegeben sowie die Gewichtsspannung über den Sourcefolger auf den Bus Richtung Komparatoren gelegt[10]. Nach zwei Takten (20 ns), die für das Einschwingen der Digitalsignale und besonders der Komparatoren veranschlagt sind, werden WEIGHT$<x>$, SOURCE_SEL und die Ausgangssignale der Komparatoren COMP$<0,1>$ von der FSM ausgelesen und in Takt drei das neue Gewicht berechnet bzw. entschieden, ob eine Gewichtsänderung stattfindet. In Takt 4 wird im Fall einer Gewichtsänderung das neue Gewicht geschrieben und $U_{weight}$ zurückgesetzt, Takt 5 dient der sicheren Beendigung des Schreibzustands der Synapse, bevor eine neue Synapse selektiert wird. Die oben angesprochene Zeitdauer, die zwischen zwei Auslesezuständen derselben Synapse in der Matrix vergeht, ist damit $\frac{1}{100\,MHz} \cdot 5 \cdot 32 \cdot 16 = 25.6\,\mu s$ (256 ms biologisch), d.h. schnell genug, um auch starke Stimulationsprotokolle (bei denen $U_{weight}$ innerhalb weniger (biologischer) Sekunden über den gesamten Dynamikbereich geführt wird) korrekt als Gewichtswirkung in der FSM zu verarbeiten und mithin auch schnell genug, um den flüchtigen Speicher $C_{weight}$ innerhalb des oben erwähnten Speicherzeitfensters auszulesen. Die Gewichtswirkung findet über einen 4-Bit strombasierten DA-Wandler statt, d.h. binär gewichtete Stromspiegel werden gemäß der WEIGHT$<x>$-Bits auf eine gemeinsame Sammelschiene $I_{weight\_out}$ geschalten. Im Fall eines präsynaptischen Pulses wird der zugehörige PSC dann gewichtet Richtung Neuronenmembran weitergegeben.

---

[10] Das SOURCE_SEL wird i.d.R. nur einmal zur Initialisierung gesetzt, jedoch zur Einsparung eines separaten Interfaces wie die Gewichtsbits immer mit ausgelesen und bei Änderungen des Gewichts auch geschrieben, allerdings im Unterschied zu WEIGHT$<x>$ mit seinem alten Wert. Die für die Synapse aus den Auswahlsignalen generierten internen Schreib- und Lesesignale $R_{syn}$ und $W_{syn}$ existieren in 1.8 V (für die digitalen Schaltungsteile) und in 3.3 V (analoge Schaltungsteile). Die WEIGHT-Bits im Strom-DAC benötigen keine Pegelwandler: Die Stromspiegel selbst sind zwar aus 3.3 V-Transistoren aufgebaut, die Gatespannungen sind jedoch so niedrig, dass als Schalter im Strom-DAC 1.8 V-Transistoren verwendet werden können.

## 5.5 Konzepte zur Implementierung von Metaplastizität

Aus neuromorpher Sicht ist zur Umsetzung von Metaplastizität jeder Parameter des MAPLE geeignet, der das Verhalten von Neuron, Synapse oder PSC beeinflusst und auf einer für Metaplastizität relevanten Zeitskala angepasst werden kann (biologisch >20 min, d.h. technisch >100 ms). Es wurde bei der Implementierung Augenmerk darauf gelegt, dass alle metaplastisch relevanten Parameter im MAPLE mit JTAG zur Laufzeit konfiguriert werden können. Bei bis zu 10 MHz möglicher JTAG-Taktfrequenz kann selbst bei entsprechendem JTAG-Overhead jeder einzelne Parameter auf Zeitskalen von 1-10 $\mu s$ (biologisch 10-100 ms) angepasst werden, für den gesamten MAPLE liegt die Konfigurationszeit bei ca. 1 ms. Es wurde deshalb im MAPLE auf eine explizite Implementierung von Metaplastizitätsmechanismen verzichtet, da diese ohnehin in ihrer genauen Form meist nicht bekannt sind (Abraham, 2008). Metaplastische Änderungen können aus den vielfältigen von außen messbaren dynamischen Größen deshalb außerhalb des MAPLE flexibel in Software abgeleitet werden und damit wiederum über verschiedenste Neuronen-, Synapsen- oder PSC-Einstellungen auf den entsprechenden Zeitskalen das neuromorphe Verhalten angepasst werden.

Der offensichtlichste Eingriffspunkt für Metaplastizität ist der für jedes postsynaptische Neuron spezifische Spannungsschwellwert $\Theta_U$, für dessen Pendant im BCM-Kontext eine langsame Nachführung nach der mittleren Rate des Neurons postuliert wird (Bienenstock et al, 1982). Mit dieser Art von Metaplastizität können beispielsweise Phänomene binokularer Dominanz oder eine Anpassung des Lernverhaltens an die mittlere Populationsaktivität erklärt werden (Abraham, 2008; Bienenstock et al, 1982). Die 8-Bit DACs, welche die $\Theta_U$ bereitstellen, erhalten ihr Referenzsignal von einer zwischen VDDA und GND liegenden Kette von 256 Polysilizium-Widerständen. Der mittlere Abgriff (Bitstand 10000000, d.h. auf der Hälfte der Widerstände) kann auf $U_{rest}$ gelegt werden, um definiert mit allen $\Theta_U$ eine Einstellung zu erreichen, bei der das LCP Grundmodell vorliegt (d.h. $\Theta_U = U_{rest}$). Die einzelnen Ausgänge bestehen aus einem Dekodierbaum und Transfergates, welche die dem jeweiligen Bitstand entsprechende Analogspannung an der Widerstandskette abgreifen und über ein RC-Glied (zur Unterdrückung von Transienten und von Crosstalk) ungepuffert weitergeben. Die ungepufferte Variante wurde gewählt:

- um Leistung einzusparen
- da die Änderungsgeschwindigkeit für die $\Theta_U$ nicht sehr hoch sein muss
- da nur statische Lasten (die Gates der differentiellen Paare in der Synapsenmatrix) betrieben werden

Die Zustände der DACs können im JTAG-Automaten einzeln oder blockweise geschrieben werden. In der für den MAPLE gewählten Dimensionierung der Widerstandskette und der RC-Tiefpässe wird ein MSB-Sprung innerhalb von 5 $\mu s$ am Ausgang bis auf ein LSB genau angezeigt. Eine Überwachung der ohnehin nach aussen gemeldeten Ausgangspulse der Neuronen, Tiefpassfilterung und LSB-weise Nachstellung über die DACs reichen demnach schon aus, um einen metaplastischen gleitenden Schwellwert nachzustellen. Alternativ können die $\Theta_U$ auch ohne Bezug zur Pulsrate verändert werden, um beispielsweise Experimente wie in Abbildung 3.8 nachzustellen.

Neben den $\Theta_U$ stellt der DAC auch die drei Parameterspannungen für die postsynaptische

Adaption bereit[11]. Metaplastische Anpassung der postsynaptischen Dynamik steht in der Forschung nicht so stark im Vordergrund wie der oben angeführte gleitende Schwellwert, jedoch finden sich Hinweise, dass beispielsweise die postsynaptische 'Spike Frequency Adaptation' ebenfalls von Metaplastizität beeinflusst wird (Le Ray et al, 2004). Generell sind im dynamischen Verhalten des Neurons Zeitkonstanten sichtbar, die für deterministische Veränderungen in Neuronen weit jenseits der üblichen Neuronenzeitskalen sprechen (Gal et al, 2010), d.h. verschiedene Aspekte des Neuronenverhaltens können als Metaplastizität interpretiert werden. Diese Effekte wären über entsprechende Anpassung der Neuronenadaption ebenfalls in Hardware testbar.

Für Untergruppen der in den Neuronen und PSCs bereitgestellten Biasströme, insbesondere solcher, die zur Konfiguration von Zeitkonstanten verwendet werden, wurden individuelle 4-Bit Einstellmöglichkeiten vorgesehen, d.h. es ist beispielsweise möglich, jede einzelne Membranzeitkonstante nachzujustieren. In diesem Zusammenhang ist ein weiterer interessanter metaplastischer Aspekt des Neuronenverhaltens die Änderung der AHP (siehe Abschnitt 4.2), welche in Zellkulturen, aber auch bereits am lebenden Tier nachgewiesen wurde (Disterhoft and Oh, 2006; Zelcer et al, 2006). Ein expliziter Mechanismus hierzu wurde in den MAPLE-Neuronen nicht vorgesehen, aber es kann, wie aus Abbildung 4.1B+C ersichtlich, diese verlängerte Refraktärszeit durch eine größere Refraktärsamplitude oder eine erhöhte Membranzeitkonstante emuliert werden, wobei der resultierende Effekt auffallend dem des gleitenden Schwellwertes ähnelt.

Über diese individuell justierbaren Stromspiegel kann auch die Stärke und Zeitkonstante der präsynaptischen Adaption eingestellt werden. Es gibt einzelne Hinweise, dass auch diese Konfigurationsmöglichkeit zur Emulierung von Metaplastizität eingesetzt werden kann. So wurde etwa in (Goussakov et al, 2000) eine Langzeitanpassung der PPD an bestimmten Synapsen festgestellt. Die Autoren postulieren dass auch weitere, beispielsweise im Quantalmodell enthaltene, präsynaptische Verhaltensweisen einer metaplastischen Anpassung unterliegen.

## 5.6 Ergebnisse

Abbildung 5.12 zeigt das Layout des MAPLE mit den Baugruppen entsprechend Abbildung 5.1 und zusätzlichen Teststrukturen. Im Folgenden werden Simulationsergebnisse dieses Entwurfs vorgestellt. Wie bereits am Anfang dieses Kapitels vermerkt, arbeiten alle Schaltungen um den Faktor $10^4$ schneller als biologische Echtzeit (vergleiche Zeitachsen in Abbildung 5.15). Die in diesem Abschnitt dokumentierten Simulationsergebnisse sind jedoch zur leichteren Vergleichbarkeit mit biologischen Messdaten mit Ausnahme von Abbildung 5.15 in die biologische Zeitskala umgerechnet.

Abbildung 5.13 und 5.14A zeigen einige Simulationsergebnisse der PSC-Schaltung aus Abschnitt 5.3 für typische biologische präsynaptische Pulsfolgen. Da bei der biologischen Messung i.d.R. nicht der PSC-Strom, sondern das PSP ausgewertet wird (siehe Kapitel 1), wurde die resultierende PSC-Wellenform zur besseren Vergleichbarkeit auf ein einfaches LIAF-Neuron (d.h. eine RC-Parallelschaltung) integriert. Wenn nur die präsynaptische Adaption aktiv ist (d.h. bei 'Nächster Nachbar' Modus), kann die PSC-Schaltung in guter Näherung

---

[11]Da $U_{\alpha att}$ dynamisch kapazitiv belastet wird, wird diese Spannung als einzige des DACs aktiv gepuffert.

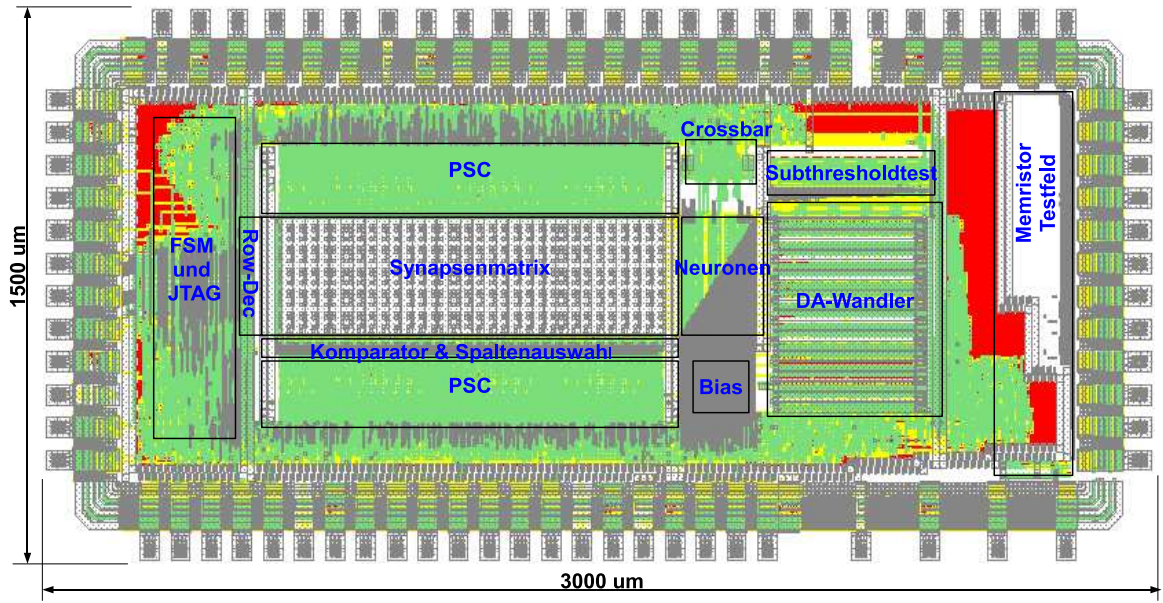

Abbildung 5.12: Layout-Tapeoutbild mit Baugruppenmarkierung (vergleiche Abbildung 5.1). Zusätzlich zu den bereits beschriebenen Baugruppen enthält der MAPLE ein Feld mit Testschaltungen für zukünftige neuromorphe ICs in Echtzeit ('Subthresholdtest') und Teststrukturen für eine Memristornachprozessierung.

von (Froemke et al, 2006) 'Paired Pulse Depression (PPD)' reproduzieren, vergleiche Abbildung 5.13. Zur Überprüfung der Robustheit wurden Simulationen über die statistische Streuungsbandbreite der Transistoren durchgeführt. Wie erkennbar beeinflussen diese die Reproduktion der biologischen Versuchsdaten nur unwesentlich.

Wie in Kapitel 2 ausgeführt wurde, zeigt sich Kurzzeitplastizität auch sehr charakteristisch im typischen Zeitverlauf einer Reihe von PSPs (siehe Abbildung 5.14). Damit kann die biologische Relevanz der vorgestellten Schaltungsimplementierung von synaptischer Kurzzeitplastizität auch durch einen Vergleich der PSP-Spannungskurven aus Abbildung 5.14A mit den vergleichbaren Kurven aus Abbildung 1 und 4 in (Markram et al, 1998) (wiedergegeben in Abbildung 5.14B) nachgewiesen werden. Diese Art von Kurzzeitplastizität ist auch ein notwendiger Bestandteil von bestimmten Formen der Langzeitplastizität (Froemke et al, 2006).

Hinsichtlich der Verwendung der PSC-Schaltung im Rahmen der Langzeitplastizität zeigt Abbildung 5.15A(oben) die sehr gute Übereinstimmung der von der PSC-Schaltung bereitgestellten Wellenform mit der idealen Formulierung der LCP-Regel in Abbildung 5.15B(oben). Die Transienten auf der PSC-Wellenform kommen durch das Zuschalten des AP-skalierten PSC zustande (siehe Abschnitt 5.4). Die von der Neuronenschaltung aus Abschnitt 5.2 bereitgestellte Membranspannung stimmt ebenfalls gut mit dem beabsichtigten Zeitverlauf überein. Am Verlauf der Membranspannung in Abbildung 5.15A(mitte) ist deutlich die Verkürzung konsekutiver AP-Flächen durch die postsynaptische Adaption zu erkennen, was durch die Verwendung von Diracimpulsen in Abbildung 5.15B(mitte) verdeckt wird. Die auf diesen Wellenformen basierende Gewichtsberechnung mittels der

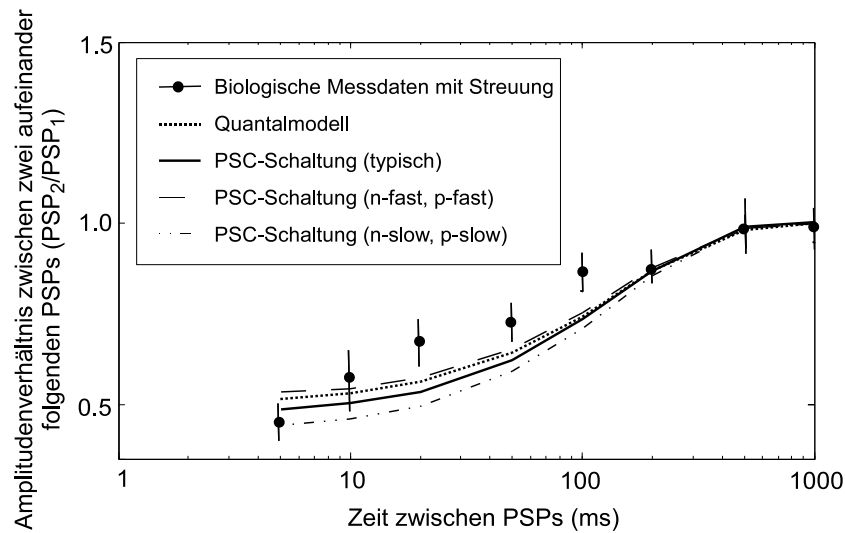

Abbildung 5.13: Simulationsergebnisse für das Versuchsprotokoll aus (Froemke et al, 2006), d.h zwei konsekutive PSCs mit unterschiedlichen Zeitdifferenzen. Gezeigt ist die Abschwächung des zweiten PSCs relativ zum ersten durch die präsynatische Adaption, aufintegriert auf ein einfaches LIAF-Neuron (d.h. es wurden die PSP-Amplituden ausgewertet), mit $\tau_{mem} = 33\,\mathrm{ms}$, $\tau_{rec} = 22.3\,\mathrm{ms}$, $\tau_g = 13.5\,\mathrm{ms}$. Im Diagramm sind die Kurven für typisches (nominales) Transistorverhalten sowie für den $6\sigma$ Streuungskorridor enthalten. Zusätzlich aufgeführt sind die zugehörigen biologischen Messdaten sowie die Simulation des idealen Quantalmodells aus Abbildung 2.8.

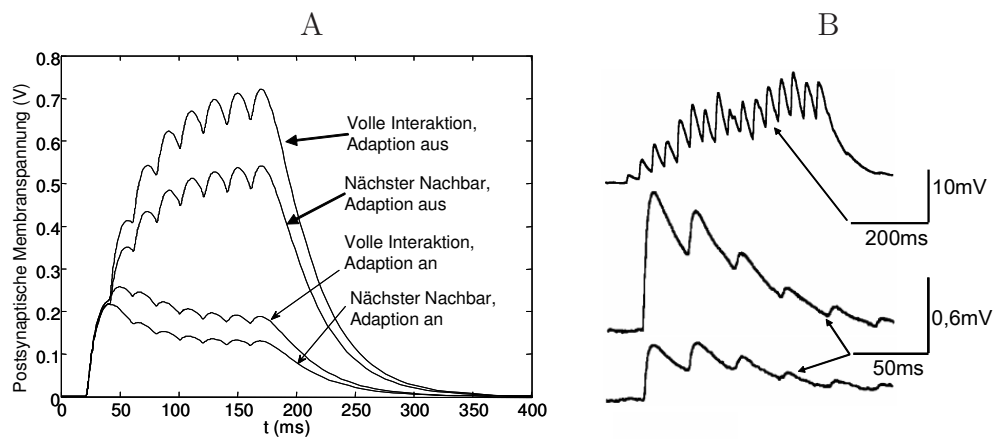

Abbildung 5.14: (A) Simulation von 8 PSCs im Abstand von 20ms (als aufintegrierte PSP-Kurven), $\tau_{mem} = 33\,\mathrm{ms}$, $\tau_{rec} = 1\,\mathrm{s}$, $\tau_g = 13.5\,\mathrm{ms}$; (B) Vergleich mit an Pyramidenneuronen des Neokortex aufgenommenen typischen PSP-Kurven, adaptiert aus (Markram et al, 1998)

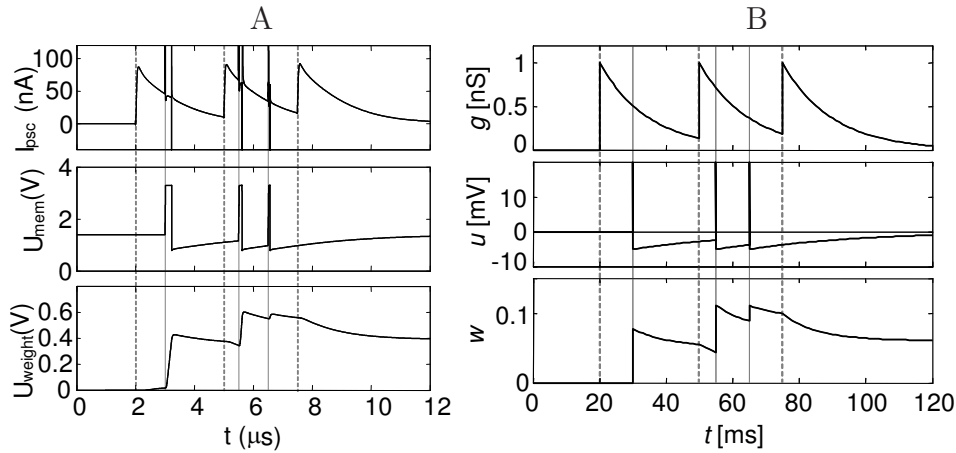

Abbildung 5.15: (A) PSC- und Membranspannungsverlauf für eine zufällige Pulsfolge sowie resultierende synaptische Gewichtsspannung (Schaltungszeit); (B) Vergleich mit idealer Version (nochmals aufgeführt aus Abbildung 3.4).

Synapsenschaltung aus Abschnitt 5.4 erreicht eine sehr gute Reproduktion von Gleichung 3.2 (vergleiche die Spannung auf $C_{weight}$ in Abbildung 5.15A(unten) mit $w$ in Abbildung 5.15B(unten)).

Die Experimentnummern im Folgenden beziehen sich auf die in Abschnitt 3.1 eingeführten Standardexperimente. STDP-Verhalten [1] (vergleiche Abbildung 3.1C) wird für die aus PSC-Schaltung, Neuron und Synapsenschaltung bestehende Implementierung der LCP-Regel in Abbildung 5.16A gezeigt[12]. Auch hier wurden Simulationen über die Streuungsbandbreite der Transistoren durchgeführt. Wie zu sehen, wird die STDP-Kurve robust reproduziert, v.a. die Zeitfenster unterliegen beinahe keiner Streuung. Die Absolutwerte der Gewichtsänderung hingegen unterliegen Schwankungen um den Faktor 4. Dies kann allerdings leicht durch Parametereinstellungen des entsprechenden Neurons bzw. PSCs kompensiert werden. Da die Kurven ausserdem die $6\sigma$ Streuungsbandbreite des Produktionsprozesses darstellen, dürften die zu erwartenden tatsächlichen Streuungen über einen einzelnen MAPLE deutlich niedriger liegen. In Erweiterung bisheriger rein auf STDP fokussierter CMOS-Implementierungen von Plastizität können, wie bereits für die theoretische Form der LCP-Regel gezeigt, mit der vorgestellten Schaltungsimplementierung auch nichtlineare Puls-Raten-Interaktionen wie etwa das bekannte Triplet-Paradigma (Froemke and Dan, 2002) nachgebildet werden (siehe Abbildung 5.16B). Triplets stellen einen 'harten' Testfall für das korrekte Funktionieren der postsynaptischen Adaption dar (vergleiche Abbildung 3.7). Insbesondere ist damit gezeigt, dass neben der reinen Adaptionsschaltung auch die Zuschaltung der zusätzliche PSC-Stromquelle in der Synapsenschaltung während eines AP (vergleiche Abbildung 5.9) hinreichend schnell und genau genug ist, um den Triplet-Effekt nachzubilden.

Ein weiteres Beispiel für ein derartiges kombiniertes Raten- und Pulsprotokoll ist frequenzabhängiges STDP [2], für die LCP-Schaltung gezeigt in Abbildung 5.17A. Als ein Beispiel für BCM-/Ratenplastizität wurde das korrelierte Ratenprotokoll [7] (Sjöström et al, 2001) mit der Schaltung simuliert, gezeigt in Abbildung 5.17B. Vergleichbar mit Abbildung 3.6 wurde zur Unterdrückung der Streuungen im Ergebnis ebenfalls über mehrere Durchläufe gemittelt. Allerdings konnten innerhalb vertretbarer Simulationszeit nur 20 Experimentinstanzen durchgeführt werden, wodurch die Kurve noch deutliche Streuungen aufweist. Um in einer derartigen Kurve den Effekt verschiedener $\Theta_U$-Einstellungen nicht im stochastischen Rauschen zu verlieren, wurden dieselben 20 Versuchsinstanzen für beide $\Theta_U$'s verwendet, wodurch die ähnlichen 'Stufen' in beiden Kurven zustande kommen.

---

[12]Da die digitale Gewichtsspeicherung nur den Aussteuerbereich des Gewichts linear erweitert, die Spannung auf $C_{weight}$ aber die eigentliche Gewichtsberechnung repräsentiert, wird zum Vergleich mit den biologischen Messdaten als Gewichtsvariable diese Spannung, d.h. $U_{weight}$, dargestellt.

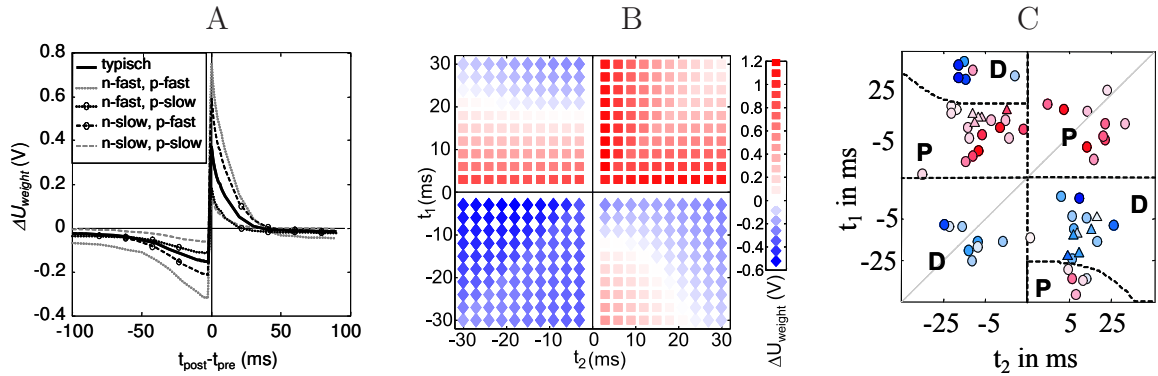

Abbildung 5.16: Ergebnisse für die Simulation verschiedener biologischer Plastizitätsexperimente mit der Schaltungsrealisierung der LCP-Regel, Gewichtsänderung als Spannungsdifferenz auf $C_{weight}$ vor und nach dem Experiment; (A) STDP [1], simuliert für verschiedene Prozessvariationen; (b) Triplets [3], Ergebnisdarstellung ähnlich wie in (Froemke and Dan, 2002); (C) zum Vergleich: biologische Messdaten aus (Froemke and Dan, 2002)

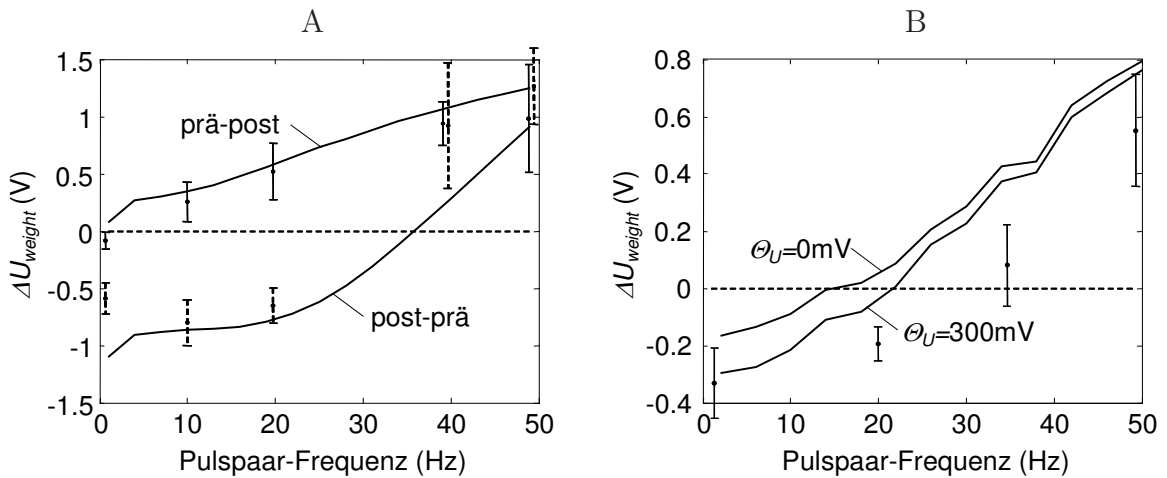

Abbildung 5.17: Weitere Plastizitätsexperimente aus (Sjöström et al, 2001): (A) frequenzabhängiges STDP [2] (B) korreliertes Ratenprotokoll [7], Mittelung über 20 Experimente; beide Darstellungen enthalten die biologischen Messdaten in Form der Fehlerschranken.

Von Interesse ist hier besonders, dass in deutlicher Erweiterung des momentanen Standes

der Technik der metaplastische gleitende Frequenzschwellwert der BCM Theorie (Bienenstock et al, 1982) gut nachgebildet wird (d.h. der Punkt, bei dem mit zunehmender Frequenz ein Wechsel von LTD zu LTP stattfindet). Wie aus Abbildung 5.17B ersichtlich, ist der Frequenzschwellwert für die biologischen Versuchsergebnisse etwas höher, was sich aber durch eine höhere Einstellung von $\Theta_U$ jederzeit kompensieren ließe. Mithin konnte gezeigt werden, dass die in den letzten Abschnitten diskutierten Einzelschaltungen in ihrer Verwendung als Gesamtplastizitätsschaltung die LCP-Regel sehr gut abbilden und damit komplexe Puls- und Ratenplastizitätsphänomene in VLSI nachgebildet werden können.

## 5.7 Ausblick 1: Auslesen von Attraktordynamiken mit Delta-Sigma-Modulatoren

In (Ellguth et al, 2009) und (Mayr et al, 2010d) wird ein Analog Frontend (AFE) vorgestellt, bestehend aus analogem Vorverstärker und einem ebenfalls vorverstärkenden Delta-Sigma-Modulator (DSM) als Analog-Digital-Wandler (ADC). Ein DSM ADC besteht aus einem Quantisierer mit geringer Auflösung, der in einer rückgekoppelten Tiefpassschleife den Unterschied zwischen jeweils anliegendem Signal und dem momentanen Zustand des Quantisierers bewertet. Somit läuft das Quantisierersignal immer dem anliegenden Analogsignal nach, wobei der Quantisierungsfehler bzw. das durch die niedrige Auflösung bedingte Quantisierungsrauschen durch die Tiefpasscharakteristik der Schleife zu hohen Frequenzen verschoben wird. Ein nachgeschaltetes Digitalfilter unterdrückt das Quantisierungsrauschen und rekonstruiert aus dem hochfrequenten, gering quantisierten Digitalsignal ein höher aufgelöstes Signal mit niedrigerer Abtastrate. Das Hauptanwendungsgebiet für DSM-basierte ADCs ist die Wandlung von sich langsam und stetig ändernden Sensor- oder Audiosignalen (Norsworthy et al, 1997). In neuromorphen Schaltungen werden ADCs bis jetzt nur sehr selten eingesetzt, da die Hauptkommunikation zwischen Neuronen über quasi-digitale Pulse verläuft und vorhandene analoge Signale des Netzwerkes (z.B. Bias, synaptische Ströme) meistens direkt eingespeist und/oder ausgelesen werden (Giulioni, 2008). Allerdings gewinnen in den letzten Jahren sogenannte Attraktornetzwerke als Modell für kognitive Verarbeitung sehr stark an Popularität (Amit, 1992). Die Grundidee ist dabei, dass ein rückgekoppeltes Netzwerk eine selbststabilisierende Dynamik aufweist, bei der in langsamem Rhythmus zwischen verschiedenen Aktivitätszuständen gewechselt wird (Freeman, 2003). Das extern anliegende Signal steuert dabei die Trajektorie dieser Zustandsübergänge und damit die dynamische Verarbeitungsfunktion des Netzwerkes (Mejias and Torres, 2009). Zeitgleich mit der zunehmenden Popularität von Attraktornetzwerken in theoretischen und simulativen Untersuchungen gibt es Anstrengungen, diese auch in neuromorpher VLSI nachzubilden (Camilleri et al, 2010; Giulioni et al, 2009). In Attraktornetzwerken lässt sich allerdings in der Regel die Verarbeitungsfunktion nur grob über die (quasi-digitalen) Ausgangspulse des Netzwerkes charakterisieren, da die Attraktorzustände eher durch kollektive analoge Zustandsvariablen der Neuronenpopulation gekennzeichnet sind, welche eine genauere Analyse ermöglichen. Eine derartige kollektive Zustandsvariable ist etwa die mittlere Membranspannung einer Gruppe von Neuronen (Amit, 1992; Freeman, 2003; Volgushev et al, 2006)(siehe auch (Brüderle et al, 2011) für eine Gegenüberstellung von puls- und membranspannungsbasiertem Auslesen von Attraktordynamiken). Hier bietet

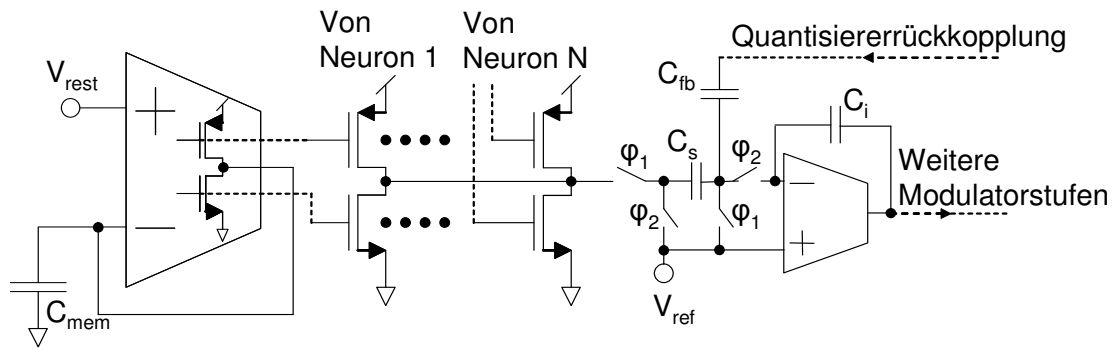

Abbildung 5.18: Blockschaltbild zur Gewinnung der mittleren Membranspannung und der nachfolgenden DSM-basierten analog-digital Wandlung zur populationsbasierten Analyse von Attraktordynamiken

sich ein bislang ungenutztes Anwendungsfeld für ADCs im Allgemeinen, wobei speziell das Auslesen der Dynamiken von Attraktornetzwerken sich sehr gut auf die Charakteristiken von DSM-basierten ADCs abbilden läßt:

- Die kollektive analoge Zustandsvariable ist i.d.R. ein kontinuierliches Signal ohne Sprünge (Galan et al, 2004), was dem kontinuierlichen wirkenden Filterverhalten eines DSM in Verbindung mit seinem Dezimationsfilter entgegenkommt (Norsworthy et al, 1997).
- Von einem DSM werden Gleichanteile im Signal fehlerbehafteter quantisiert als von konventionellen ADCs. Diese können jedoch bei der Analyse von Attraktordynamiken vernachlässigt werden (Galan et al, 2004).
- Hingegen kann ein DSM-basierter Wandler deutlich besser als konventionelle ADCs die genaue spektrale Zusammensetzung des Eingangssignals abbilden, was z.B. für die Charakterisierung des Netzwerkzustandes über die Populationsmembranspannung äußerst wichtig ist (Freeman, 2003).
- Ein DSM stellt insbesondere für die beim Auslesen benötigte mittlere bis hohe Auflösung (12-16Bit) eine energieeffiziente Alternative zu anderen Wandlerkonzepten dar (Murmann, 2008). Die maximale Wandlerrate ist zwar inhärent langsamer als bei Direktwandlern, für die Analyse von Attraktordynamiken jedoch völlig ausreichend, wie unten am konkreten Beispiel ausgeführt wird. Seine Energieeffizienz prädestiniert den DSM für Überwachungszwecke in künftigen implantierbaren neuromorphen Schaltkreisen, in die er ohne große zusätzliche Leistungsaufnahme integriert werden kann.

Für eine praktische Umsetzung dieser DSM-basierten Analyse von Attraktordynamiken auf einer zukünftigen Iteration des MAPLE muss zuerst ein Mittelwert der Membranspannung über eine Neuronenpopulation aufgebaut werden. Eine direkte Spannungsmittelung verbietet sich hier aufgrund der Schaltungskomplexität für die benötigte n-fache Spannungsaddition. Eine äquivalente Stromaddition ist demgegenüber trivial auszuführen, würde allerdings die zusätzliche Integration eines U/I-Wandlers ($V_{mem} \rightarrow I_{mem}$) für jedes Neuron voraussetzen. Eine einfachere Methode ist die Verwendung des von der Leckstromschaltung in Abbildung 5.3 vom synaptischen Strom abgezogenen Membranleckstromes, da dieser Strom ein direktes Maß dafür darstellt, wie weit die Membranspannung $V_{\text{mem}}$ positiv oder negativ von der Ruhespannung $V_{\text{rest}}$ abweicht. Wie aus der linken Hälfte von Abbildung 5.18 ersichtlich wird, besteht die Leckstromschaltung im Kern aus einem rückgekoppelten OTA, dessen gm als Widerstand wirkt (Koickal et al, 2007). Die Gatespannungen der beiden Stromquellentransistoren am Ausgang des OTA werden nun auf ein weiteres Paar baugleicher Transistoren gegeben, die zusammen mit den Pendants der anderen Neuronen an einem gemeinsamen Stromknoten hängen (siehe die Mitte von Abbildung 5.18). Der Gesamtleckstrom repräsentiert die mittlere Membranspannung für alle Neuronen oder (über eine konfigurierbare analoge Crossbar) für bestimmte Unterpopulationen. Der nachfolgende DSM wird nicht im üblichen spannungsabtastenden Modus betrieben (Mayr et al, 2010d), sondern stromgesteuert. Der Gesamtleckstrom verursacht während der Taktphase $\varphi_1$ auf der Abtastkapazität $C_s$ einen Spannungshub, der durch die Modulatorschleife und ein nachgeschaltetes Dezimationsfilter in ein digitales Ausgangssignal gewandelt wird. Da die Ladung auf $C_s$ während $\varphi_2$ jedesmal gelöscht wird, ergibt ein gleichbleibender Leckstrom einen konstanten Spannungshub und damit ein konstantes ADC-Ergebnis. Zu beachten ist nur, dass das vom DSM gelieferte Signal invertiert ist, d.h. wenn $V_{\text{mem}}$ über $V_{\text{rest}}$ liegt, führt dies zu einem Stromfluß durch den NMOS-Transistor des Leckstrom-OTA, und damit zu einem negativen Signal auf $C_s$.

Das in Mayr et al (2010d) vorgestellte Konzept für eine variable Vorverstärkung im DSM eignet sich dabei insbesondere, um unterschiedlich große Unterpopulationen bzw. Attraktoren mit wenig mittlerer Aktivität (d.h. geringem Signalpegel) abzubilden bzw. zuverlässig auflösen zu können. Somit kann der Vollausschlag des DSM optimal an den zu erwartenden Dynamikbereich der Amplitude des Attraktors angepasst werden.

Aus Freeman (2003); Galan et al (2004); Volgushev et al (2006) ist ersichtlich, dass das Membranpotential eine langsame Evolution mit ca. 0.5-10 Hz von Zustand zu Zustand durchläuft. Mithin wäre unter Berücksichtigung eines Faktors 2 (Nyquist) und des Beschleunigungsfaktors $10^4$ des MAPLE je nach Attraktordynamik eine Wandlerrate von 10-200 kSamples/s zur Auflösung der einzelnen Attraktoren nötig. Die momentanen Kenndaten des DSM sind 20 kSamples/s bei einer Auflösung von 14-16 Bit (Mayr et al, 2010d), womit der niedrigfrequente Bereich der Attraktordynamiken erreichbar wäre. Da zur Analyse des Attraktorzustandes jedoch vermutlich nur eine deutlich geringere Auflösung des Wandlers vonnöten ist (Volgushev et al, 2006), kann über eine Verringerung der Überabtastung des DSM Auflösung gegen Wandlerrate ausgetauscht werden. Bei einer Verringerung der Auflösung von 16 Bit (96 dB SNR) auf 10 Bit (60 dB SNR) kann aufgrund der Struktur des DSM in Mayr et al (2010d) (Rauschunterdrückung vierter Ordnung) die Überabtastung von momentan Faktor 64 auf Faktor 9 verringert werden. Damit wäre eine 7-fach höhere Abtastrate möglich, d.h. 140 kSamples/s (bei gleichbleibendem Systemtakt, Wandlerstruktur und Leistungsaufnahme).

Wenn zusätzlich leichte Designänderungen an dem DSM aus (Mayr et al, 2010d) vorgenommen werden, ist noch eine weitere Erhöhung der Samplerate möglich. Da aufgrund der gesunkenen Anforderungen an die Auflösung auch das kTC-Rauschen der Abtast- und Integrationskapazitäten im DSM höher ausfallen kann, ist es möglich, diese Kapazitäten um einen Faktor 8 zu verkleinern. Dadurch kann bei ansonsten identischer Wandlerstruktur und Auslegung der OTAs in den Integratoren der Systemtakt um ebenfalls einen Faktor 8 höher gewählt werden, wodurch eine Wandlerrate von 1MSample/s möglich ist, mithin können auch sehr hochfrequente Wechsel zwischen verschiedenen Attraktorzuständen aufgelöst werden. Der DSM aus (Mayr et al, 2010d) könnte zudem leicht mit der adaptiven Bias aus (Ellguth et al, 2009) umgerüstet werden, wodurch die in (Mayr et al, 2010d) verwendeten Klasse A OTAs in Klasse AB umgewandelt werden. Damit besteht eine einfach umsetzbare Möglichkeit zur Geschwindkeitserhöhung des DSM bei gleichzeitiger Reduktion der Leistungsaufnahme.

## 5.8 Ausblick 2: Anwendung von ADPLLs in neuromorphen Schaltkreisen

Phase-Locked-Loops (PLL) werden in HF- und Digitalschaltungen zur Takterzeugung, -vervielfältigung und -synchronisation eingesetzt. Sie bestehen aus einem in seiner Phasenlage und Frequenz regelbaren Oszillator und einem Phasenfilter, das den Oszillator so nachregelt, dass die Phasenlage des Oszillators (oder eines daraus über einen Frequenzteiler gewonnenen Signals) der eines Referenzsignals entspricht. In der Regel ist der Frequenzteiler einstellbar, so dass eine der Hauptanwendungen von PLLs in der Synthese eines einstellbaren hochfrequenten Signals aus einem festen niedrigfrequenten Takt liegt. PLLs enthalten in der Regel analoge Bauelemente, etwa das Schleifenfilter, und werden deshalb im Handentwurf an die jeweiligen Anforderungen der zu versorgenden Baugruppen hinsichtlich Jitter, Einschwingverhalten, etc. angepasst. Um den nötigen Aufwand bei einem Technologiewechsels bzw. einer Änderung des PLL-Designs möglichst gering zu halten, verwendet eine Untergruppe der PLLs nur digitale Gatter (All-Digital-Phase-Locked-Loop, ADPLL). Der Oszillator wird dabei über eine abgreifbare rückgekoppelte Verzögerungskette (Ringoszillator) aus Digitalgattern aufgebaut. Das Phasenfilter besteht in der Regel aus einer Torschaltung, welche den Zeitpunkt einer Flanke des Signals des Ringoszillators mit dem der Flanke des Referenztaktes vergleicht (Shu and Sinchez-Sinencio, 2005).

In (Eisenreich et al, 2009) wird eine ADPLL-Architektur vorgestellt, welche die für eine bestimmte Frequenz nötige Kettenlänge über eine sukzessive Approximation, d.h. eine Intervallschachtelung der möglichen Kettenlängen, erreicht. Dadurch ist die Einschwingzeit gegenüber vergleichbaren Konzepten stark reduziert (Eisenreich et al, 2009). Im eingerasteten Zustand wechselt eine Zähleinheit nach einem bestimmten Verhältnis zwischen den der vorgegebenen Frequenz nächstliegenden Kettenlängen ab, um den Langzeitmittelwert der Ausgangsfrequenz möglichst nahe an die gewünschte Frequenz zu bringen. Das Schaltverhältnis zwischen beiden Kettenlängen wird in der ADPLL aus (Eisenreich et al, 2009) ebenfalls über sukzessive Approximation ermittelt.

Neben der o.a. Frequenzsynthese werden ADPLLs in verteilten Digitalsystemen häufig zur Taktsynchronisation eingesetzt, d.h. es wird ein niedrigfrequentes Taktsignal über die einzelnen Untersysteme verteilt, aus dem in jedem Untersystem das eigentliche (höherfrequente) Taktsignal gewonnen wird. Somit lässt sich eine globale Synchronisation erreichen, unter Vermeidung der Schwierigkeiten, die eine direkte Verteilung des eigentlichen (Hochfrequenz-)Taktsignal verursachen würde. Eine der Anwendungen von ADPLLs in neuromorphen VLSI-Systemen dient ebenfalls diesem Zweck, da insbesondere bei pulsbasiertem Lernen global die Zeitabfolge zwischen Pulsereignissen eingehalten werden muss, d.h. die Zeitmarken von Pulsen müssen global synchronisiert sein. Beispielsweise wird eine

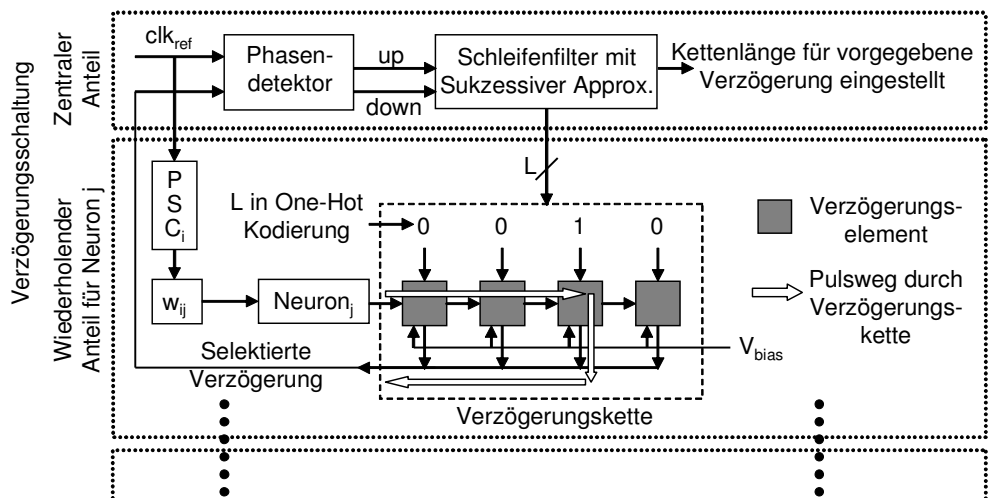

Abbildung 5.19: Blockschaltbild zum Einsatz des ADPLL-Prinzips aus Eisenreich et al (2009) zur Generierung einstellbarer neuronaler (dendritischer/axonaler) Verzögerungen.

Weiterentwicklung der ADPLL aus (Eisenreich et al, 2009) bereits in verschiedenen ASICs des Facets Waferscale Systems für die (Zeitmarken-)Synchronisation eingesetzt, etwa im DNC (Scholze et al, 2011a) oder im HICANN (Schemmel et al, 2008).

Eine weitere Anwendung von ADPLLs in neuromorphen Systemen ist die Erzeugung dieser Zeitmarken, d.h. ein hochfrequentes Ausgangssignal einer ADPLL wird in einem sogenannten Time-to-Digital-Converter (TDC) für die Digitalisierung von Pulszeitpunkten (Schemmel et al, 2007) eingesetzt. In einer zukünftigen hochskalierten Variante des MAPLE wird die Bandbreite des gegenwärtigen Scanprinzips (siehe Abschnitt 5.2) relativ zum Beschleunigungsfaktor der Neuronen i.d.R. unzureichend sein, um einzelne Pulsereignisse noch hinreichend genau aufzulösen. Deshalb ist dort eine Anwendung der ADPLL aus (Eisenreich et al, 2009) vorgesehen, die den TDC beispielsweise mit einer Ausgangsfrequenz von 250 MHz versorgt. Für einen Beschleunigungsfaktor von $10^4$ entspräche dies einer biologischen Zeitauflösung von 40 $\mu s$, ausreichend für eine hinreichend genaue Auflösung der Pulsereignisse (Morrison et al, 2007). Verglichen mit dem momentanen Scanprinzip kann so mit digital codierten Pulsereignissen die vorhandene Bandbreite besser ausgenutzt werden (Scholze et al, 2010a).

In der Analyse dynamischer Vorgänge in neuronalen Netzen spielt die axonale und/oder dendritische Pulsverzögerung eine zentrale Rolle, etwa für Oszillationen (Deco et al, 2009), Populationssignalübertragung (Riecke et al, 2007), Attraktorverhalten (Meyer et al, 2008) oder als Verzögerungsleitung in der Verarbeitung von Zeitdifferenzen (Carr and Friedman, 1999). Die in einer ADPLL verwendete Kette aus Verzögerungselementen kann aufgrund ihrer Ähnlichkeit zu einer Verzögerungsleitung deshalb auch als unmittelbarer Bestandteil einer neuromorphen Schaltung verwendet werden. Abbildung 5.19 stellt eine mögliche Variante dar, basierend auf der ADPLL mit sukzessiver Approximation im Schleifenfilter aus (Eisenreich et al, 2009). Die Baugruppen Phasendetektor und Schleifenfilter können zentral verwendet werden, während die Verzögerungskette selbst und ihre Speichereinstellung in jedem Neuron ausgeführt ist. Mit $V_{\text{bias}}$ kann über stromlimitierte Inverter in den

Verzögerungselementen der grobe Arbeitsbereich der Kette eingestellt werden. In einer Kalibrierungsphase vor dem eigentlichen Experiment wird über clk$_{ref}$ als Taktsignal die zu konfigurierende Verzögerung angelegt, die über die gesamte Kette aus PSC, synaptischem Gewicht und Neuron wirksam werden soll (vergleiche Abbildung 5.1). Es wird periodisch ein PSC ausgelöst, der das Neuron zum Pulsen bringt und damit die Verzögerungskette aktiviert. Das entstehende periodische Signal wird über das Schleifenfilter bzw. die Länge der Verzögerungskette an clk$_{ref}$ angepasst.

Die einzelnen Verzögerungselemente sind über $V_{bias}$ typisch von 5 ns bis 200 ns durchstimmbar, mithin beträgt die biologisch äquivalente Zeitschrittauflösung bei einem Beschleunigungsfaktor von $10^4$ 50 $\mu$s bis 2 ms. Mit 64 Kettengliedern (d.h. 6 Bit Stellwort L) sind damit (biologisch äquivalente) Verzögerungszeiten von 3.2 ms bis 128 ms erreichbar. Es wird somit ein breiter Bereich an typischen biologische Verzögerungszeiten abgedeckt (Riecke et al, 2007). Durch die sukzessive Approximation kann eine Verzögerungskette mit 64 Elementen innerhalb von 6 Zyklen von clk$_{ref}$ konfiguriert werden (Eisenreich et al, 2009). Selbst im ungünstigsten Fall (d.h. längste Einzelverzögerung und gesamte Kettenlänge geschaltet) benötigt dieser Einstellvorgang nur 76.8 $\mu$s, d.h. es können bei z.B. einem Systemstart eine große Anzahl Neuronen innerhalb kurzer Zeit nacheinander auf ihre gewünschte Pulsverzögerung eingestellt werden.

Gegenüber einer globalen Einstellung der Pulsverzögerung beispielsweise nur über $V_{bias}$ (bei Nutzung einer festen Kettenlänge) hat der in Abbildung 5.19 gezeigte individuelle Einmessansatz mehrere Vorteile. Die Verzögerung wird über die gesamte Kette an neuromorphen Baugruppen eingestellt, wodurch fabrikationsbedingte Schwankungen z.B. bei der PSC-Generierung (Abschnitt 5.3), in der Komparatorverzögerung des Neurons (Abschnitt 5.2) oder in der Verzögerungskette selbst ausgeglichen werden können. Des weiteren ist die globale Verzögerungseinstellung über $V_{bias}$ stark nichtlinear, so dass bei einer rein vorwärtsgerichteten Einstellung ohne Nachjustieren über den Regelkreis die resultierende Gesamtverzögerung nur sehr unpräzise festgelegt werden könnte. Der zusätzliche Flächenbedarf für Verzögerungskette und Speicher für jedes Neuron beträgt ca. $30*20\mu m^2$. Dieser nicht unwesentlichen Flächenzunahme steht jedoch als Kompensation eine deutlich erweiterte neuromorphe Funktionalität gegenüber. Die individuelle Verzögerungseinstellung ermöglicht beispielsweise eine genaue Wahl der statistischen Verzögerungsverteilung, wie sie etwa in (Meyer et al, 2008) gefordert wird. Die biologische Akkuratheit einer Netzwerkimplementierung wird erhöht, da die Verzögerungen individuell anhand einer vorgegebenen Netzwerkgeometrie konfiguriert werden können (Deco et al, 2009). Für die beispielsweise im auditiven Kortex postulierte, auf der Verzögerung von individuellen Axonen aufbauende, Verarbeitung von Zeitdifferenzen (Carr and Friedman, 1999) können diese über die Verzögerungskette in feiner Abstufung axonspezifisch präzise festgelegt werden.

# 6 Diskussion

Die Modellierung immer komplexerer neurobiologischer Systeme und Verarbeitungsfunktionen bedingt eine zeitgleiche Größenzunahme der korrespondierenden Hardwarerealisierungen, d.h. der neuromorphen VLSI-Systeme[1]. Dies erfordert flächeneffiziente Schaltungsrealisierungen speziell der Synapsen, da Biologie-abgeleitete Topologien eine sehr viel größere Anzahl an Synapsen als an Neuronen verwenden (Binzegger et al, 2004), was Synapsen zum bestimmenden Faktor der gesamten IC-Komplexität macht (Mitra et al, 2009; Schemmel et al, 2007) (Noack et al, 2010). Zugleich soll die biologische Relevanz der synaptischen Lernfunktion idealerweise zunehmen, um mit dem fortwährend verfeinerten Verständnis der Biologen und Modellierer für kognitive Funktionen Schritt zu halten (Buesing and Maass, 2010; D'Souza et al, 2010; Lazar et al, 2007).

Diese widerstreitenden Zielsetzungen, d.h. eine komplexitätsreduzierende effiziente Schaltungsumsetzung und eine komplexitätserhöhende biologische Relevanz, wurden bisher beim Entwurf der Lernschaltungen nur wenig beachtet. Üblicherweise wird beim Entwurf zwischen zwei separaten Stufen unterschieden, d.h. es werden zuerst mathematische Modelle der verschiedenen an Synapsen gefundenen Lernvorgänge abgeleitet, danach Schaltungsumsetzungen dieser Abstraktionen entworfen (Koickal et al, 2007; Mitra et al, 2009). Dieser zweistufige Prozess führt aufgrund der unterschiedlichen Zielsetzungen von Modellierern und neuromorphen Chipdesignern zu relativ komplexen Schaltungen (Koickal et al, 2007; Schemmel et al, 2007)[2]. Ein weiterer zur Komplexität beitragender Faktor gegenwärtiger Schaltungsrealisierungen synaptischer Plastizität ist die Tendenz, ein Gesamtmodell aus einzelnen Baublöcken zusammenzustückeln, d.h. es wird beispielsweise die präsynaptische Kurzzeitplastizität aus einer Filtermodellierung entnommen, die aus prä- und postsynaptischen Mustern zu bildende Langzeitplastizität aus einem Lernmodell, und die postsynaptische Kurzzeitplastizität aus einer Neuronenmodellierung.

Demgegenüber stellt das in dieser Arbeit vorgestellte Konzept einen ganzheitlichen Entwurf dar. Es wird auf konsequente Optimierung eines mehrere Zeitskalen umfassenden Plastizitätsmodells sowohl hinsichtlich biologischer Relevanz als auch hinsichtlich des Schaltungsaufwandes gesetzt. Inspiration findet sich hierbei in der Beobachtung, dass biologische Synapsen ihre Plastizität auf den lokal an der Synapse verfügbaren Zustandsvariablen aufbauen (Koch, 1999), d.h. die Lernfunktion kann in der technischen Realisierung ebenfalls auf Basis von ohnehin vorhandenen Zustandsvariablen aufgebaut werden.

Beispielsweise kann, wie für die LCP-Regel in Abschnitt 3.3 ausgeführt wird, etwa der in einem neuromorphen System durch die Neuronenrealisierung bereits vorhandene charakteristische Spannungsverlauf an der Neuronenmembran direkt zum Aufbau einer Hälfte des klassischen STDP-Zeitfensters verwendet werden. Da diese Vorgehensweise näher am biologischen Vorbild ist als ein rein phänomenologischer Aufbau der Zeitfenster (Debanne et al, 1994; Lisman and Spruston, 2005), erhöht sich als Nebeneffekt zugleich die biologische Relevanz, d.h. es können z.B. Membranspannungsbeeinflussungen der STDP-Zeitfenster modelliert werden (Mayr and Partzsch, 2010). Ähnliche Argumentationsketten lassen sich für die postsynaptische Adaption sowie die Wellenform und die Adaption des präsynaptischen Stromes (PSC) bilden, d.h. in jedem Fall wird bereits vorhandene Funktionalität für die Bildung der Gesamtplastizität wiederverwendet und damit sowohl unnötige Schaltungsredundanz vermieden als auch die biologische Relevanz erhöht (Mayr et al, 2010a; Noack et al, 2010).

---

[1] Dieser Trend ist insbesondere während der letzten 10 Jahre zu beobachten, vergleiche beispielsweise die zunehmenden Systemgrößen jeweils innerhalb einer Systemfamilie in (Fusi et al, 2000) und (Camilleri et al, 2010) oder (Schemmel et al, 2004) und (Schemmel et al, 2010)

[2] Eine lobenswerte Ausnahme stellt hier die Fusi-Lernregel dar (Fusi et al, 2000; Indiveri et al, 2006). Allerdings wurde diese Regel zur Nachbildung von Informationsverarbeitung entworfen, nicht im Hinblick auf tatsächliche biologische Experimente wie die LCP-Regel (Brader et al, 2007).

Die in den Kapiteln 2, 3 und 4 eingeführten Plastizitätsbaublöcke sind jedoch nicht nur im Hinblick auf ihre Funktion innerhalb der Gesamtplastizität und innerhalb des MAPLE interessant. Beispielsweise wurde für das bisher nur in iterativer Form vorliegende Quantalmodell der präsynaptischen Kurzzeitplastizität (Markram et al, 1998) für modulierte Pulsfolgen eine geschlossene Übertragungsfunktion hergeleitet. Die in der Übertragungsfunktion manifestierte Vereinfachung gegenüber der ursprünglichen iterativen Formulierung ermöglicht eine weitaus vollständigere Exploration der Filtercharakteristiken als bisherige Ansätze (vergleiche Abbildung 2.5 bzw. (Mayr et al, 2009a) mit (Natschläger and Maass, 2001)). Aus der Übertragungsfunktion wurden zudem 'effektive' Zeitkonstanten $\tau_{u,\lambda}$ and $\tau_{R,\lambda}$ extrahiert, welche einen sehr viel direkteren Rückschluss auf das Verhalten des Quantalmodells zulassen als die originalen iterativen Parameter (siehe Abbildung 2.7).

Die LCP-Formulierung für Langzeitplastizität zeichnet sich durch einen hohen Grad an reproduzierten biologischen Plastizitätsphänomenen aus (Mayr and Partzsch, 2010), welcher auf einer Stufe mit neuesten, speziell für diesen Zweck entworfenen, Plastizitätsmodellen steht (siehe Tabelle 3.1). Aufgrund der kompakten Formulierung der LCP-Regel sind für diese Experimentreproduktion nur sehr wenige Parameter erforderlich (Mayr et al, 2010c), was den Bezug zwischen Parameteränderungen und dem LCP-Regelverhalten sehr einfach gestaltet, siehe Gleichung 3.9 und 4.4. Dies steht im positiven Kontrast zu dem großen Parameterraum und parameterabhängig ambivalenten Verhalten anderer Plastizitätsmodelle (Badoual et al, 2006; Shah et al, 2006). Wie bereits oben erwähnt, zeichnet sich die LCP-Lernregel zudem dadurch aus, dass sie sehr leicht mit anderen Formen der Plastizität kombiniert werden kann (siehe u.a. Abbildung 3.7 bzw. die Experimente in Abschnitt 3.4).

Die LCP-Regel beinhaltet keine direkten Mechanismen für Metaplastizität, d.h. für eine Modulation der Langzeitplastizität auf Zeitskalen von mehreren Stunden bis Monaten. Jedoch wird in Kapitel 4.1 diskutiert, dass LCP mehrere Parameter beinhaltet, die über einfache zeitliche Mittelwerte (vergleiche etwa Benuskova and Abraham (2007)) metaplastisch nachgeführt werden können. Die durch die Parameternachführung hervorgerufenen Änderungen am Lernverhalten stimmen gut mit am lebenden Tier gemessenen Zusammenhängen überein (Disterhoft and Oh, 2006; Lebel et al, 2001; Zelcer et al, 2006). Im Vergleich dazu sind insbesondere phänomenologisch begründete Plastizitätsmodelle generell nicht in der Lage, diese Korrespondenz herzustellen, da die dortigen Parameter (z.B. $A_+$ bei STDP) keine Entsprechung in biologischen Zustandsvariablen haben. Bei Neuronenkulturen sind zwei Ausdrücke von Metaplastizität experimentell gesichert, eine Verschiebung

der Spannungsschwellwerte in 'voltage clamp'-Experimenten (Ngezahayo et al, 2000) oder eine Änderung des Frequenzschwellwertes in Ratenexperimenten (Abraham et al, 2001). Die LCP-Regel bietet den Kontext, diese separaten Metaplastizitätserscheinungen zu vereinen, d.h. Spannungs- und Frequenzschwellwert sind mathematisch ineinander überführbar (Abbildung 4.1).

Für den MAPLE wurde eine flächen-/komplexitätsoptimierte Topologie entworfen. Diese erreicht zum einen eine gute Redundanzvermeidung durch Auslagerung gemeinsam genutzter Funktionalität in separate Baugruppen (z.B. präsynaptisch konzentrierte Rekonstruktion des PSC und der Adaption). Zum anderen werden die durch eine Konzentrierung von Funktionalität eingebüßten Konfigurationsmöglichkeiten durch die gewählte Topologie weitgehend kompensiert, d.h. es wird trotzdem eine beinahe optimale Konfigurierbarkeit hinsichtlich typischer Netzwerkmodelle erreicht (siehe Abbildung 5.2).

Die auf dem MAPLE beinhaltete Form des Quantalmodells ist neben ihrer Relevanz für die Langzeitplastizität auch dahingehend interessant, dass Modelle für neuronale Netze sich zunehmend der dynamischen Informationsverarbeitung widmen, die wiederum entscheidend von der Form und Komplexität der präsynaptischen Kurzzeitplastizität abhängt (Mejias and Torres, 2009; Sussillo et al, 2007). Allerdings stehen diesen theoretischen und simulativen Analysen bisher nur sehr wenige Schaltungsrealisierungen gegenüber (Camilleri et al, 2010). Dies ist vermutlich darauf zurückzuführen, dass die meisten neuromorphen Schaltungsrealisierungen nur Teilmengen und/oder stark vereinfachte Varianten der beispielsweise im Quantalmodell vorhandenen komplexen Dynamiken abbilden (Bartolozzi and Indiveri, 2006; Boegerhausen et al, 2003). Im Gegensatz dazu wurde in Abschnitt 2.4 eine Methode entwickelt, um die Beschreibungsgleichungen des Quantalmodells schaltungseffizient im MAPLE zu realisieren. Für den Einsatz neuromorpher Hardware in dynamischer Verarbeitung ist neben der Kurzzeitplastizität die Realisierung von kontrollierbaren dendritischen bzw. axonalen Verzögerungen vorteilhaft (Dahlem et al, 2009; Meyer et al, 2008). Eine zukünftige Erweiterung des MAPLE in dieser Richtung wird in Abschnitt 5.8 diskutiert. Desweiteren wird in Abschnitt 5.7 eine technisch vorteilhafte Methode zur Analyse dieser dynamischen Verarbeitung vorgestellt.

Insgesamt ist in dieser Arbeit ein wichtiger Schritt im Hinblick auf die Anwendung von biologienahen, zeitskalenübergreifenden Lerndynamiken in großflächigen neuromorphen Schaltungssystemen gemacht worden. Es wurde ein sehr robustes Lernverhalten implementiert (Abschnitt 5.6), welches als erste Schaltungsumsetzung signifikant über das STDP-Niveau hinausgeht, d.h. es können verschiedenste aktuelle Plastizitätsexperimente nachgestellt werden, etwa kombinierte Zeitpunkts- und Rateneffekte, spannungsbasiertes Lernen und multiple Formen von Metaplastizität. Die unterschiedlichen Zeitskalen des Lernverhaltens (Millisekunden bis Wochen in biologischer Zeit) operieren auf dem MAPLE kombiniert und zeitgleich in einer Form, die in der Literatur so nicht zu finden ist. Dabei ist die Schaltungsumsetzung dieses Lernens zugleich um den Faktor 2-3 flächeneffizienter gegenüber den im Vergleich mit deutlich geringerer Funktionalität ausgestatteten Umsetzungen in der Literatur (Koickal et al, 2007; Bofill-i Petit and Murray, 2004).

Bezüglich des praktischen Einsatzes des geschilderten Konzeptes ist die Professur 'Hochparallele VLSI-Systeme und Neuromikroelektronik' beispielsweise am EU-Projekt CORONET beteiligt, das zum Ziel hat, in CMOS biologienahe neuronale und synaptische Dynamiken zu realisieren und bidirektional mit biologischem Nervengewebe zu koppeln (z.B. für Neu-roprothesen). Der MAPLE stellt durch den hohen Grad an biologischer Detailtreue einen wichtigen Entwicklungsschritt hin zu diesem ambitionierten Ziel dar. Ein weiteres EU-Projekt mit Beteiligung der Professur, BrainScaleS, befasst sich mit der auf ganze Wafer skalierten VLSI-Umsetzung von zeitskalenübergreifenden kortikalen Verarbeitungsfunktionen. Durch die erreichbare hohe Packungsdichte und die multiplen Plastizitätszeitskalen ist der MAPLE bzw. das hier geschilderte Gesamtkonzept ebenfalls sehr gut für dieses Projekt geeignet. Die in BrainScaleS aus der Biologie abgeleitete nicht-von-Neumann Informationsverarbeitung auf multiplen Zeitskalen kann somit effizient auf CMOS übertragen und in Folge nutzbar gemacht werden.

# References: External